\documentclass[preprint]{aastex}
\usepackage{array,wrapfig}
\usepackage{graphicx} 
\usepackage{natbib}
\defcitealias{aarps}{Paper~I}
\usepackage{amsmath}
\usepackage{color}

\newcommand{\moments}[1]{${\cal M}$(#1)}
\newcommand{\nci}{{\tt NCI}}

\newcommand{\maxtss}{{\rm Max(TSS)}}

\newcommand{\BSS}{{\rm B\,S\,S}}
\newcommand{\ROCSS}{{R\,O\,C\,S\,S}}
\newcommand{\CCL}{C1.0+/24\,hr}
\newcommand{\MML}{M1.0+/24\,hr}
\newcommand{\CCS}{C1.0+/6\,hr}
\newcommand{\MMS}{M1.0+/6\,hr}

\newcommand{\CC}{C1.0+}
\newcommand{\MM}{M1.0+}

\received{}
\revised{}
\accepted{}
\submitjournal{ApJ}

\shorttitle{Flare-Imminent {\it vs.} Flare-Quiet Corona through Chromosphere II: NCI Results}
\shortauthors{Leka et al.}

\begin{document}

\title{Properties of Flare-Imminent versus Flare-Quiet Active Regions from the Chromosphere
through the Corona II: NonParametric Discriminant Analysis Results from the NWRA Classification 
Infrastructure (NCI)}

\correspondingauthor{K.~D.~Leka}
\correspondingauthor{Karin Dissauer}
\email{leka@nwra.com; dissauer@nwra.com}

\author[0000-0003-0026-931X]{K.D. Leka}
\affiliation{NorthWest Research Associates, 3380 Mitchell Lane, Boulder, CO 80301 USA}
\affiliation{Institute for Space-Earth Environmental Research, Nagoya University, \\
Furo-cho Chikusa-ku, Nagoya, Aichi 464-8601 JAPAN}

\author[0000-0001-5661-9759]{Karin Dissauer}
\affiliation{NorthWest Research Associates, 3380 Mitchell Lane, Boulder, CO 80301 USA}

\author[0000-0003-3571-8728]{Graham Barnes}
\affiliation{NorthWest Research Associates, 3380 Mitchell Lane, Boulder, CO 80301 USA}

\author[0000-0002-7709-723X]{Eric L. Wagner}
\affiliation{NorthWest Research Associates, 3380 Mitchell Lane, Boulder, CO 80301 USA}

\begin{abstract}

A large sample of active-region-targeted time-series images from the
Solar Dynamics Observatory / Atmospheric Imaging Assembly, 
the AIA Active Region Patch database (``AARPs'', Paper I: \citet{aarps}) is
used to investigate whether parameters describing the coronal, transition
region, and chromospheric emission can differentiate a region that will
imminently produce a solar flare from one that will not.  Parametrizations
based on moment analysis of direct and running-difference images provide for
physically-interpretable results from nonparametric discriminant analysis.
Across four event definitions including both 24\,hr and 6\,hr validity
periods, 160 image-based parameters capture the general state of the atmosphere,
rapid brightness changes, and longer-term intensity evolution.  We find
top Brier Skill Scores in the 0.07\,--\,0.33 range,  True Skill
Statistics in the 0.68\,--\,0.82 range (both depending on event definition),
and Receiver Operating Characteristic Skill Scores above 0.8.
Total emission can perform notably as can steeply
increasing or decreasing brightness, although mean brightness measures do
not, demonstrating the well-known active-region-size/flare-productivity
relation.  Once a region is flare productive, the active-region coronal
plasma appears to stay hot.  The 94\,\AA\ filter data provides the most
parameters with discriminating power, with indications that
it benefits from sampling multiple physical regimes.  In particular,
classification success using higher-order moments of running difference
images indicate a propensity for flare-imminent regions to display
short-lived small-scale brightening events.  Parameters describing the
evolution of the corona can provide flare-imminent indicators, but at
no preference over ``static'' parameters.   Finally, all parameters and
NPDA-derived probabilities are available to the community for additional
research.
\end{abstract}

\keywords{methods: statistical -- Sun: flares -- Sun: corona -- Sun: chromosphere}

\section{Introduction}\label{sec:intro}

In \citet[][hereafter Paper I]{aarps} we briefly introduce the goal for this study: to 
quantitatively characterize the brightness distributions, their temporal
variations and implied kinematics, and eventually a more complete physical 
state  of the chromosphere and corona, for two populations of 
solar active regions: those that are flare-productive on specified time-scales {\it vs.}
those that are not.  We are addressing this goal with a large sample
of data from the Atmospheric Imaging Assembly \citep[AIA;][]{aia_Lemen} 
onboard the Solar Dynamics Observatory \citep[SDO;][see Section~\ref{sec:data}]{sdo}.  There has not
yet been such a characterization in the context of flare productivity.
The approach we invoke explicitly avoids focusing on ``pre-flare''-specific  phenomena, and 
instead examines more general behaviors.

Recently, the dominant use of large-sample coronal image data in 
the context of solar energetic phenomena has been for 
machine learning tools to try and predict solar flares
\citep[][although see \citep{deft}]{Nishizuka_etal_2017,Jonas_etal_2018,Alipour_etal_2019}.
Generally, these statistical tools have not yet provided
``interpretable'' results in terms of a physics-based outcome,
although they have demonstrated some added classification success
when combining coronal data with, {\it e.g.}, photospheric magnetic
field data from the Helioseismic and Magnetic Imager \citep[HMI;][]{hmi,hmi_pipe,hmi_sharps}.

Case-study analyses of the pre-event solar corona have found evidence of loop
formation, energization and increased dynamic behavior \citep[``crinkles'';][and
references therein]{SterlingMoore2001b,Joshi_etal_2011,
SterlingMooreFreeland2011,ImadaBambaKusano2014}, an
increase in chromospheric non-thermal velocities and
high blueshifts \citep{Cho_etal_2016,Harra_etal_2013,
Woods_etal_2017,Seki_etal_2017}, very localized chromospheric
heating \citep{Li_etal_2005,Bamba_etal_2014}, and coronal dimming
\citep{ImadaBambaKusano2014,Zhang_etal_2017,QiuCheng2017} in the hours
prior to energetic events.

The present study attempts to do for the solar corona and chromosphere
what was done for the photosphere in a previous series of papers
\citep{params,dfa,dfa2,dfa3,nci_daffs}: test the ability to statistically 
differentiate between
flare-quiet and flare-imminent active regions through analysis of 
photospheric magnetic field data.  Here
we begin to test the same question but with a focus on the chromosphere, transition
region, and corona.  Guided by the previous series of papers, we use here
active regions as defined by the HMI Active Region 
Patches \citep[HARPs:][]{hmi_pipe} but now use time-series images of the upper solar
atmosphere in the UV and EUV (Section~\ref{sec:aarps}; see also \citetalias{aarps}).  
We introduce
human-constructed parametrizations (Section~\ref{sec:params}) designed to 
provide insights into the physical state of the upper atmosphere in
a manner parallel to what the ``SHARP parameters'' \citep{hmi_sharps} and especially
the extended parameter list examined in \citet{dfa3,nci_daffs} provide for the
photosphere \citep[see also][and references therein]{flarecast}.  Without focusing 
on forecasting {\it per se}, here we extend
insights gained by prior case studies to a large sample, to statistically
test (Section~\ref{sec:nci}) whether we can differentiate the state of 
active region atmospheres that are flare-imminent from those that are not.

Employing a large sample size provides a broad picture not only of the
standard workings of the corona over all sizes and activity levels of
active regions, but to what extent there is such a thing as standard
workings.  In other words, what is important for our understanding of the
Sun is not only the mean of some characteristics, but the more nuanced
nature of the distributions of those characteristics, their degree of
overlap, {\it etc}.  Here we quantify some characteristic behaviors
between defined groups, setting empirically-derived standards to which
models may then need to speak.

\section{The Data}
\label{sec:data}

The observational data used in this study are described in this section,
both the AIA timeseries data (Section~\ref{sec:aarps})
and the data used (Section~\ref{sec:goes}) to construct the solar flare event 
lists for analysis (Section~\ref{sec:events}).

\subsection{The AIA Active Region Patches (AARPs)}
\label{sec:aarps}

The AIA Active Region Patches (AARPs) database is described in full in \citet[Paper I;][]{aarps}.
Broadly speaking, they consists of curated UV- and EUV-image timeseries counterparts to the
photospheric magnetic field time-series data deployed in \citet{nci_daffs}.

The primary data source used in constructing the AARPs is
\textit{SDO}/AIA, supplemented with meta-data from 
the Helioseismic and Magnetic Imager \citep[HMI;][]{hmi,hmi_pipe}
{\tt hmi.Mharp\_720s} series.  The latter provides the coordinates and
bounding-box of the HMI Active Region Patches \citep[HARPs;][]{hmi_pipe},
which are the basis for defining the areas extracted from the AIA full-disk
images.  Of note, however, the AARP boxes are larger by 20\% than the HARP
definitions in order to accommodate the larger projected extent of 
the 3-D coronal structures, especially when a region is located near a limb,
and the bounding-box is extended further in the limb-direction to include 
the AR loops
(see Paper I for details).  There is no spatial binning applied to the images.

For each numbered HARP on any particular day, there is one corresponding
AARP consisting of seven hourly samples each containing 13\,min of
data sampled at 72\,s (11 images), across each of eight AIA bands.
To match the database of HMI vector
magnetic field extractions already in place at NWRA, the seven hourly
samples span 15:48 TAI -- 21:48 TAI. FITS files are produced for
each of seven EUV filters (94, 131, 171, 193, 211, 304, and 335 \AA), 
and the UV 1600\,\AA.
This approach provides information on both short-term and longer-term evolution
of all magnetic patches at chromospheric, transition region, and coronal heights and
temperatures.  The NWRA AARP database, which is available at the Solar
Data Analysis Center \citep{aarp_data} is
summarized in Table~\ref{tbl:AARPS};
here the number of samples is the total number of AARP datasets available
over the full date range.  The AARPs provide the data for
parametrization (Section~\ref{sec:params}), so the number of samples
in Table~\ref{tbl:AARPS} is the total
sample size available for statistical analysis for the present study.
There is no further down-selecting for AR size, complexity, location, or activity level.

\begin{table}[h!]
\begin{center}
\caption{Summary of AARP Data Set \label{tbl:AARPS}}
\begin{tabular}{|ccccc|}
\hline \hline
Date Range & AARP Range & NOAA AR Range & Number of ``AARP-Day'' Samples & Archive Size \\ \hline
06/2010 -- 12/2018 & 36 -- 7331 & 11073 -- 12731 & 32,067  & $\approx$ 9.5\,TB  \\\hline
\end{tabular}
\end{center}
\end{table}

\subsection{GOES Data and Source for Event Lists}
\label{sec:goes}

The event lists are constructed following \citet{nci_daffs}, using
events as recorded by NOAA using the Geostationary Operational
Environmental Satellite X-Ray Sensor \cite[``GOES''/XRS][]{goes_xrs}.
The dataset used is consistent with regards to flux calibration
\citep{2017AGUFMSH42A..06V,Machol_2022_PC}.  Only those events
associated with NOAA-assigned Active Regions are included.  
Flare lists based on GOES 1--8\AA\ peak emission from the GOES/XRS
sensors are available through either the National
Center for Environmental Information (NCEI) or by way of the ``edited event lists''
from NOAA/Space Weather Prediction Center.  In the present study
we used the latter by which to construct the event lists used 
(see Section~\ref{sec:events}).

\section{Analysis}
\label{sec:analysis}

The question posed here is, ``for solar active regions, are flare-imminent 
epochs distinguishable
from flare-quiet epochs on the basis of chromospheric and coronal
emission and kinematics?''  
Specifically we ask this using UV and EUV
intensity images and HMI-defined active regions, without the added 
benefit of spectroscopy \citep{PanosKleint2020},
but with the explicit use of time-series analysis \citep{Cinto_etal_2020}
in order to enhance physical interpretation of the results.  We answer the question
through statistical classification, multiple event definitions, and quantitative
metrics to evaluate how well the samples can differentiate the two populations.

\subsection{Parametrization}
\label{sec:params}

Parametrization allows both spatial and temporal information to be
summarized succinctly and in a manner conducive to physical interpretation
upon statistical analysis.  Moment analysis through the fourth moment
is used on the spatially-sampled target $x$: mean $\mu(x)$\footnote{To
avoid confusion, we use here $\mu(x)$ for mean(x) which breaks with our
previous use of $\overline{x}$; we also refer explicitly to the cosine
of the observing angle $\cos(\theta)$ without invoking $\mu$ in that
context.}, standard deviation $\sigma(x)$, skew $\varsigma(x)$, and
kurtosis $\kappa(x)$.  The lower-order moments capture bulk differences
whereas the higher-order moments are much more sensitive to subtle
differences in distribution wings, but are also more susceptible to
errors when image sizes are small.  The odd moments detect offsets or
asymmetries as related to a normal distribution, whereas the even moments
are sensitive to deviations in width or peakedness.  Previous research of
magnetic field distributions \citep{dfa,dfa2,dfa3,SWJ,nci_daffs} shows
the power of 3rd and 4th-order moments to capture subtle differences in
distribution {\it tails} that can signal significant, but very localized,
changes -- such as from a small emerging flux region.

The moment-analysis parametrizations produce a selection of {\it
intensive} variables that do not scale directly with active region size
\citep{Welsch_etal_2009};  these are complemented by {\it extensive}
parameters (such as totals over the field of view), which {\it do}
scale with region size.  It is important to note that the moment
analysis is not intended to provide a basis for image decomposition 
\citep{Raboonik_etal_2017} and as such, while the resulting parametrizations
may not be unique, they readily allow interpretation of the 
image intensity behavior.  The parametrization is applied to the images
by themselves, what we call the ``direct'' images (``I''), as well as the
running-difference images (``$\Delta$I'').  For this analysis, the parameters
target the following (for each wavelength separately, indicated by
``{\tt \_*}''):

\begin{list}{$\bullet$ }{
\setlength{\topsep}{0cm}   
\setlength{\partopsep}{0cm} 
\setlength{\itemsep}{0cm}   
\setlength{\parsep}{0cm}    
\setlength{\leftmargin}{2cm} 
\setlength{\rightmargin}{0cm}
\setlength{\listparindent}{0cm} 
\setlength{\itemindent}{-1.0cm} 
\setlength{\labelsep}{0cm}   
\setlength{\labelwidth}{0cm} 
}
\item The total brightness of an image, $\Sigma({\rm I_*})$, and of the running-difference 
image $\Sigma(\Delta{\rm I_*})$.
\item The moments of the brightness distribution \moments{${\rm I_*}$} which summarizes the 
mean $\mu(\rm I_*)$, 
standard deviation $\sigma({\rm I_*})$, 
skew $\varsigma({\rm I_*})$, and 
kurtosis $\kappa({\rm I_*})$.
\item The moments of the running-difference image distributions \moments{$\Delta{\rm I_*}$}, which
summarizes individually the 
mean $\mu(\Delta{\rm I_*})$, 
standard deviation $\sigma(\Delta{\rm I_*})$, 
skew $\varsigma(\Delta{\rm I_*})$, and 
kurtosis $\kappa(\Delta{\rm I_*})$.
\item The cosine of the central observing angle $\cos(\theta)$; this is essentially used as a 
control since flare activity should not have preferred locations.
\end{list}

\begin{figure}[t]
\centerline{
\includegraphics[width=0.4\textwidth, clip, trim = 0mm 0mm 14mm 10mm]{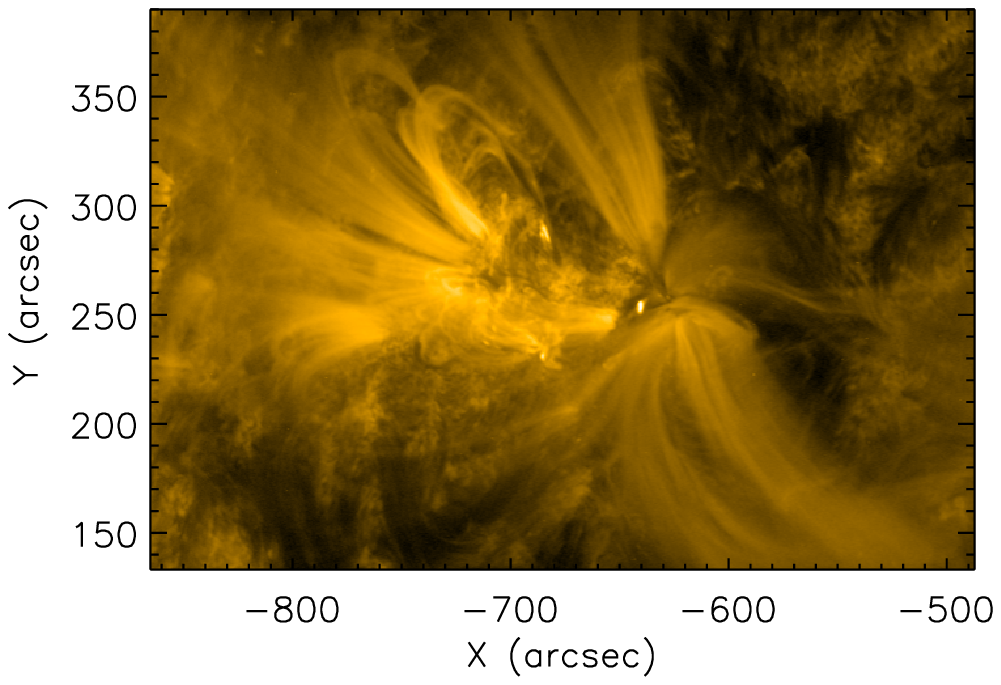}
\includegraphics[width=0.4\textwidth, clip, trim = 0mm 0mm 14mm 10mm]{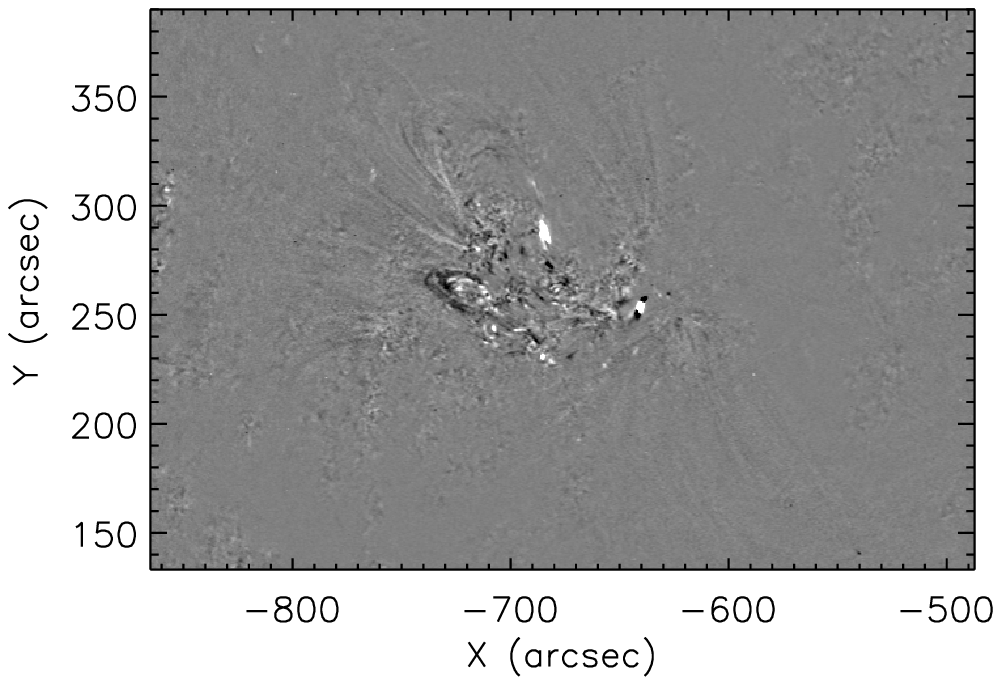}}
\centerline{
\includegraphics[width=0.4\textwidth, clip, trim = 0mm 0mm 14mm 10mm]{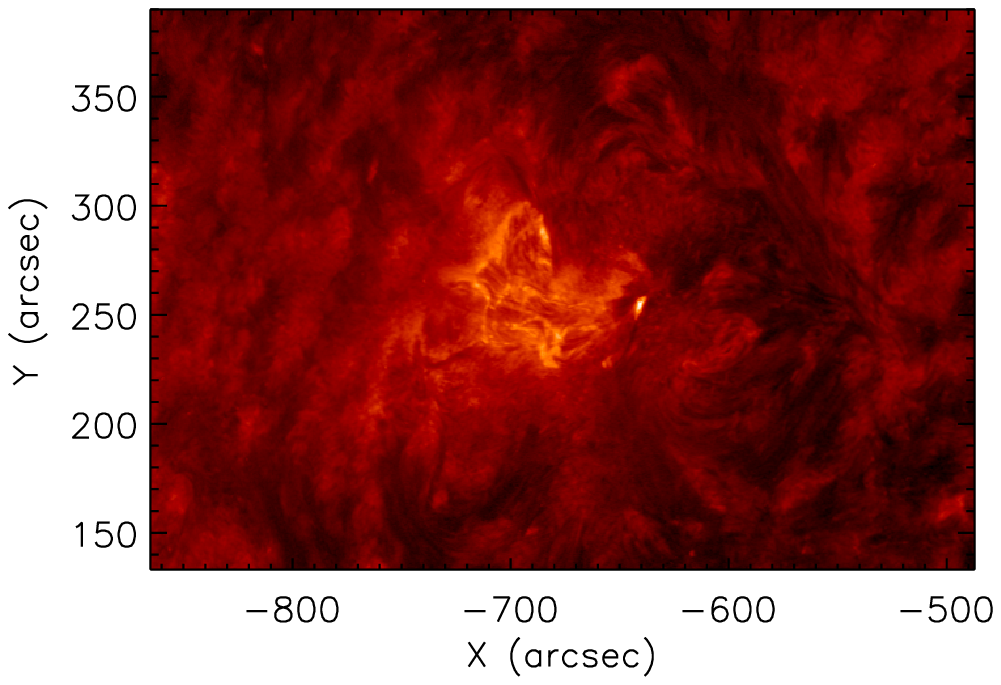}
\includegraphics[width=0.4\textwidth, clip, trim = 0mm 0mm 14mm 10mm]{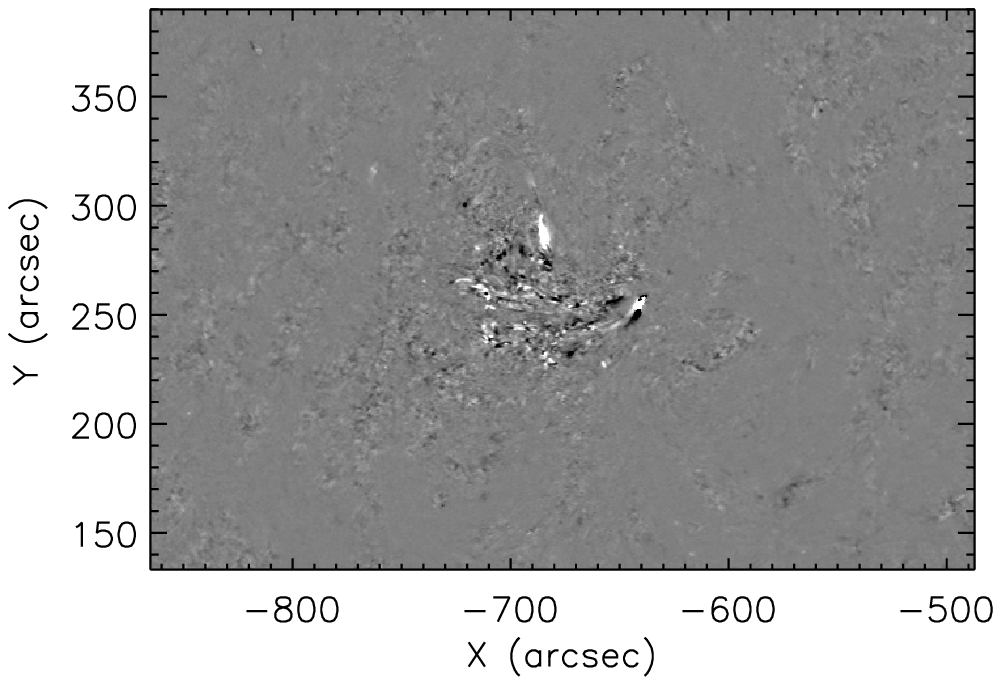}}
\caption{
Demonstration of parametrizing AIA 171\,\AA\ (top row) and AIA 304\,\AA\
(second row) intensity (left) and running-difference (right) timeseries,
for AARP\,746, NOAA\,AR\,11260; direct images: 2011-07-26T17:45:38Z, running-difference
images are 2011-07-26T17:45:38Z - 2011-07-26T17:44:26Z.
The running-difference variations are
similar between the two but there is more structure in the AIA 171\,\AA\
data that could provide additional information, or could be construed
as noise by \nci.  The procedure demonstration continues in Figure~\ref{fig:aia_params_plots}.}
\label{fig:aia_params_images}
\end{figure}

\begin{figure}[t]
\centerline{
\includegraphics[width=0.40\textwidth, clip, trim = 5mm 2mm 5mm 6mm]{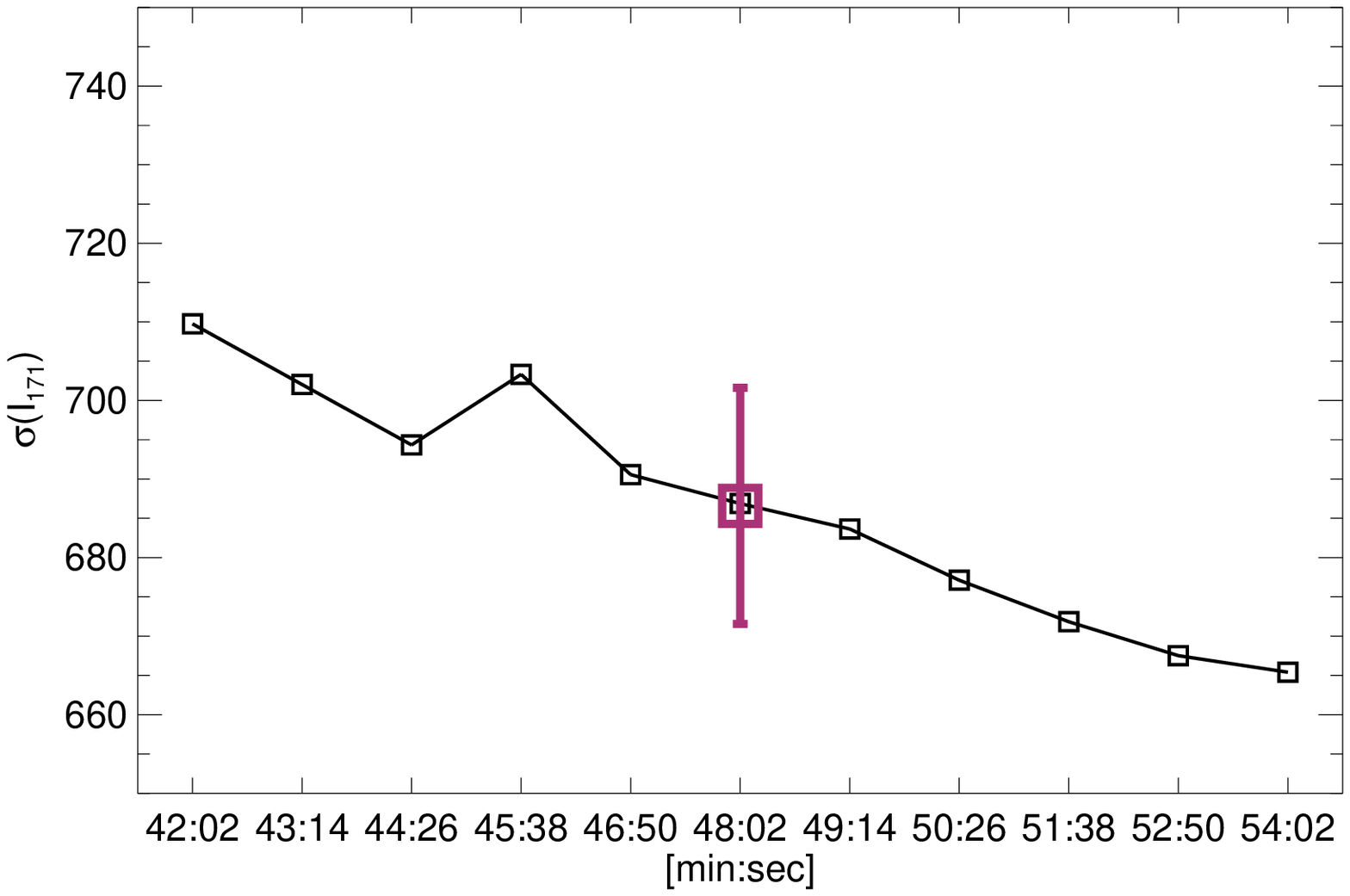}
\includegraphics[width=0.40\textwidth, clip, trim = 5mm 2mm 5mm 6mm]{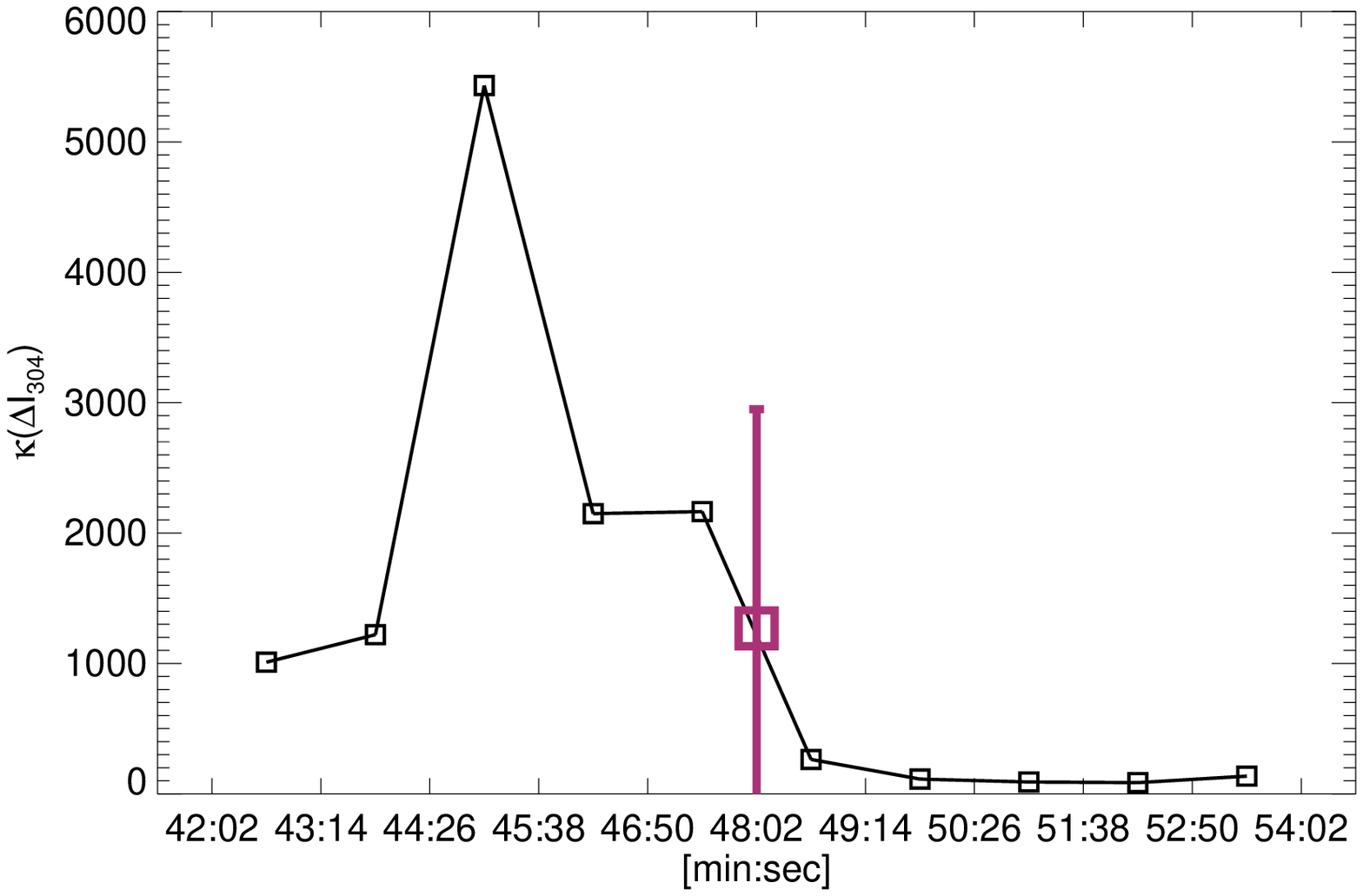}}
\centerline{
\includegraphics[width=0.40\textwidth, clip, trim = 5mm 2mm 5mm 6mm]{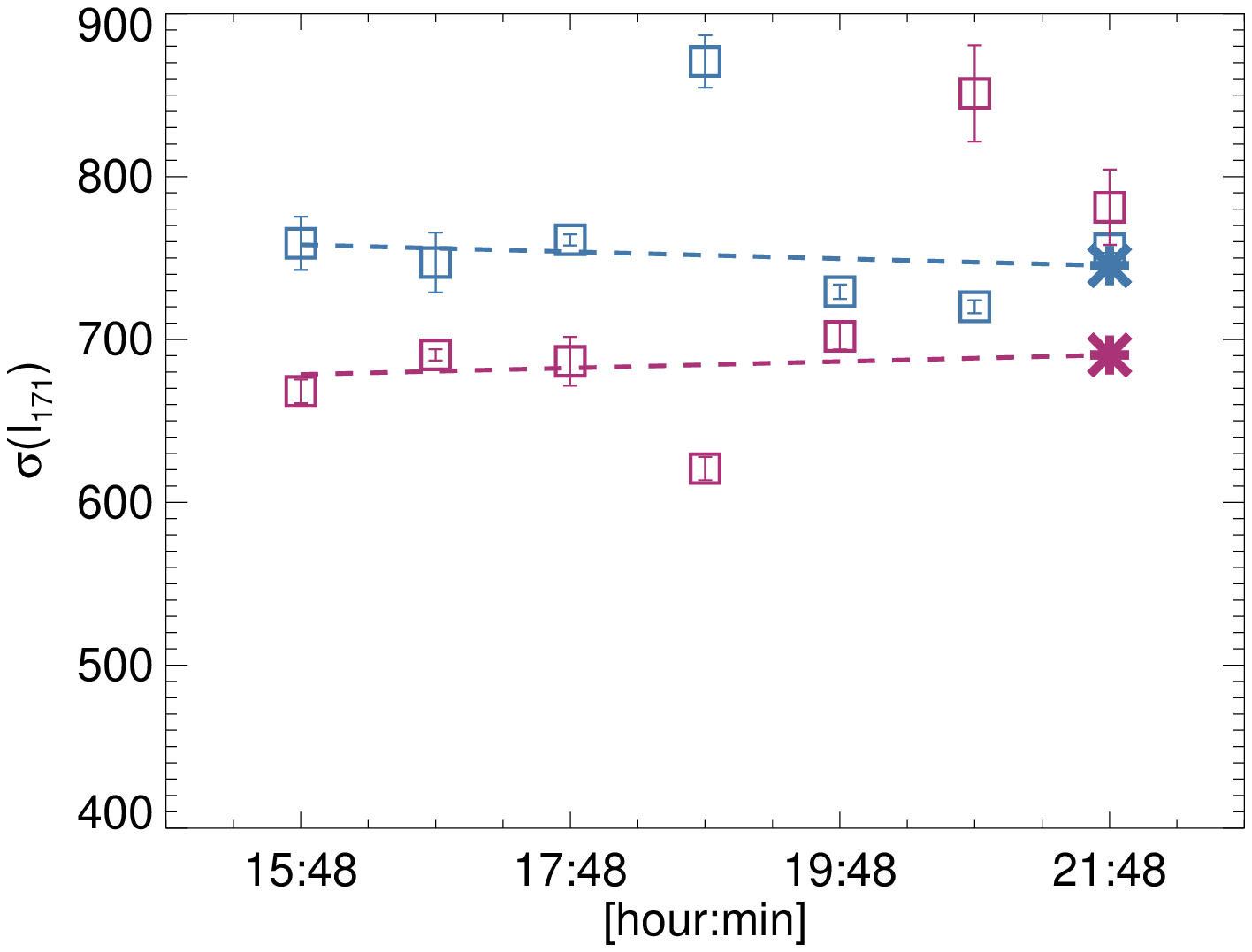}
\includegraphics[width=0.40\textwidth, clip, trim = 5mm 2mm 5mm 6mm]{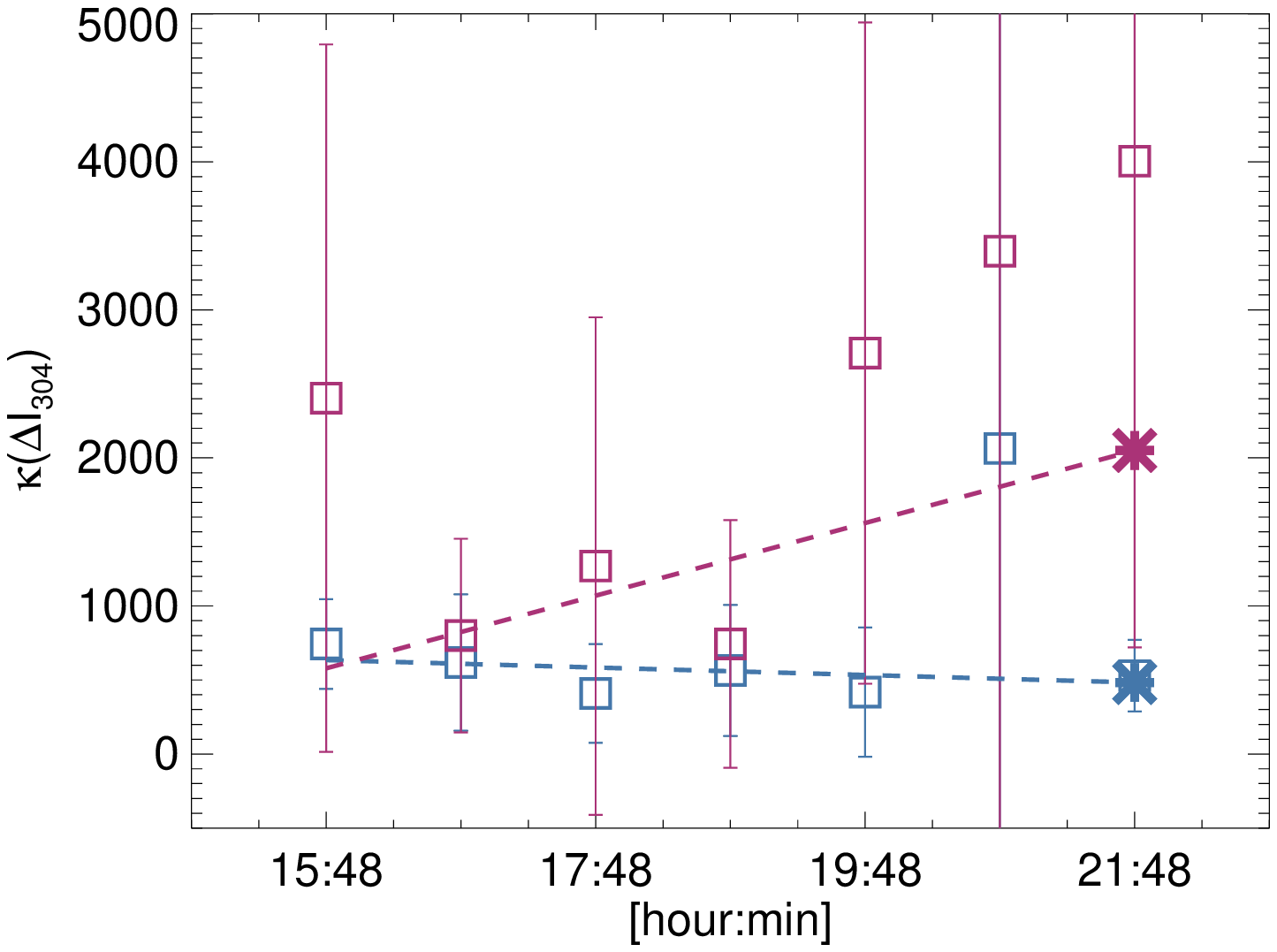}}
\caption{
Following from Figure~\ref{fig:aia_params_images}, the standard deviation of the brightness 
images of 171\,\AA\ is calculated for the 13\,min (11 images) (``$\sigma({\rm
I_{171}})$'', left, top) centered at 2011.07.26 17:48\,TAI, from which the mean and 
standard deviation
are shown (thick point with error bar); these become the data points
for each of the 7 samples covering 6 hours inclusive (left, bottom),
from which the linear slope and last-data (21:48\,TAI) intercept
(thick asterisk) provide the final variables that are analyzed in \nci.
Shown are the results for an \MML\ ``yes-event'' sequence sample on 2011.07.26 (red)  
and a ``no event'' sample time period on 2011.07.25 (blue).  The same sequence
is shown for the kurtosis of the running-difference images of 304\,\AA
(``$\kappa(\Delta{\rm I_{304}})$'', right plots).}
\label{fig:aia_params_plots}
\end{figure}

\noindent
In all, 80 base parameters are defined plus the observing angle (the
same for all wavelengths).  A parameter $X$ is computed for each of 11 images
(or each of 10 running-difference image) within the 13-minute sample 
(see Figure~\ref{fig:aia_params_images}).  
The average and standard deviation of these 11 (10) is assigned to 
the mid-time (the ``{\tt :48}'') that matches the {\tt hmi.Mharp\_720s} data
(see Figure~\ref{fig:aia_params_plots}, top panels), the standard
deviation being used as an estimate of the uncertainty of that parameter over
the 13\,min.  This procedure is performed for each of the 7 hourly samples (see 
Figure~\ref{fig:aia_params_plots}, bottom panels).

A parameter $X$'s ``static'' state and its temporal behavior $d{\rm X}
/ d{\rm t}$ are finally described using the slope and intercept (at the
last data sample's central time (using {\tt T\_REC}), 21:48\,TAI) of a
linear fit over the 7 hourly samples (Figure~\ref{fig:aia_params_plots},
bottom panels), following the magnetic field analysis in \cite{nci_daffs}.
Of note, parameters that are by definition positive- or negative-
definite are limited in the ``static'' parameter to the appropriate
sign; if the inferred value by the intercept of the fit does imply a
crossing in sign, the returned parameter is set to 0.0.  Data outages
exist; at minimum, 2 data points are required, for which only the mean
is returned as the static parameter, and the $d{\rm X} / d{\rm t}$ is
returned as a NaN.  To fit the slope, we require a minimum of 3 data
points.  We have found that a linear fit is sufficient to describe the
general behavior without over-fitting for short-timescale fluctuations.
We (de-)weight the fits by the uncertainties at each time, and one or a
few outlier data points rarely corrupt the linear fits, especially if they
include large uncertainties.  Flares occur during the data acquisition
(Figure~\ref{fig:events_schematic}) but rarely do their influence persist
more than 2\,--\,3\,hr, and they are usually extremely variable on short
timescales (resulting in large uncertainties in the hourly means of the
parameters).  As such, the linear fits generally all but ignore them.
That being said, there exist ``perfect storm'' situations that will
introduce outlier points.  One example is 2016.01.20, AARP\#6281 where two
B-class flares occurred between 15:48--17:48\,TAI, after which there was a
data outage, so that only three points were available.  The parameters for
this AARP on this day were severely influenced ({\it e.g.} $d(\kappa({\rm
I_{131}}))/d{\rm t}$).  This situation can influence both the static and
$d{\rm X} / d{\rm t}$ parameters, but the latter may be more susceptible.
That being said, we have examined the frequency of such outliers and
have found that they typically occur no more than 0.1\% of the time,
which should not influence the final metrics beyond that level.

Thus, the final number is 160 image-based parameters plus the $\cos(\theta)$ variable,
for 161 independent parameters to be analyzed.  These parametrizations
are chosen to be physically interpretable.  For example, one can expect
that the appearance of new bright loops will enhance overall brightness
levels of, for example, 171\AA\ images ($\Sigma({\rm I_{171}}))$ and the
mean brightness levels ($\mu({\rm I_{171}})$), but also possibly produce
a distinct positive skew in the associated running-difference images
($\varsigma(\Delta{\rm I_{171}})$) as the new loops appear.  The brightness
of coronal structures can also change due to heating or cooling \citep{ViallKlimchuk2012}
especially for 171\AA.  On the
other hand, we could expect that increased kinematic activity such as
enhanced loop motion {\it without} significant brightness enhancements
or new structures appearing will be signaled by broader distributions
in running-difference images {\it without} an accompanying increase in
the total, mean, or skew.

We do not, here, consider parameters that use base-difference or base-ratio analysis.
The event definitions employed (Section~\ref{sec:events}) mean that
the data sampling is agnostic as to the time of any event.  Base-difference
and similar approaches are most relevant when the base image refers to 
a known or specified state against which changes are 
measured \citep{Plowman2016}.  The running-difference images
used here focus instead on evaluating the degree of variability of the atmosphere,
by way of the intensity images, at the sampled times only.

\subsection{The NWRA Classification Infrastructure}
\label{sec:nci}

The NWRA Classification Infrastructure \citep[\nci; ][]{nci_daffs} is
a well-established statistical classifier system based on Nonparametric
Discriminant Analysis (NPDA).  There are four components at work in this
facility: the input parameters, the event definitions and event lists,
the statistical package, and the evaluation metrics.  We described the
input parameters that will be used here, in Section~\ref{sec:params},
above.  A general description of \nci\ is given in the referenced work,
and below we describe the particulars as employed here.

\subsubsection{Event Definitions and Event Lists}
\label{sec:events}

\begin{figure}[h]
\centerline{
\includegraphics[width=0.950\textwidth,clip, trim = 0mm 0mm 0mm 0mm, angle=0]{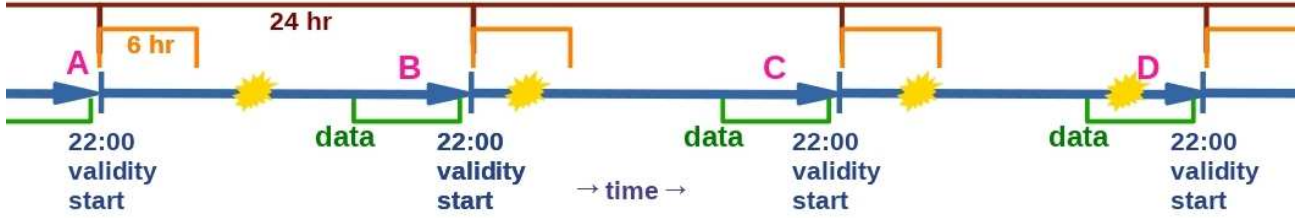}}
\caption{Schematic illustrating the relationship between the AARP data
acquisition periods, the two validity periods invoked, and a few flares
(events).  Time proceeds to the right.  24\,hr days are marked out by
the blue arrows, with an implied start/stop time of 22:00\,TAI.  The data are
acquired at the same time each day (green).  The validity periods, both
6\,hr (orange) and 24\,hr (red) are indicated, all starting at 22:00\,TAI.
The first event (yellow graphic) would be a ``yes-event'' for the 24\,hr
validity period based on the data acquired by ``A'' but a ``non-event''
for the 6\,hr period, whereas the data collected during ``B'' leads to
a classification of the second flare as an ``event'' entry for both
validity periods.  The third and fourth events are classified according to the data
collected in ``C'' even though it occurs during the ``D'' data collection,
and would be designated an ``event'' for both the 6\,hr and 24\,hr definitions, 
even though there are two qualifying events for the latter within its validity period.}
\label{fig:events_schematic}
\end{figure}

The ``event definition'' includes all relevant characteristics to
what defines ``an event'', such as details on timing, event size,
event characteristics, {\it etc.}  In this context, an event is when at least one
flare above a specified threshold occurs during a specified validity period.
A data point ({\it e.g.} a parameter for one AARP) will be assigned to
the flaring population
in this case (Figure~\ref{fig:events_schematic}), and assigned to the
flare-quiet population if no such events occurred.  The assignments of
AARPs to populations change according to the event definitions.  We invoke
{\tt NCI} in its standard ``prediction'' mode which describes the timing
definitions (see Figure~\ref{fig:events_schematic}).  Specifically, there
is no explicit coordination between the time of the events and the data
acquisition time  \citep[as is the case for super-posed epoch analysis,
{\it e.g.} ][]{MasonHoeksema2010,BobraCouvidat2015,Jonas_etal_2018}.

The solar flare specific event definitions used here are described by
(1) lower- and upper- peak intensity thresholds of peak GOES 1--8\AA\
flux (here upper-thresholds are set to infinity), (2) the validity period
during which an event is predicted to occur, (3) the latency period that
defines the interval between the end of the data and the beginning of the
validity period.  The event definitions considered here are summarized
in Table~\ref{tbl:event_defs}.  Some reflect standard definitions used
for flare-prediction research, but some are more focused on shorter-term
chromospheric and coronal behavior in the present context.  

Of note, for \MML\ and \MMS\ definitions, C-class and smaller flares are considered
``non-events''.  Additionally, for all definitions, multiple qualifying
flares within the validity window are considered together as a single
positive event, so that the number of events may be smaller than the total
number of flares during the period.  Finally, a data point assigned to the 
``non-event'' population
may have previously or may subsequently flare -- a ``flare-quiet region'' in 
the context of this analysis is a ``flare-quiet epoch'', or a time of no 
events, regardless of past or future activity.

\begin{table}[b!]
\begin{center}
\caption{Event Definition Summary}
\label{tbl:event_defs}
\begin{tabular}{|ccccc|} \hline
Label & GOES lower limit & Validity Period & Latency Period & \# Events, \\ 
	& 10$^{-6}$\,W\,m$^{-2}$ & hr & hr & (Event Rate $\mathcal{R}$) \\ \hline
\CCL\ & 1.0 & 24 & 0.2 &  2752 (0.086) \\
\MML\ & 10.0 &24 & 0.2 & 450 (0.014) \\
\CCS\ & 1.0 & 6 & 0.2 & 1262 (0.039) \\
\MMS\ & 10.0 & 6 & 0.2 & 155 (0.005) \\ \hline
\end{tabular}
\end{center}
\end{table}

One difference from earlier work on magnetic field-based analysis \citep{nci_daffs}
is the start time for the validity periods.  We matched the AARPs to the 
HMI-based database already in place (see \citetalias{aarps}).  
That database was constructed with anticipation
to the delay in acquiring the near-real-time vector data for a true forecasting
system that would produce forecasts starting at 00:00\,UT \citep{nci_daffs}.   We have
no such constraints here except the desire to match the HMI dataset.

Hence, the start time of the validity periods moved to 22:00\,TAI
for all event definitions.  For the ``24\,hr'' definitions, the validity
time then runs from 22:00\,TAI the day of the data acquisition, to 21:59:59\,TAI the 
next day; in the case of the ``6\,hr'' definitions, it runs from 22:00\,TAI the day 
of the data acquisition to 03:59:59\,TAI the next day.   The ``6\,hr'' definitions thus
have significantly smaller event sample sizes, but the analysis becomes
closer to ``precursor'' parameter evaluation.

\subsubsection{NonParametric Discriminant Analysis}
\label{sec:npda}

Discriminant Analysis (DA) in general classifies input as belonging to one
of two (or more) populations by dividing parameter space into two regions
based on where the probability density of one population ({\it e.g.}
flare-imminent regions) exceeds the other ({\it e.g.} not flare-imminent
regions) so as to best separate the two samples.  Discriminant Analysis
does not simply look for correlations; a statistical classifier such as
DA or Random Forest \citep{Breiman2001} divides parameter-space from
samples of known populations, in the same mathematical ``spirit'' as
machine-learning algorithms.

In NonParametric Discriminant Analysis (NPDA), no assumptions are made
about the functional form of the distributions; instead, the probability
density function is estimated directly from the data.  Since it was
described in \cite{nci_daffs}, we have added the capability of using
adaptive kernel density estimation to \nci. This technique, used here,
starts with a pilot density estimate from the Epanechnikov kernel
and a fixed smoothing parameter determined by reference to a standard
distribution \citep[normal in this case;][]{Silverman86,dfa3}, which
works well for sufficiently large sample sizes, but tends to under-smooth
the tails of a distribution and over-smooth the peak. This pilot density
estimate is then used to estimate local bandwidth factors which determine
the local width of the Epanechnikov kernel in combination with an overall
sensitivity parameter, taken here to be $\alpha=0.5$.

\begin{figure}
\centerline{
\includegraphics[width=0.30\textwidth, clip, trim = 6mm 0mm 0mm 0mm]{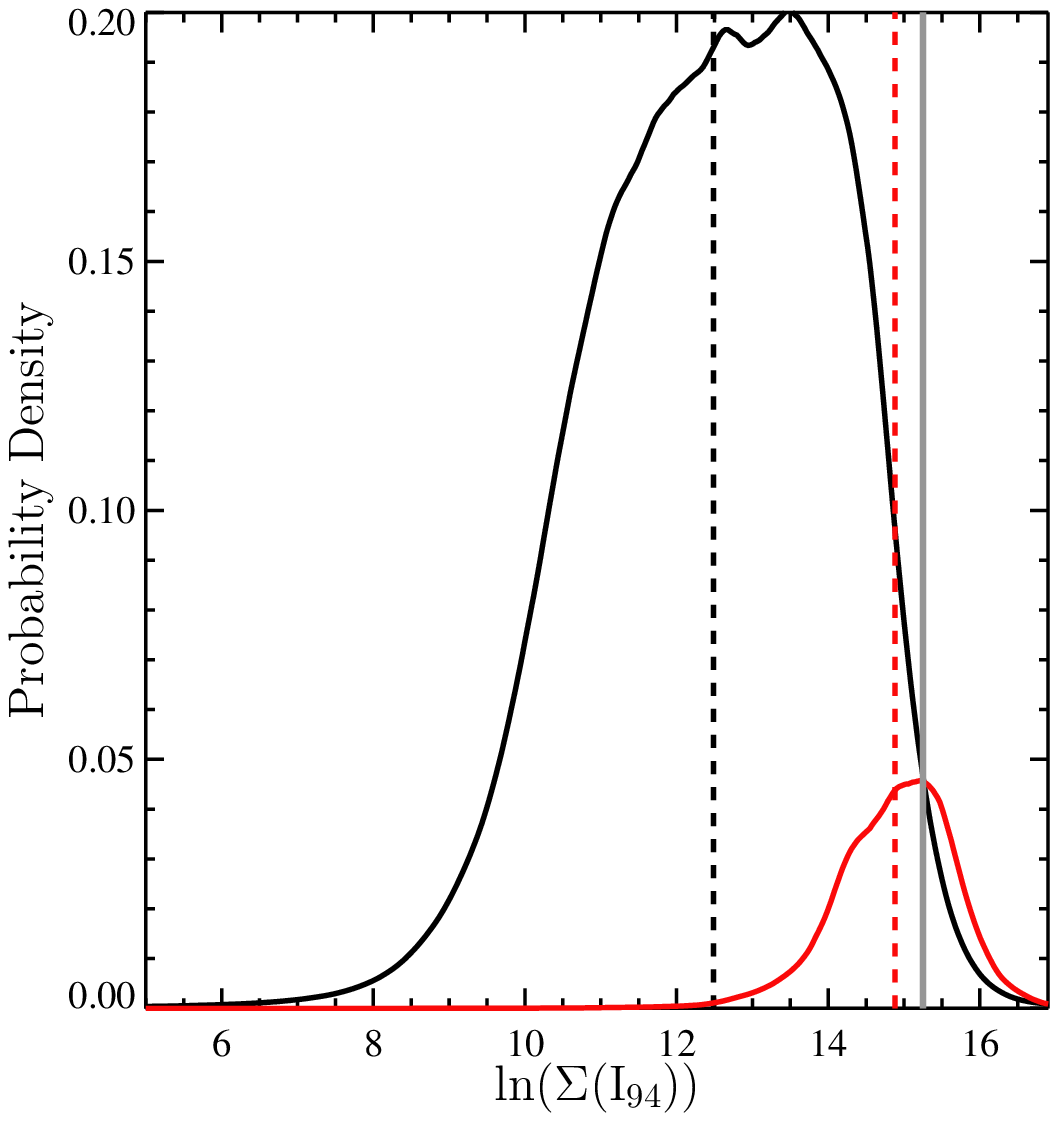} 
\includegraphics[width=0.30\textwidth, clip, trim = 6mm 0mm 0mm 0mm]{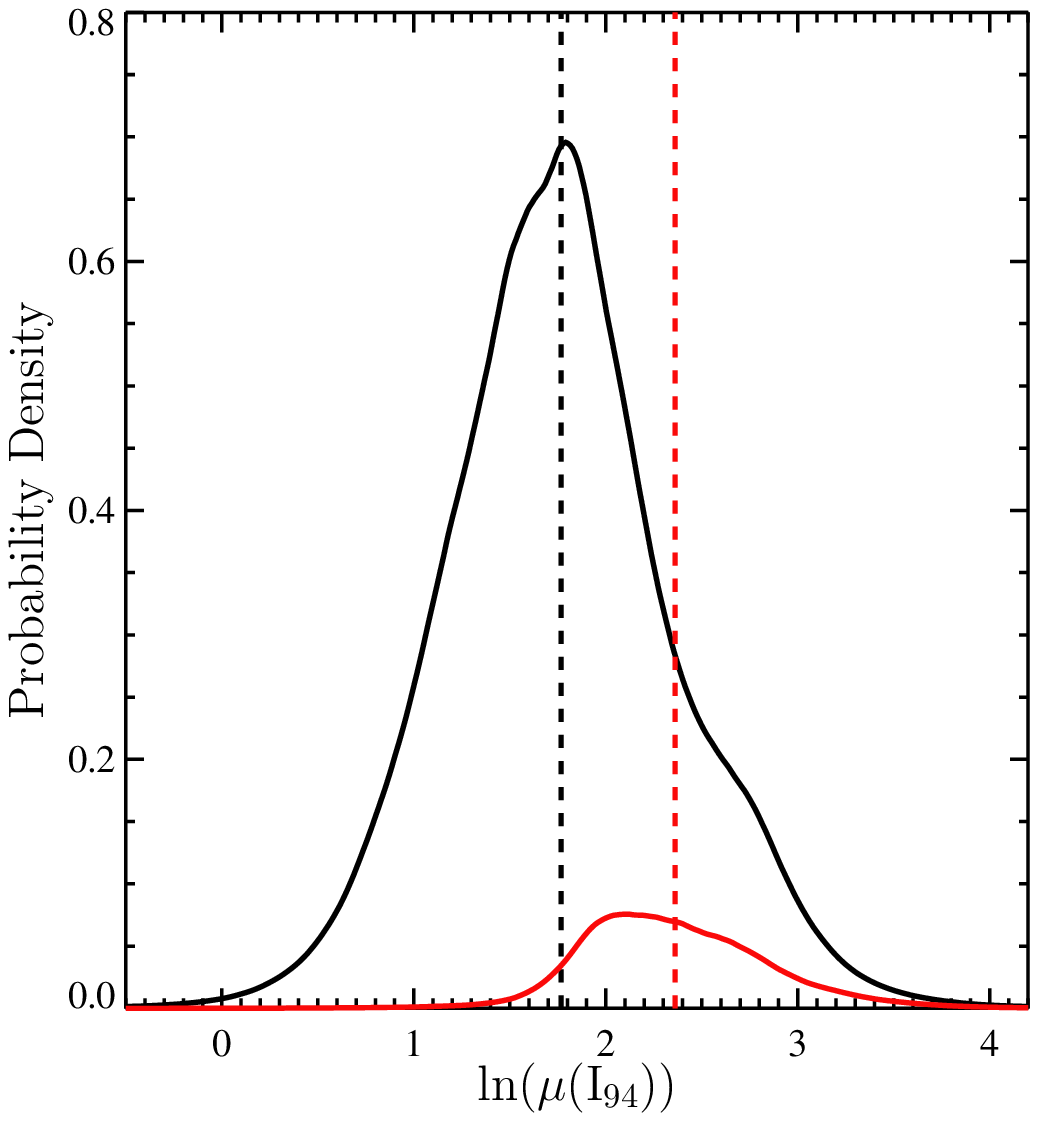}
\includegraphics[width=0.30\textwidth, clip, trim = 6mm 0mm 0mm 0mm]{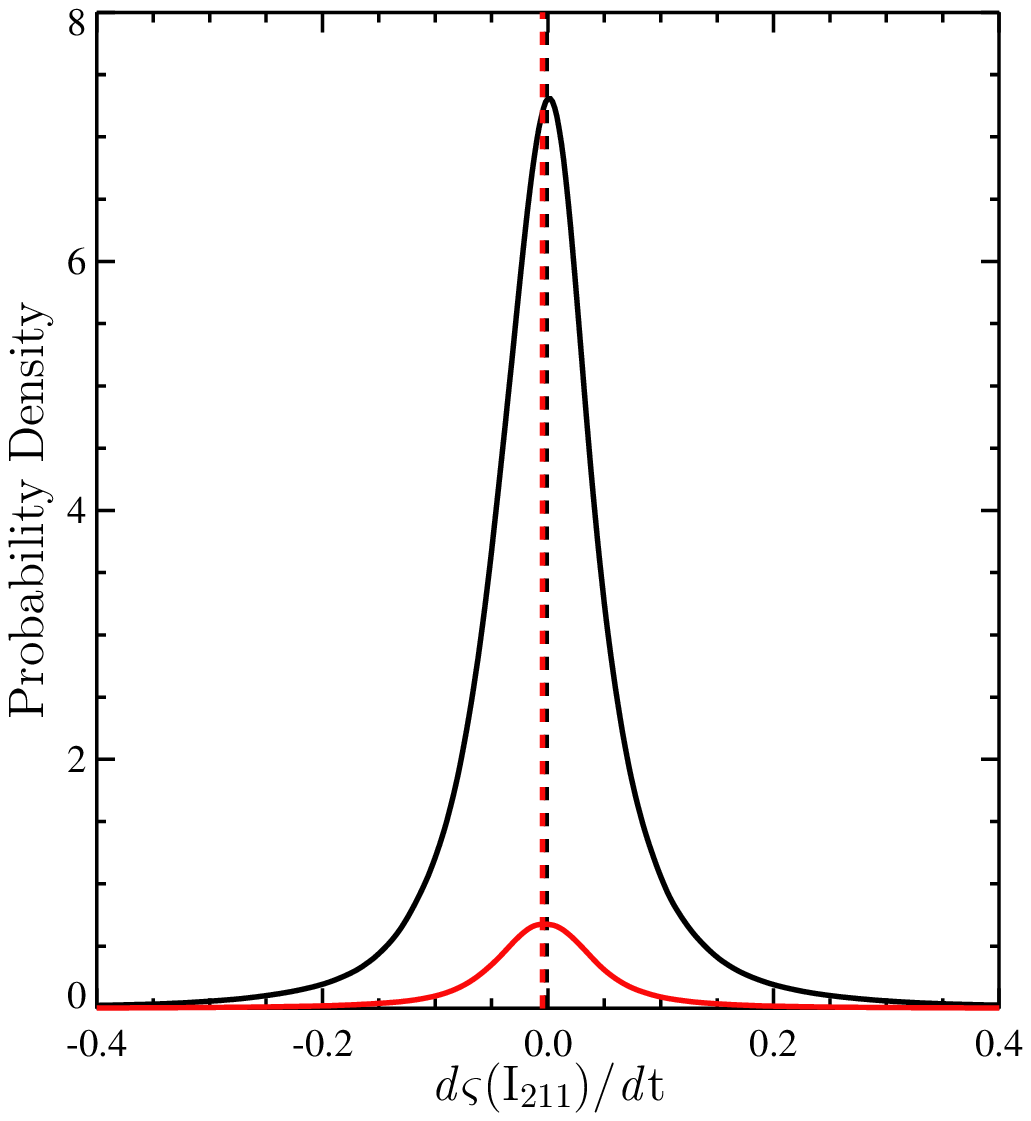}}
\centerline{
\includegraphics[width=0.30\textwidth, clip, trim = 6mm 0mm 0mm 0mm]{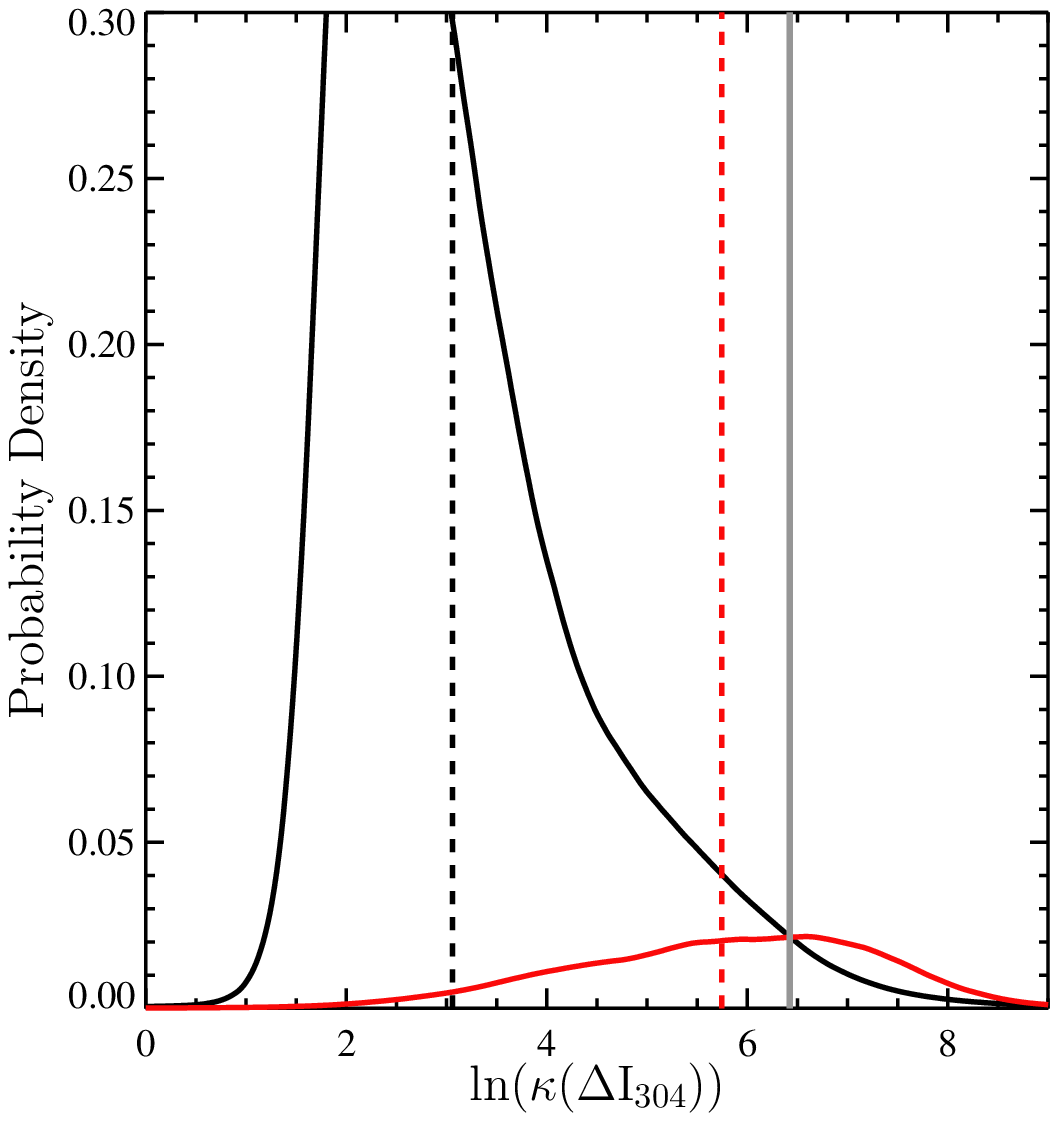}
\includegraphics[width=0.30\textwidth, clip, trim = 6mm 0mm 0mm 0mm]{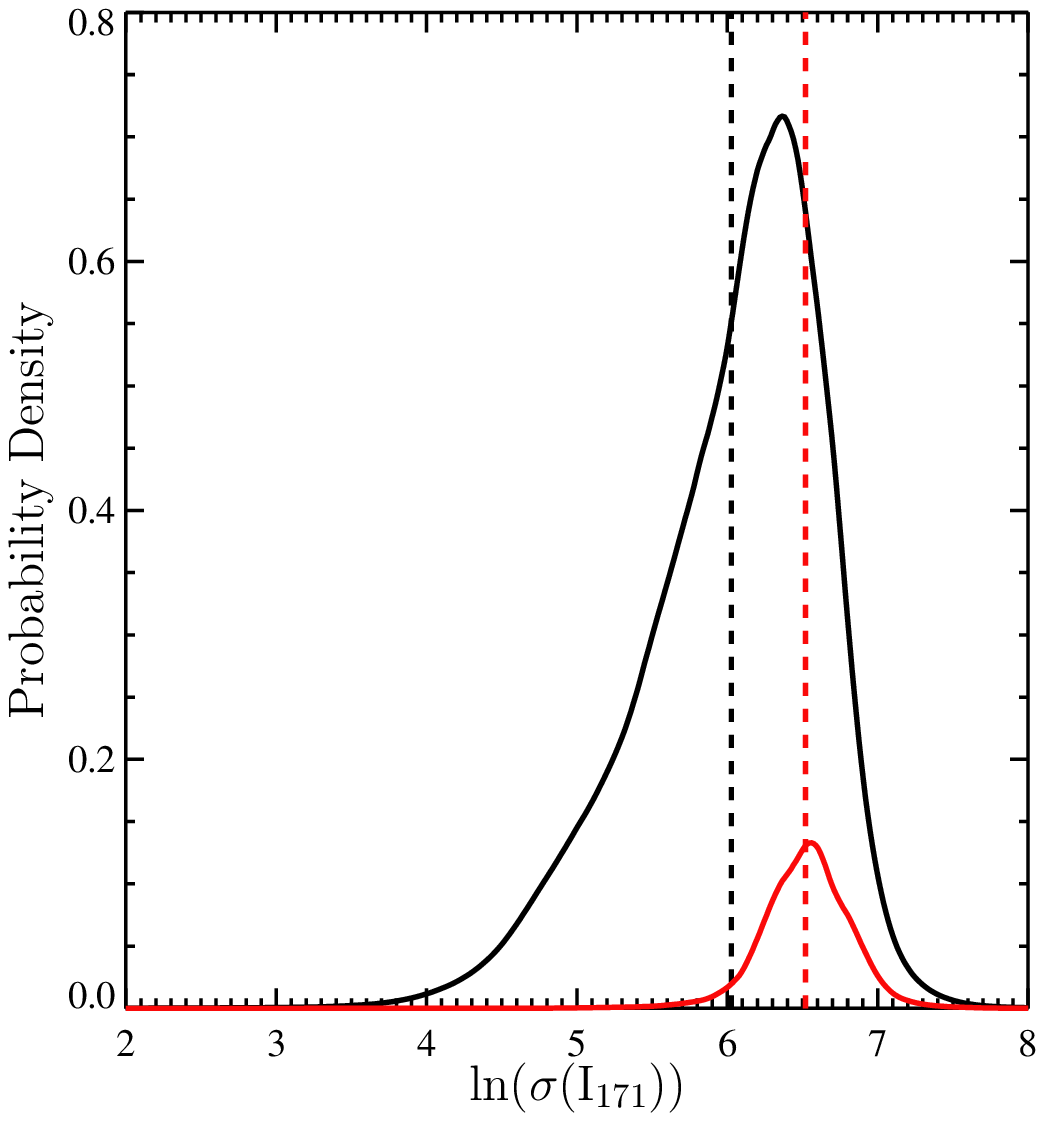}
\includegraphics[width=0.30\textwidth, clip, trim = 0mm 0mm 0mm 0mm]{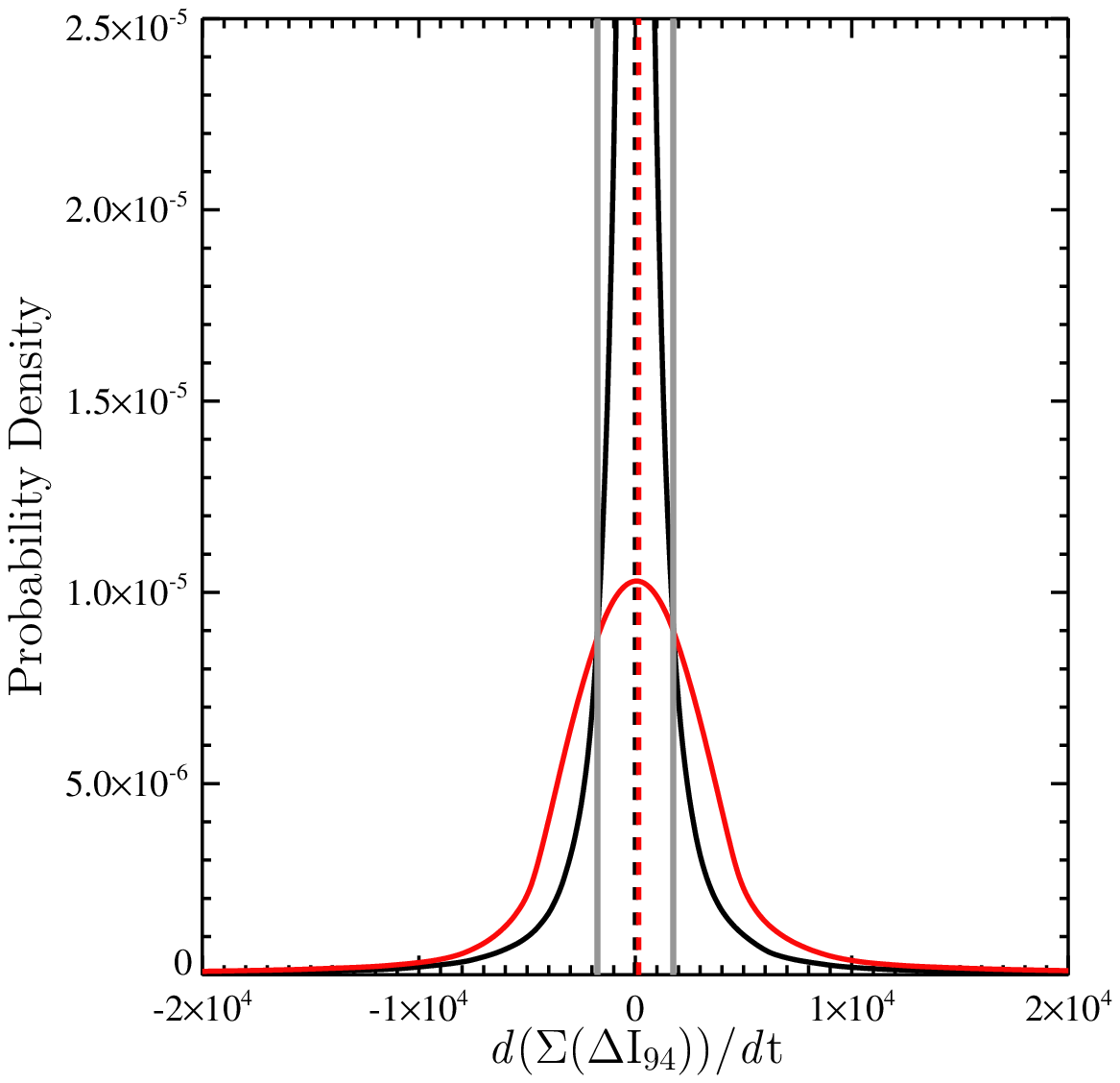}}
\caption{NonParametric Probability Density Functions of six parameters
for the \CCL\ event definition: (Top, left-to-right): the natural log of the total 
of the 94\AA\ emission ($\ln\Sigma(I_{\rm 94})$), the natural log of the 
mean of the 94\AA\ emission ($\ln\mu(I_{\rm 94})$), and the change with time
of the skew of the 211\AA\ emission ($d{\varsigma({\rm I_{211}})} / d{\rm t}$).  
(Bottom, left-to-right): the natural log of the kurtosis of the running-difference of 
304\AA\ images ($\ln\kappa(\Delta I_{\rm 304})$),
the natural log of the standard deviation of the 171\AA\ emission ($\ln\sigma(I_{\rm 171})$, 
{\it c.f.} Figure~\ref{fig:aia_params_plots}),
and the change with time of the total of the 94\AA\ running-difference images 
($d{\Sigma(\Delta{\rm I_{94}})} / d{\rm t}$).  For
all,  {\bf {\color{red}{event}}}, {\bf non-event}
non-parametric density estimates are shown, their means ({\bf {\color{red}{- -
-}}/{- - -}}), and the {\bf{\color{gray}\bf{discriminant boundary(ies)}}}
which may not be present within the range shown (which itself always encompasses
all but the most extreme outliers, if any).  See text for discussion. }
\label{fig:npda_examples}
\end{figure}

Although \nci\ with NPDA can be used for multi-variable
analysis (multiple parameters simultaneously creating a higher-dimension
parameter-space), we focus here on single-variable NPDA and strive
for statistically-significant sample sizes for each event definition
(Section~\ref{sec:events}) and a first-look set of results that can 
be physically interpretable.  Example density functions and NPDA boundaries
are given for select parameters in Figure~\ref{fig:npda_examples}, and discussed
in Section~\ref{sec:results}, below.

\nci\ generates probabilities that a datapoint will belong to one or
the other population based on the ratio of probability density function
estimates from the samples plus the populations' prior probabilities.
Note that as described in \citet{nci_daffs}, \nci\ treats ``null'' data
and ``bad'' data differently.  Additionally, in cases where a parameter
is positive- (negative-) definite, \nci\ automatically works with the
natural logarithm of the variable (absolute value of the variable).
This practice guarantees that the density estimate is zero for negative
(positive) values of the parameter, as it should be.  The result is
typically a slight improvement in the evaluation metrics.

\nci\ provides unbiased estimates of the table entries using
cross-validation \citep{hill66,dfa,nci_daffs}; previously \nci\
relied upon ``n-1'' method but now performs cross-validation based on
active-region number.  For the results here, the last digit of the AARP
number is used to define 10 groups, with which 10-fold cross-validation
is performed.  This approach is invoked in recognition that for any
given AARP, some parameters may not evolve significantly over a day or
longer. The goal then of AARP-based cross-validation is to avoid using
samples of the same AARP to both construct the probability density
functions and then use them to predict a sample from the same AARP.

\subsubsection{Evaluation Metrics}
\label{sec:metrics}

The classifications made by \nci\ are evaluated using standard quantitative
metrics \citep{JolliffeStephenson2012}, to answer
the question, ``how well did the classifier separate the samples drawn from
the two known populations?''.  \nci\ reports a large selection of 
metrics and graphical tools for interpretation; here we focus on 
a few that are most informative for the present study.

The native results from \nci\ are the probabilities for each data point of
belonging to one or the other population, hence evaluation metrics based
on probabilities are most appropriate.  The Brier skill score (\BSS)
quantifies the performance by normalizing the mean square error of the
probability that a point belongs to its true population by the mean square
error for the probability based on the ``climatology'', or ratios of the
two population sizes to the total sample size. It is normalized so that
``perfect'' is 1.0, no skill against the reference is 0.0, and can be
negative.  \BSS\ effectively summarizes the Reliability Plot (``attributes
diagram'') that is conditioned on the forecast (classification), and
by which sharpness and resolution can be judged; we report the \BSS\
and present Reliability plots in Section~\ref{sec:results}.

With the assignment of a Probability Threshold (${\rm P}_{\rm thr}$)
above/below which the resulting probability is deemed to belong to one
or the other population, categorical metrics are available \citep[see
the discussions in][]{allclear,ffc3_1}.  For these, a classification
table is first constructed according to the assigned probability that
a data point belongs to one or the other populations, given an assigned
${\rm P}_{\rm thr}$).  Four entries (for 2-option classification) then
comprise the classification table: True Positive (TP), True Negative
(TN), False Positive (FP) and False Negative (FN).  As we are not
providing any kind of custom forecasts, we use ${\rm P}_{\rm thr}=0.5$
by default, which maximizes the number of correct classifications when
the prior probabilities are set proportional to the sample sizes, and
is appropriate for physics-interpretable research.

The popular True Skill Statistic (TSS), also known as the Peirce
Skill Score (PSS) or Hanssen \& Kuiper Skill Statistics (H\&KSS)
\citep[see ][for discussions]{Bloomfield_etal_2012,allclear,ffc3_1}
is the difference between the probability of detection (hit rate) and
the probability of false detection (false alarm rate).  As with all
skill scores, it is normalized such that for perfect differentiation
TSS$=1.0$, while no power to discriminate the populations produces
TSS$=0.0$.  Changing the sample sizes does not impact TSS provided the
samples have been drawn from the same populations.  ``Optimal TSS'' or
``Maximum TSS'' scores are often reported, and are generally earned
by setting ${\rm P}_{\rm thr} \approx$ the event rate $\mathcal{R}$
(Table\,\ref{tbl:event_defs}) where $\mathcal{R}=n_{TP} + n_{FN} / N$
and $N$ is the sample size \citep{Bloomfield_etal_2012,allclear,Kubo2019}.
We report here \maxtss\ with ${\rm P}_{\rm thr}=\mathcal{R}$.

Finally, by calculating the hit rate (POD) and false alarm rate (POFD),
the two components of the TSS, through the range of ${\rm P}_{\rm thr}$
one builds a Receiver (Relative) Operating Characteristic Curve (ROC)
plot \citep[see examples and discussion in ][]{ffc3_1}.  The ROC plot
illustrates the ability of a forecast (or classification) to differentiate
between events and non-events, and is observation-conditioned.
This plot is then summarized by the ROC Skill Score (\ROCSS; or the
Gini Coefficient) that is related to the ROC area or popular Area Under
the Curve (AUC) metric: $\mathcal{G}=2*AUC - 1.0$ where $\mathcal{G} =
1.0$ denotes a perfect score and $\mathcal{G} < 0$ indicates worse than
zero-skill performance.  We report here $\mathcal{G}$ but also present
ROC plots for a few examples.

We sort the parameters based on the \BSS\ metric, the only
metric for which we perform 100-draw bootstrap with replacement
\citep{EfronGong1983,JolliffeStephenson2012,nci_daffs}, also based on
the last digit of the AARP number, to provide an estimate of the uncertainty in the metric.
That is, for each draw independently, the probability density estimates
for each population are calculated (Figure~\ref{fig:npda_examples}) and used to
generate a probability of an event occurring. This probability varies (usually
only slightly) between the different draws, leading to a range of values for
the \BSS, and slightly moving the location of the discriminant boundary,
sometimes leading to different classification tables. The standard deviation
of the \BSS\ values is used as an estimate of the uncertainty.
The other metrics are calculated directly from the probabilities for each data
point, computed using cross-validation but no bootstrap.  The rank order of the
different metrics does not follow identically, but is generally close
(Tables~\ref{tbl:resultsCCL}--\ref{tbl:resultsMMS}).  Our previous
investigation on photospheric magnetic field parameters \citep{nci_daffs} found
that, for a given event definition and parameter, the uncertainty across a
range of skill scores was relatively constant. Thus, the uncertainties quoted
for the \BSS\ are likely to be a reasonable representation of the uncertainty
in the \maxtss\ and \ROCSS.

\subsection{Sample Size and Statistical Flukes}
\label{sec:stats2}

With the large number of parameters being considered, it is possible
that a few parameters may falsely appear to be successful at classifying the data 
solely by happenstance of this particular sample.
The likelihood of this happening is diminished with large sample sizes,
but for the \MML\ and especially the \MMS\ event definitions, it may
become a concern.  

In \citet{trt_emerge3}, a Monte Carlo experiment was described that 
draws two random samples from the same population with sizes equal to the
sample sizes in question ({\it e.g.} of the event and non-event samples).
In the experiment, the same analysis is performed as on the actual parameters
for fifty times as many parameters as were in the investigation, to more
accurately capture the range of possible outcomes.  The experiment was
performed where the population was a normal distribution, a Cauchy
distribution, and a cosine distribution.  The resulting distributions
of skill scores for the experiment, where no difference is expected,
were then compared to the distribution found for the real experiment, and
the probability of finding outliers was estimated.  In other words,
this approach determines the number of statistical outliers that
may be expected were there {\it no} difference in the two underlying
populations.  

When this experiment was applied to the AARP-matched HARP-based magnetic
field parameters in a similar context as the present study and with a
similar sample size \citep[but indeed for a larger number of parameters
than is being tested here][]{nci_daffs}, we estimated there would be
$<\!\!1\%$ chance of a resulting ${\rm BSS}\!>\!0.001/0.002/0.003$
by chance alone for single variable NPDA for {\tt C1.0+/M1.0+/X1.0+}
flares, respectively.  Hence we are confident that the results shown
here are not particularly susceptible to statistical flukes.

Additionally, the bootstrap provides an uncertainty for the \BSS.  As
discussed in Section~\ref{sec:bestworst}, for the top performing results
and indeed for most parameters across the \CC\ and \MML\ event lists, the
reported \BSS\ are at the $5\,\sigma$, $10\,\sigma$ or higher detection
level.  For \MMS\ which is the experiment with the smallest ``yes-event''
sample size and the smallest event rate, the \BSS\ scores are smaller,
barely above 0.0, although the bootstrap-derived uncertainties are only a
factor of 2 larger (see Section~\ref{sec:bestworst}).  Even with almost a
solar-cycle's worth of data, the sample of larger events that occur within
6 hours of any given time of day is, statistically speaking, very small.

\section{Results}
\label{sec:results}

In these sections we highlight some examples and call out the best
and the worst performing parameters in order to give an overview of
the results.  All computed parameters, and resulting probabilities are
available \citep{nci_aia_data}, so readers can examine the distributions
for other parameters of interest, and (for example) compute additional
skill scores or apply other analysis methods to the data.

\subsection{NonParametric Density Estimates}

We show in Figure~\ref{fig:npda_examples} the nonparametric density
estimates for a selection of parameters, all for the \CCL\ definition
primarily because the distributions of both populations are clearly
visible; the class imbalance between events and non-events for the other
definitions (Table~\ref{tbl:event_defs}) simply make presentation more
challenging.

There is quite a range of distribution shapes amongst the parameters.
For one of the most intuitive parameters, $\Sigma(I_{\rm 94})$
(Figure~\ref{fig:npda_examples} top left), the density estimates are
distinctly offset from each other, and there is a single discriminant
boundary to the right of which the events have a higher probability
than the non-events.  In the next two parameters $\mu(I_{\rm 94})$ and
$d{\varsigma({\rm I_{211}})} / d{\rm t}$ (Figure~\ref{fig:npda_examples}
top middle and right, respectively) there is no discriminant boundary;
for the former, even though the distributions are distinctively offset
from each other (the means are visibly different), the low event rate
(large class imbalance) means that the event probability never exceeds
the non-event probability whereas in the latter, there is almost no
difference in the event {\it vs.} non-event distribution means or shapes.
Despite the lack of a discriminant boundary, $\mu(I_{\rm 94})$ still
has significant skill as measured by the \BSS\ (\BSS$=0.084\pm0.006$),
while $d{\varsigma({\rm I_{211}})} / d{\rm t}$ does not. 

The first two parameters in the bottom row of
Figure~\ref{fig:npda_examples} show similar behavior to the corresponding
parameters in the top row: $\kappa(\Delta I_{\rm 304})$ provides a
single clear discriminant boundary and very different distributions,
while the distributions for the $\sigma(I_{\rm 171})$ samples are
reminiscent of the $\mu(I_{\rm 94})$ distributions, again the population
distributions are distinguishable (the means are well separated), there
is significant skill, but there is no discriminant boundary.  Finally,
the $d{\Sigma(\Delta{\rm I_{94}})} / d{\rm t}$ distributions are centered
exactly the same, however, unlike $d{\varsigma({\rm I_{211}})} / d{\rm
t}$, there are two discriminant boundaries because the event population
is wider than the non-event population.

\subsection{Metrics Scores and Evaluation Plots for AARP-based Parameters}
\label{sec:bestworst}

The results are sorted on \BSS, and we present the
top-10 and bottom-5 \BSS-scoring parameters in Tables~\ref{tbl:resultsCCL}
-- \ref{tbl:resultsMMS}; the full results are available in
machine-readable format.  For each of the parameters we
also compute the ``\maxtss'' (with ${\rm P}_{\rm thr}=\mathcal{R}$) and
the \ROCSS\ or $\mathcal{G}$.  The order of the parameters based on the
latter scores does not exactly follow the ordering of the \BSS, but does
so loosely, especially considering the bootstrap-based uncertainties for
the \BSS.

\begin{figure}[t]
\centerline{
\includegraphics[width=0.30\textwidth,clip, trim = 8mm 0mm 0mm 5mm, angle=0]{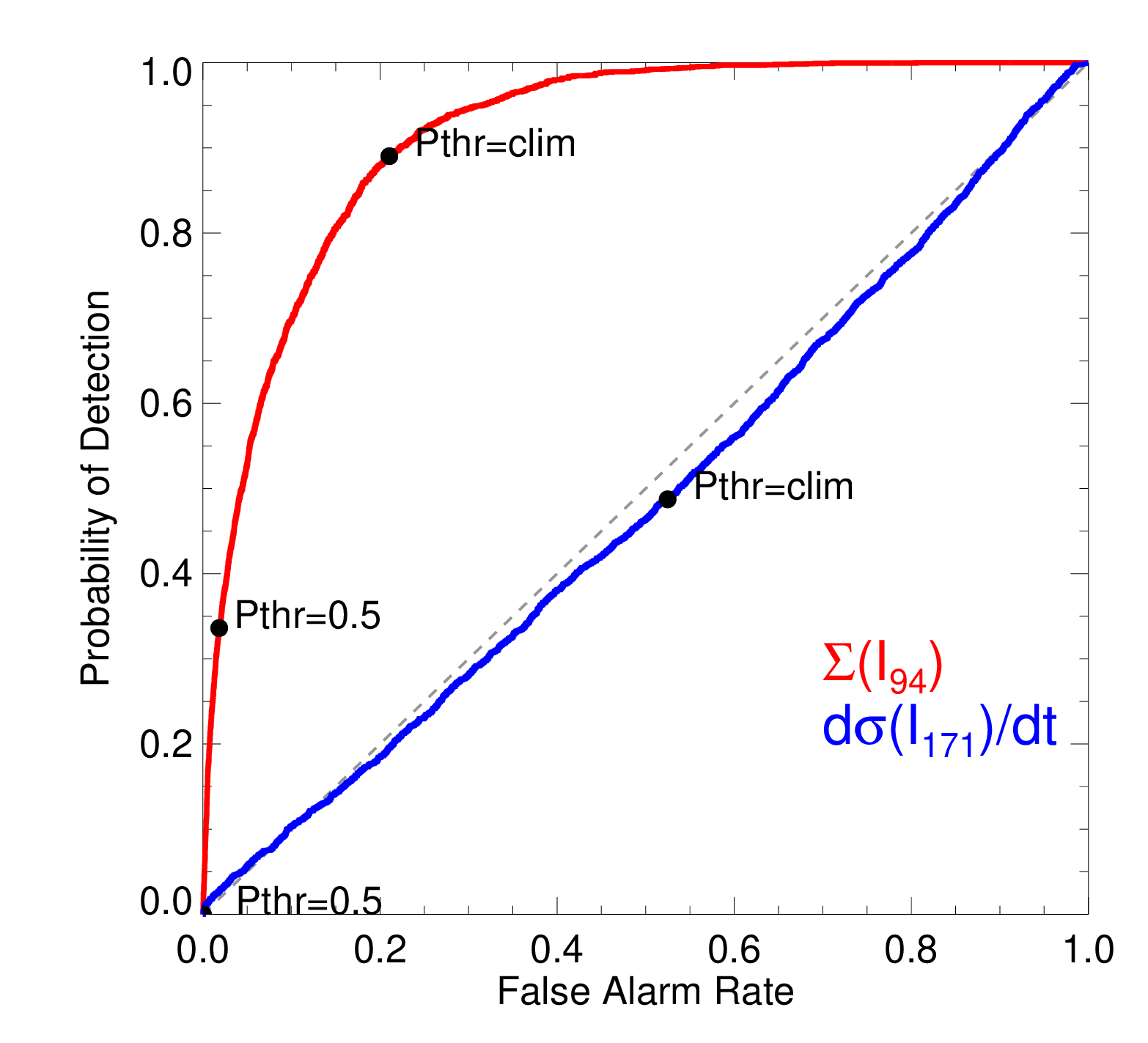}
\includegraphics[width=0.30\textwidth,clip, trim = 8mm 0mm 0mm 5mm, angle=0]{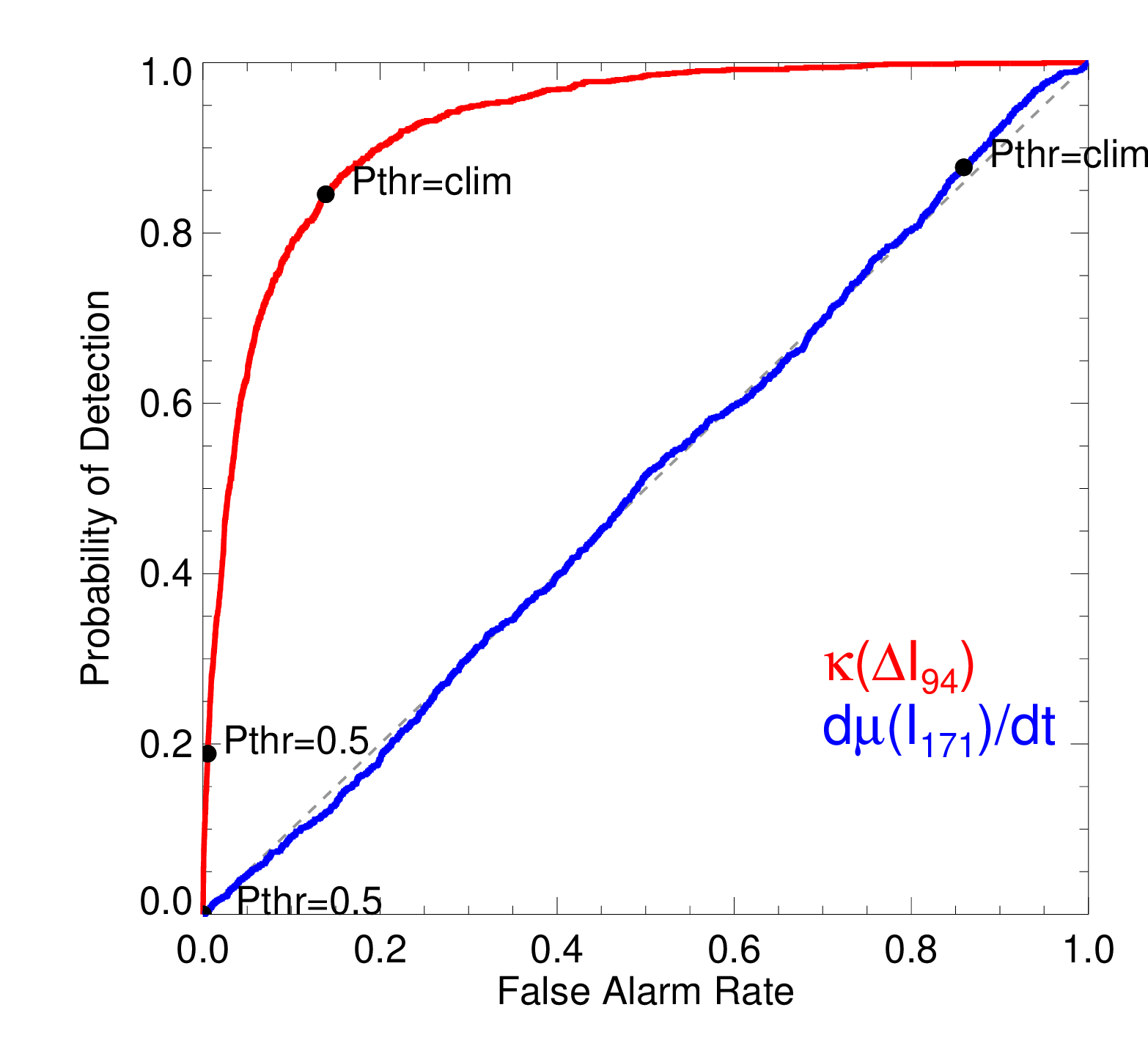}}
\centerline{
\includegraphics[width=0.30\textwidth,clip, trim = 8mm 0mm 0mm 5mm, angle=0]{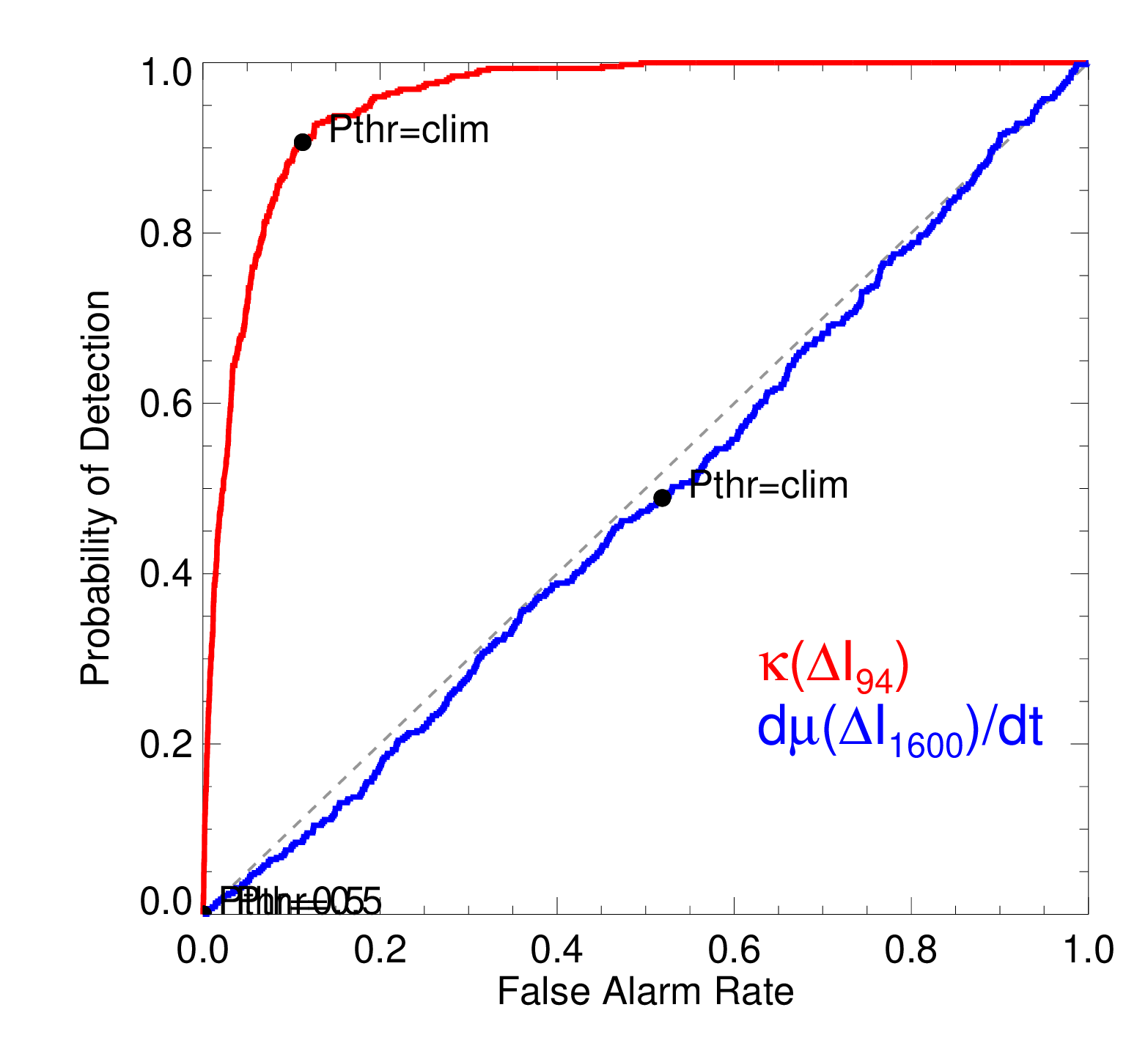}
\includegraphics[width=0.30\textwidth,clip, trim = 8mm 0mm 0mm 5mm, angle=0]{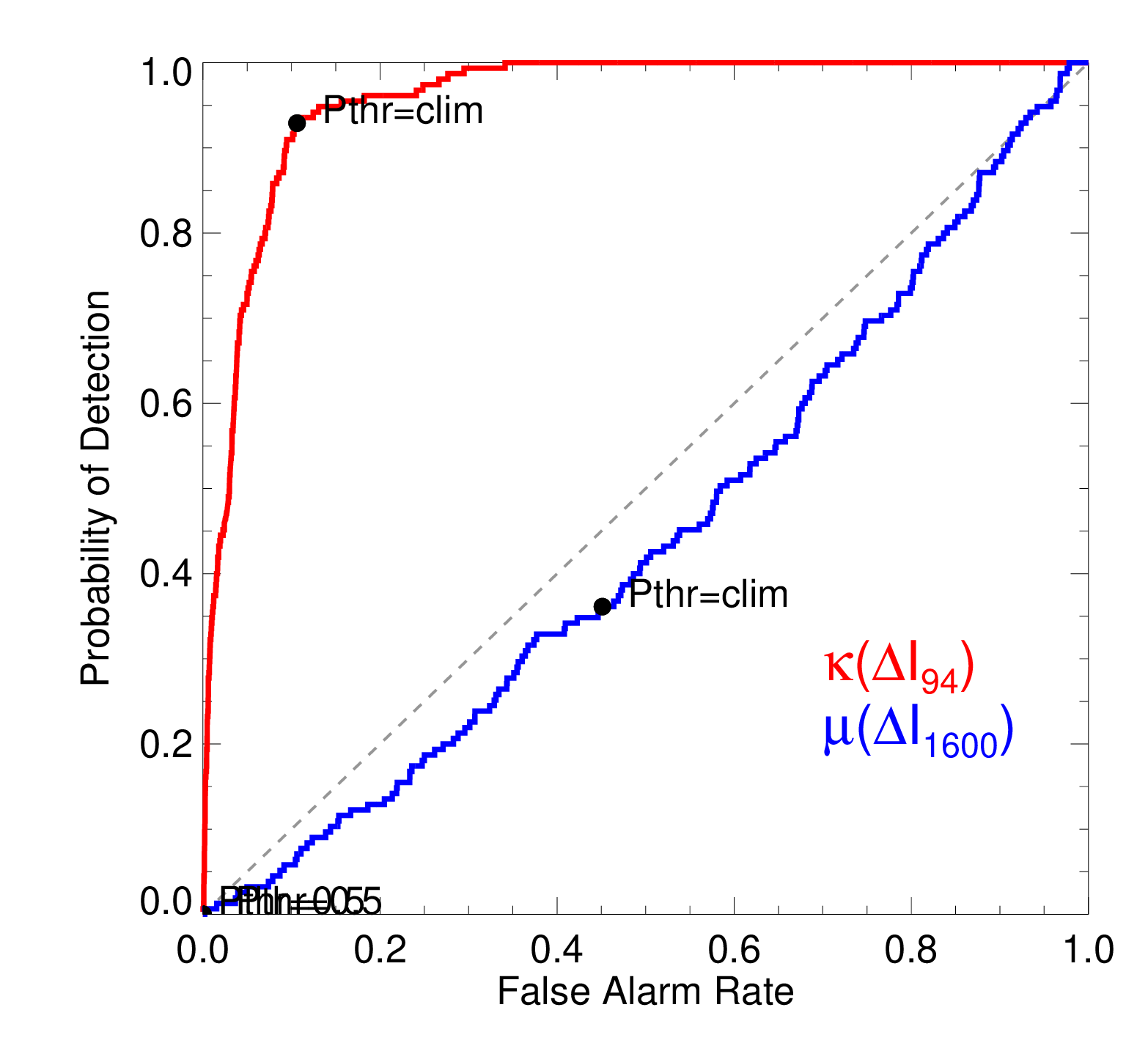}}
\caption{Receiver Operating Characteristic (ROC) plots for (Top, Left/Right) \CCL, \CCS\ 
(Bottom Left/Right)
\MML, \MMS, for the parameters as indicated.  For all, the `climatology' probability threshold is 
as listed in Table~\ref{tbl:event_defs}.  The parameters shown are generally the top- and 
bottom-scoring parameters by $\mathcal{G}$, {\it c.f.}
Tables~\ref{tbl:resultsCCL}--\ref{tbl:resultsMMS}.}
\label{fig:roc}
\end{figure}

\begin{figure}[t]
\centerline{
\includegraphics[width=0.250\textwidth,clip, trim = 8mm 0mm 2mm 5mm, angle=0]{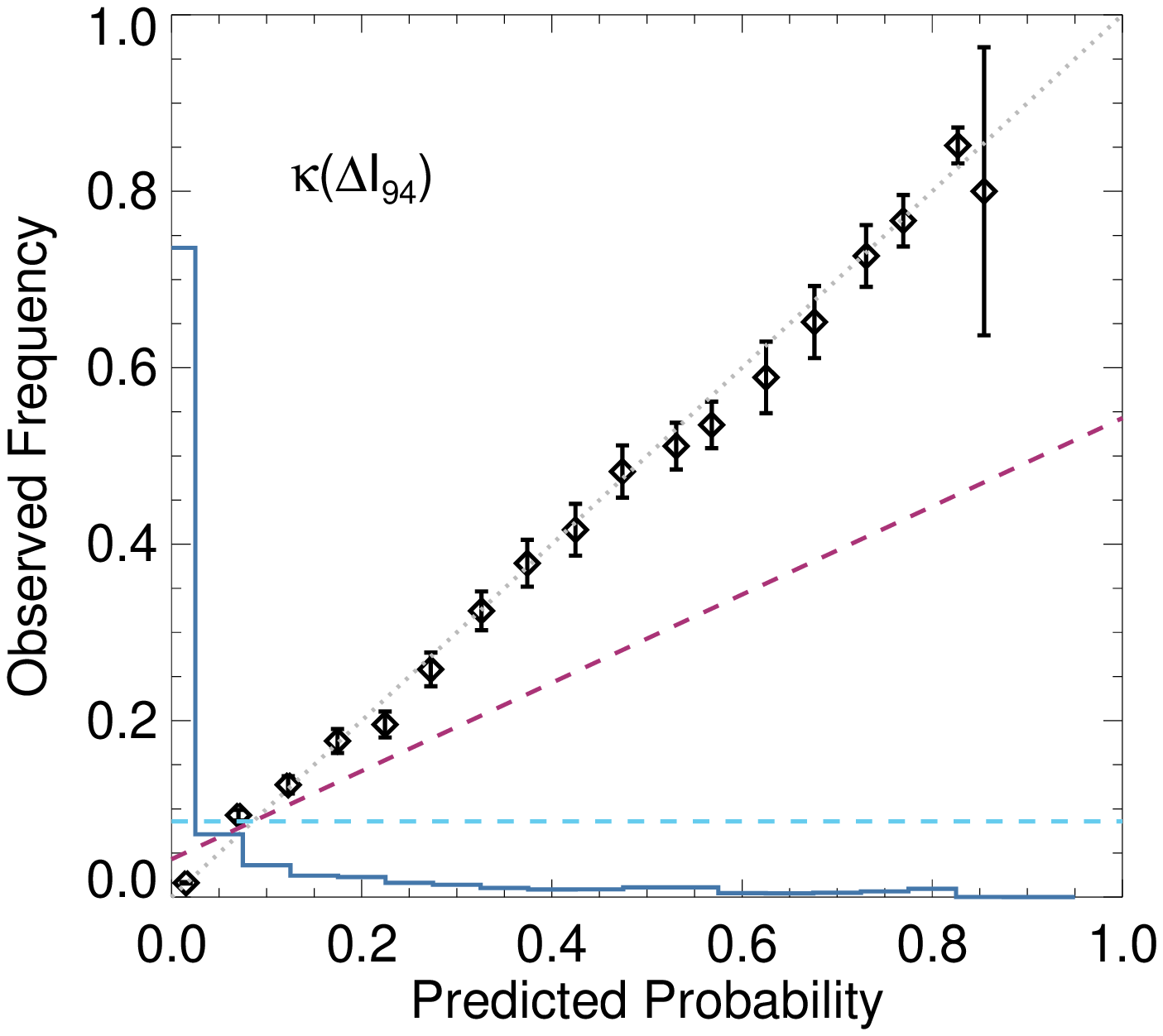}
\includegraphics[width=0.250\textwidth,clip, trim = 8mm 0mm 2mm 5mm, angle=0]{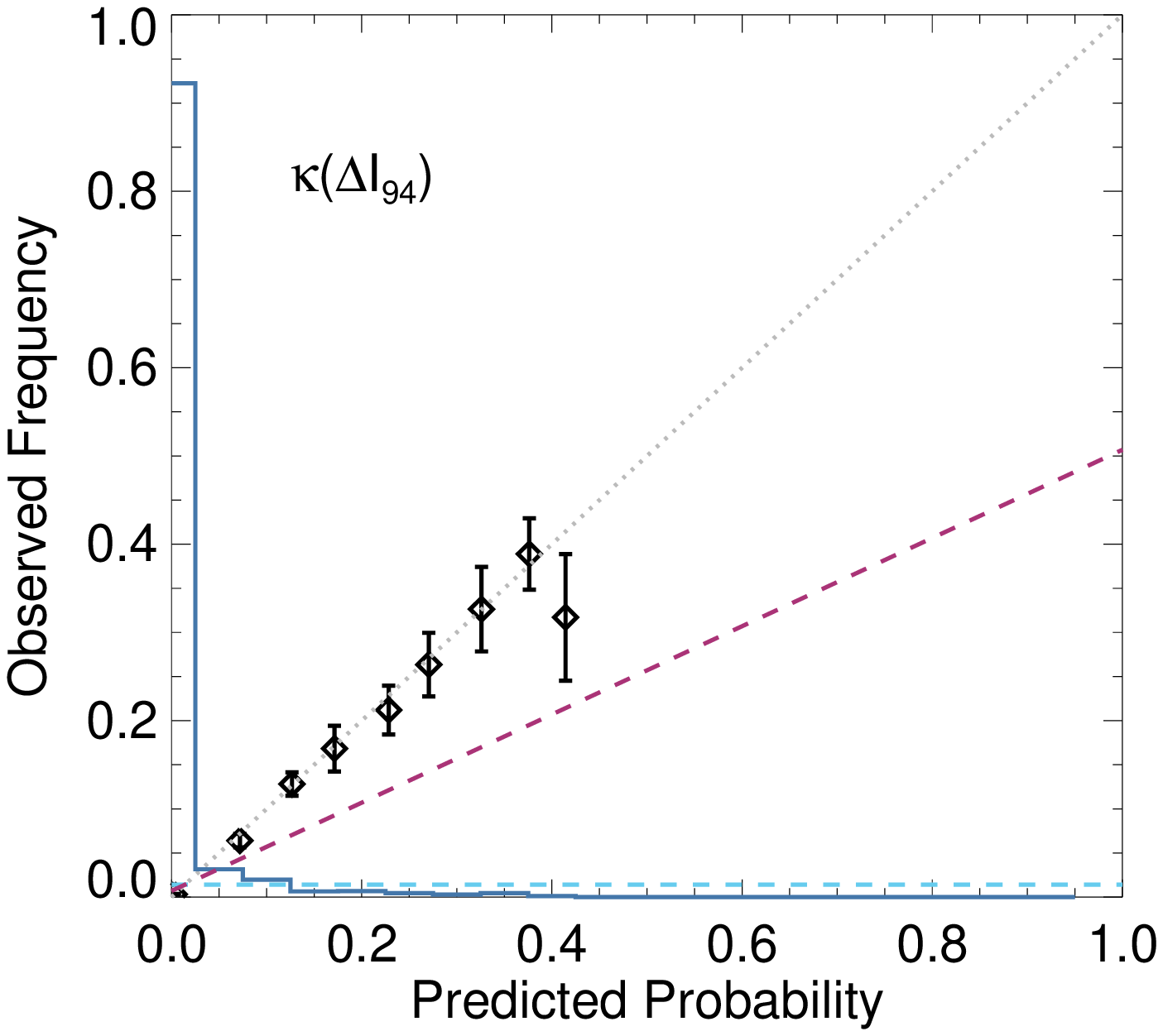}
\includegraphics[width=0.250\textwidth,clip, trim = 8mm 0mm 2mm 5mm, angle=0]{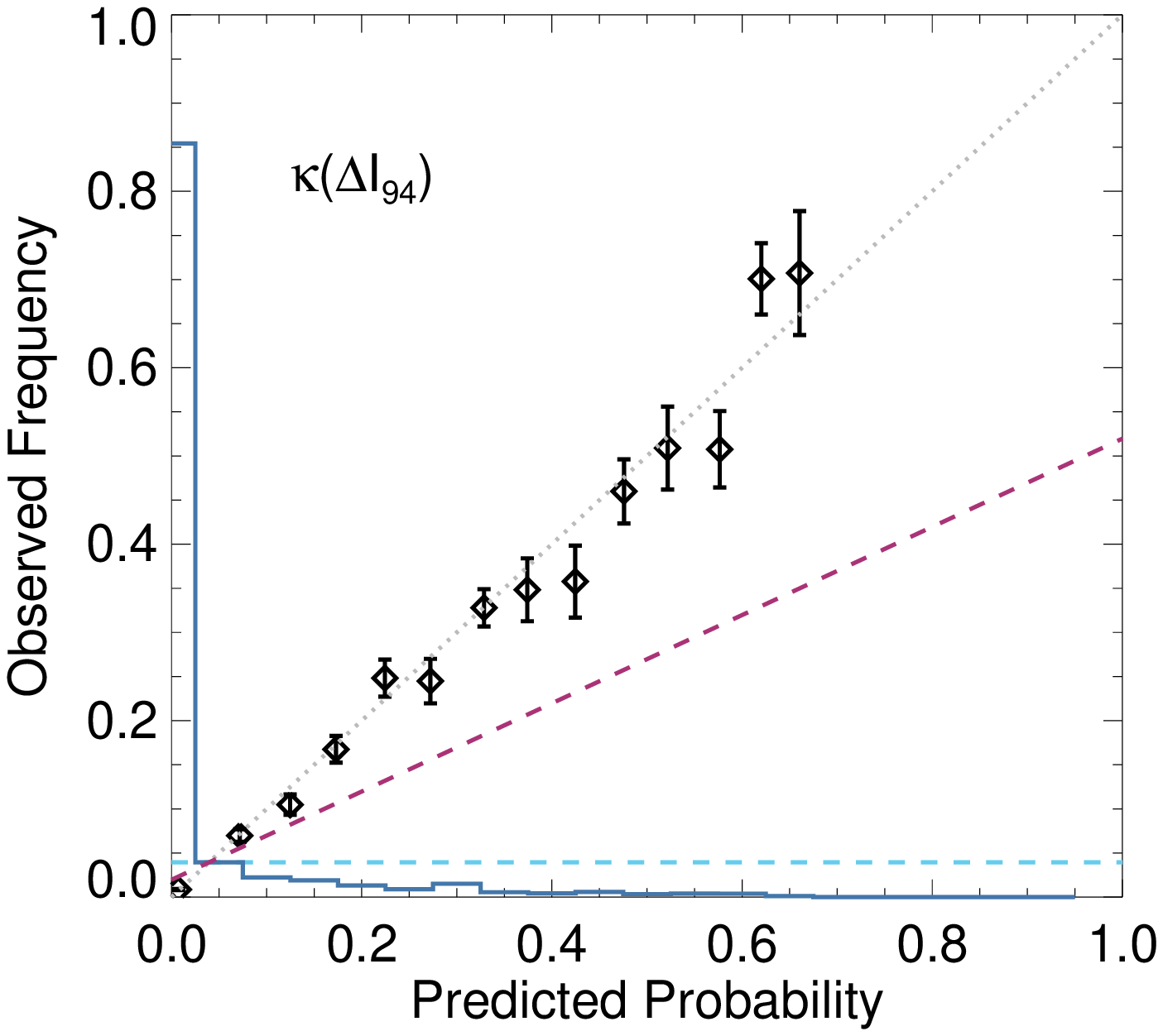}
\includegraphics[width=0.250\textwidth,clip, trim = 8mm 0mm 2mm 5mm, angle=0]{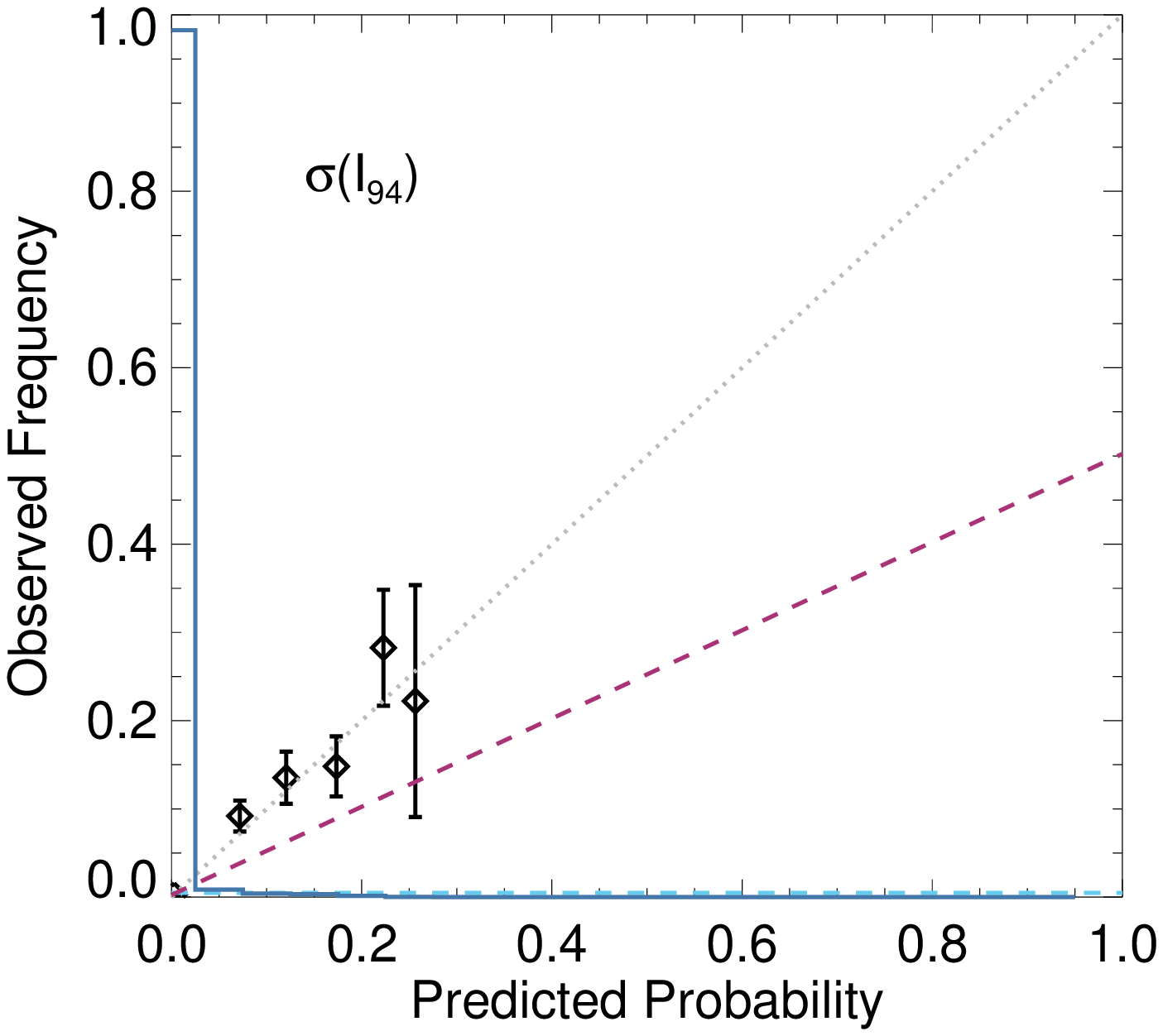}}
\centerline{
\includegraphics[width=0.250\textwidth,clip, trim = 8mm 0mm 2mm 5mm, angle=0]{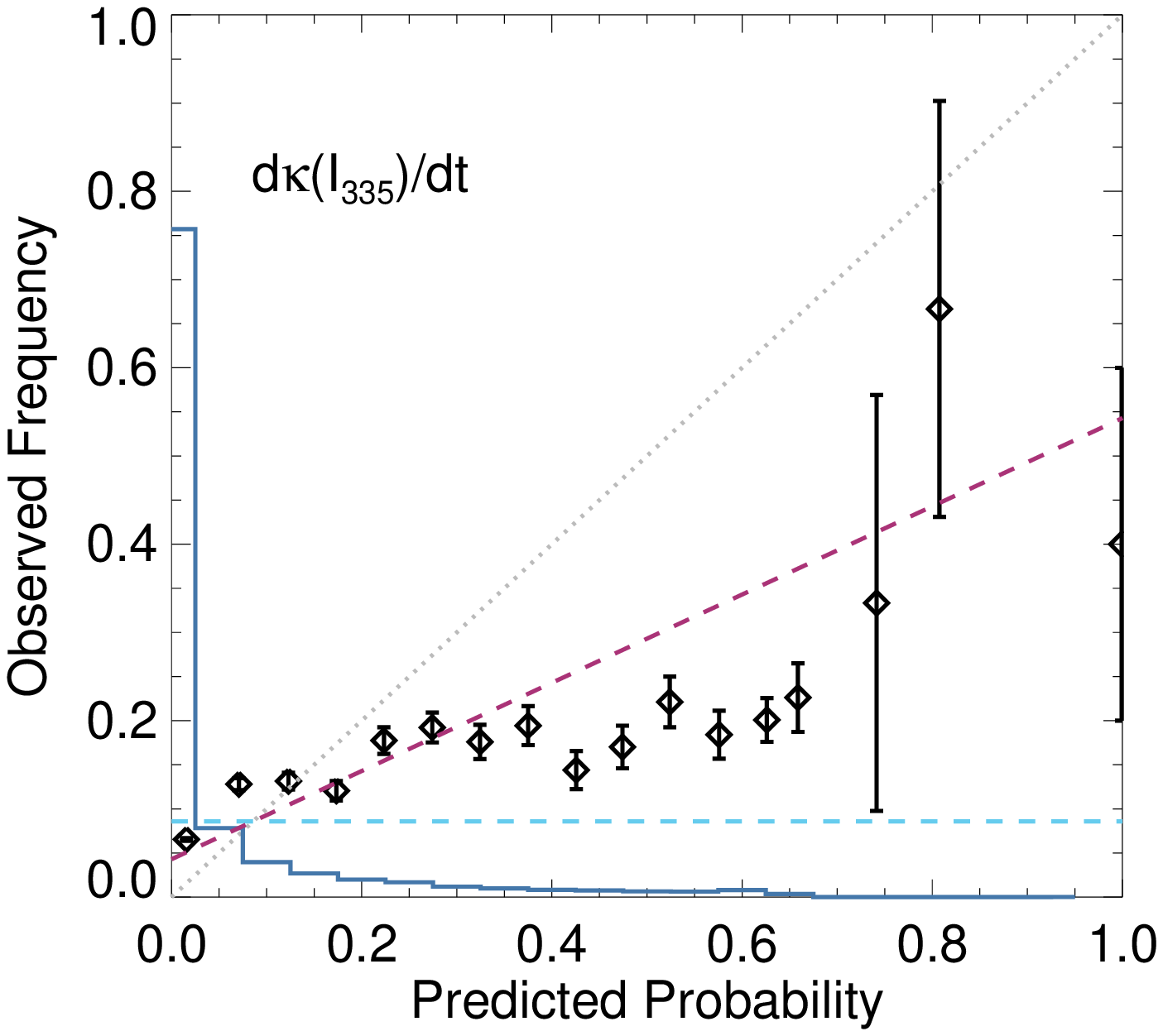}
\includegraphics[width=0.250\textwidth,clip, trim = 8mm 0mm 2mm 5mm, angle=0]{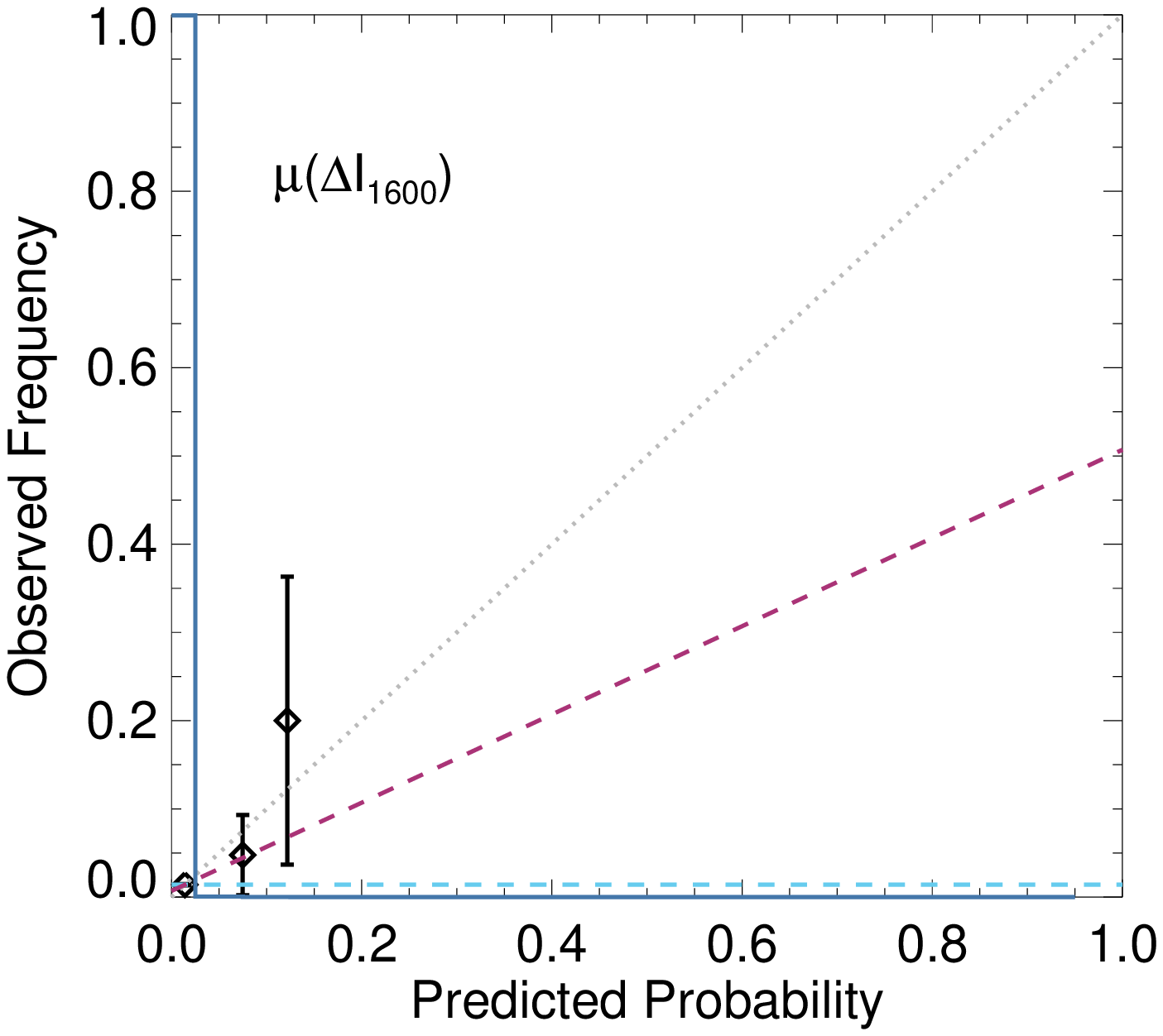}
\includegraphics[width=0.250\textwidth,clip, trim = 8mm 0mm 2mm 5mm, angle=0]{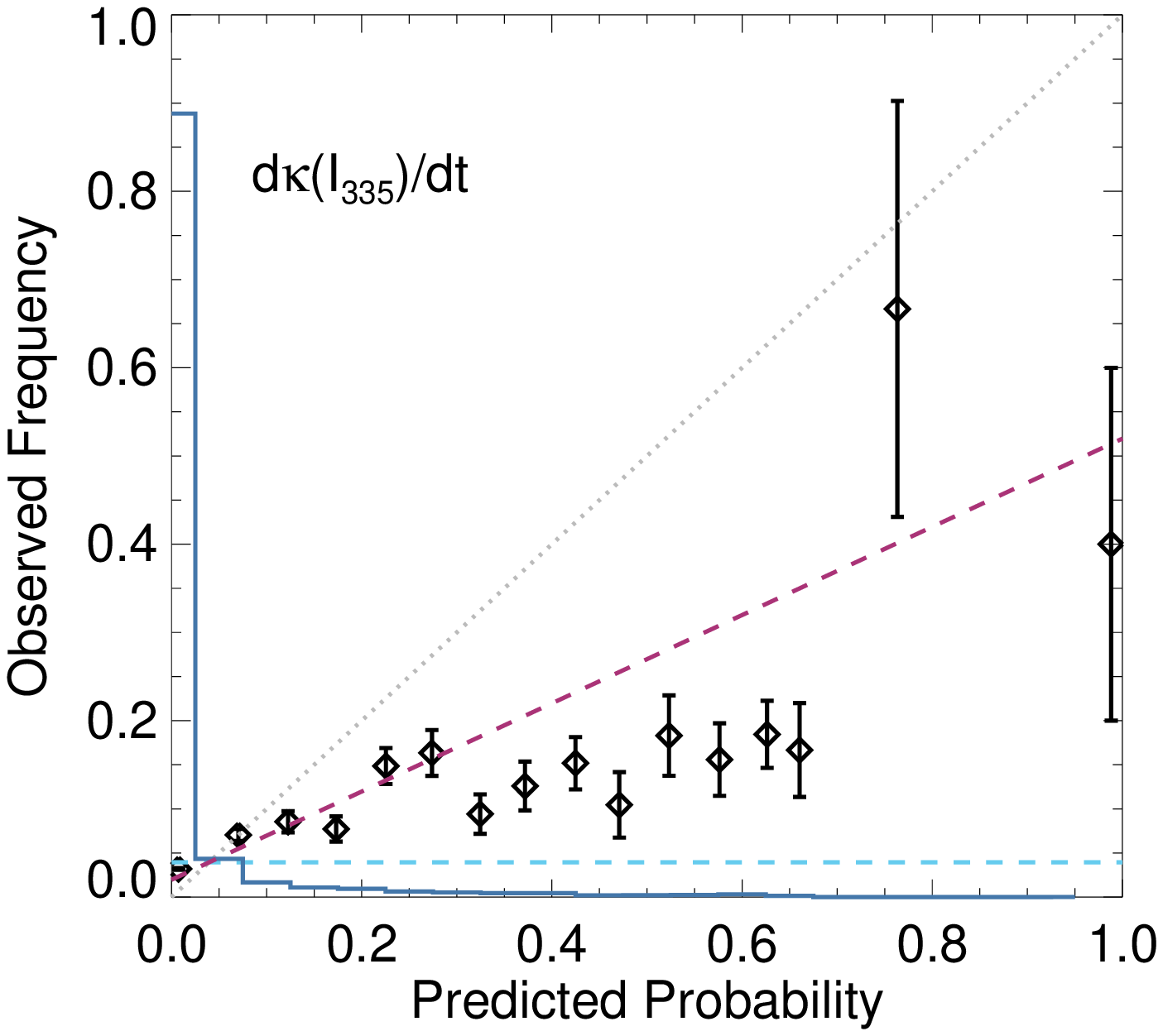}
\includegraphics[width=0.250\textwidth,clip, trim = 8mm 0mm 2mm 5mm, angle=0]{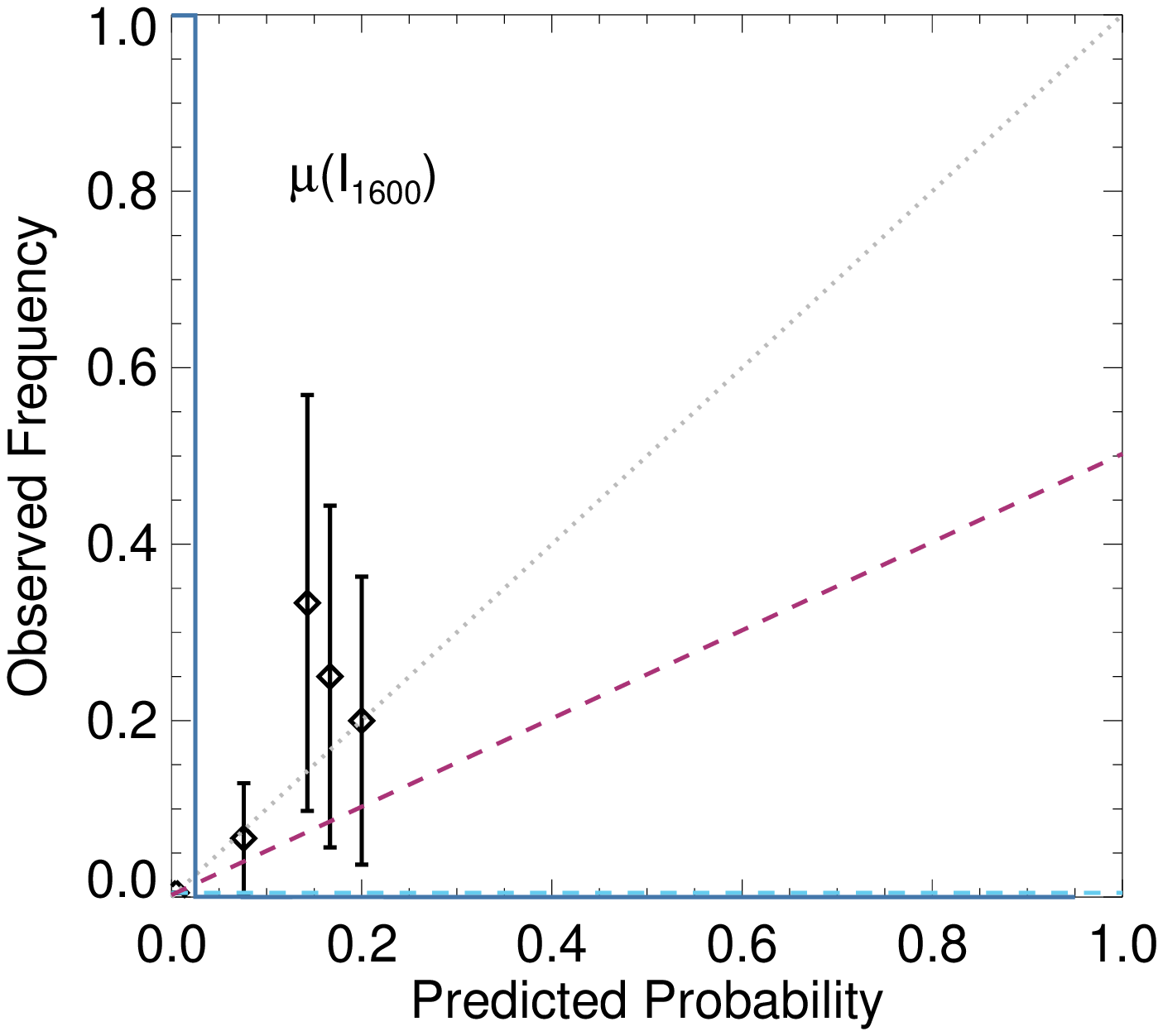}}
\caption{Reliability plots for (Left to Right) \CCL, \MML, \CCS, \MMS\ for top-performing 
parameters (Top) and low-performing parameters (Bottom), according to \BSS,
as indicated.  The $x=y$ line indicates perfect reliability, the histogram (blue) is the 
frequency of occurrence for each prediction bin, the horizontal line (light-blue dashed) indicates the 
climatology (no resolution) and the ``no skill'' line is also plotted (red dashes).  The $1\,\sigma$ error bars 
are shown, and reflect the sample size in each bin.}
\label{fig:reliability}
\end{figure}

While the \BSS\ and $\mathcal{G}$ summarize the Reliability and ROC
plots respectively, it is instructive to see the behaviors explicitly
by which to judge bias, {\it etc}.  ROC plots (Figure~\ref{fig:roc})
and Reliability plots (Figure~\ref{fig:reliability}) are shown for one
of the best and one of the worst-scoring parameters each (according
to \BSS, as per Tables~\ref{tbl:resultsCCL} -- \ref{tbl:resultsMMS}),
for each event definition.

\begin{table}[h]
\begin{center}
\caption{Results: \CCL \label{tbl:resultsCCL}}
\begin{tabular}{|llll|} \hline
\multicolumn{4}{|c|}{Top 10 Scoring Parameters: \CCL} \\
Parameter & Brier Skill Score & \maxtss\ & $\mathcal{G}$ or \ROCSS\ \\ \hline
$\kappa(\Delta{\rm I_{94}})$ & $0.332 \pm 0.011$ & 0.650  & 0.816    \\
$\kappa(\Delta{\rm I_{131}})$  & $0.315 \pm 0.011$ & 0.658 & 0.810\\
$\kappa(\Delta{\rm I_{171}})$ & $0.312 \pm 0.012$ & 0.670 & 0.809  \\
$\kappa(\Delta{\rm I_{304}})$  & $0.310 \pm 0.010$ & 0.668  & 0.812  \\
$\Sigma({\rm I_{94}})$& $0.302 \pm 0.013$ & 0.680*  & 0.830* \\
$\kappa(\Delta{\rm I_{193}})$ & $0.301 \pm 0.011$ & 0.657  &  0.814 \\
$\kappa(\Delta{\rm I_{211}})$ & $0.291 \pm 0.011$ & 0.651 &  0.794 \\ 
$\Sigma(\Delta{\rm I_{94}})$  & $0.286 \pm 0.011$ & 0.626 & 0.788  \\
$\Sigma({\rm I_{335}})$  & $0.280 \pm 0.014$ & 0.672  & 0.822  \\
$d\Sigma(\Delta I_{\rm {94}})/d{\rm t}$ & $0.273 \pm 0.011$ &  0.597 & 0.761\\ \hline

\multicolumn{4}{|c|}{Bottom 5 Scoring Parameters: \CCL} \\
Parameter & Brier Skill Score & \maxtss\ & $\mathcal{G}$ or \ROCSS\ \\ \hline
$d\mu(I_{\rm {193}})/d{\rm t}$ & $ 0.001 \pm 0.000$ &   0.035 &   0.046 \\ 
$\mu(\Delta I_{\rm {171}})$ & $0.001 \pm 0.001$ &   0.052 &   0.047 \\ 
$d\sigma(I_{\rm {171}})/d{\rm t}$ & $0.000 \pm 0.000$ &   0.023 &   0.015 \\ 
$d\kappa(I_{\rm {131}})/d{\rm t}$ & $ -0.011 \pm  0.009$ &    0.203 &    0.283 \\ 
$d\kappa(I_{\rm {335}})/d{\rm t}$ & $ -0.068 \pm   0.028$ &    0.160 &    0.262 \\ \hline
\end{tabular} 
\end{center}
\flushleft{*: Top or Bottom score for \maxtss\ and for $\mathcal{G}$.  In this 
case the worst \maxtss=\,-0.038, and $\mathcal{G}$=\,-0.034 both for $d\varsigma(I_{\rm {211}})/d{\rm t}$
which has \BSS=\,$ 0.001 \pm  0.001$.  }
\tablecomments{Table~\ref{tbl:resultsCCL} is published in its entirety in machine-readable format.
A portion is shown here for guidance regarding its form and content.}
\end{table}

\begin{table}[h]
\begin{center}
\caption{Results: \MML \label{tbl:resultsMML}}
\begin{tabular}{|llll|} \hline
\multicolumn{4}{|c|}{Top 10 Scoring Parameters: \MML} \\
Parameter & Brier Skill Score & \maxtss\ & $\mathcal{G}$ or \ROCSS\ \\ \hline
$\kappa(\Delta{\rm I_{94}})$ & $0.160  \pm 0.015$ & 0.794*  & 0.909* \\
$\kappa(\Delta{\rm I_{131}})$  & $0.132 \pm 0.010$ & 0.734 & 0.862 \\
$\Sigma(\Delta{\rm I_{94}})$  & $0.131 \pm 0.015$ & 0.704 & 0.840 \\
$d\kappa(\Delta{\rm I_{94}}) / d{\rm t}$ & $0.125 \pm 0.018$ & 0.680 & 0.837 \\
$\Sigma(\Delta I_{\rm {131}})$ & $0.118 \pm 0.021$ & 0.640 & 0.786 \\ 
$\kappa(\Delta{\rm I_{211}})$ & $0.117 \pm 0.008$ & 0.750 & 0.863 \\ 
$d\varsigma(\Delta {\rm I_{94}}) / d{\rm t}$ & $0.116 \pm 0.009$ & 0.640 & 0.802 \\
$\kappa(\Delta{\rm I_{304}})$ & $0.116 \pm 0.010$ & 0.725 & 0.851 \\
$d\varsigma(\Delta {\rm I_{131}}) / d{\rm t}$ & $0.110 \pm 0.016$ & 0.658 & 0.812 \\
$d\Sigma(\Delta{\rm I_{131}}) / d{\rm t}$ & $0.109 \pm 0.017$ & 0.626 & 0.764 \\ \hline
\multicolumn{4}{|c|}{Bottom 5 Scoring Parameters: \MML} \\
Parameter & Brier Skill Score & \maxtss\ & $\mathcal{G}$ or \ROCSS\ \\ \hline
$d\sigma(I_{\rm {171}})/d{\rm t}$ & $0.000 \pm 0.000$ &  0.049 &   0.038\\ 
$d\mu(I_{\rm {171}})/d{\rm t}$ & $ 0.000 \pm 0.000$ &  0.028 &   0.019 \\ 
$d\mu(\Delta I_{\rm {1600}})/d{\rm t}$ & $0.000 \pm 0.000$ & -0.031* & -0.036*\\ 
$\mu(\Delta I_{\rm {1600}})$ & $0.000 \pm 0.000$ & -0.031 & -0.024 \\ 
$d\mu(I_{\rm {131}})/d{\rm t}$ & $-0.002 \pm 0.004$ & 0.157 & 0.197 \\ \hline
\end{tabular} 
\end{center}
\flushleft{*: Top or Bottom score for \maxtss\ and for $\mathcal{G}$.}
\tablecomments{Table~\ref{tbl:resultsMML} is published in its entirety in machine-readable format.
A portion is shown here for guidance regarding its form and content.}

\end{table}

\begin{table}[h]
\begin{center}
\caption{Results: \CCS \label{tbl:resultsCCS}}
\begin{tabular}{|llll|} \hline
\multicolumn{4}{|c|}{Top 10 Scoring Parameters: \CCS} \\
Parameter & Brier Skill Score & \maxtss\ & $\mathcal{G}$ or \ROCSS\ \\ \hline
$\kappa(\Delta{\rm I_{94}})$  & $0.247 \pm 0.010$ & 0.703* & 0.853* \\
$\kappa(\Delta{\rm I_{131}})$ & $0.214 \pm 0.010$ & 0.684 & 0.828 \\ 
$\Sigma(\Delta{\rm I_{94}})$  & $0.207 \pm 0.012$ & 0.675 & 0.816 \\
$\kappa(\Delta{\rm I_{211}})$  & $0.203 \pm 0.008$ & 0.669 & 0.818 \\
$\kappa(\Delta {\rm I_{171}})$ & $0.199 \pm 0.008$ & 0.685 & 0.828\\
$\kappa(\Delta {\rm I_{304}})$ & $0.199 \pm 0.007$ & 0.681 & 0.820\\
$\kappa(\Delta {\rm I_{193}})$ & $0.199 \pm 0.008$ & 0.676 & 0.829 \\
$d\Sigma(\Delta {\rm I_{94}})  / d{\rm t}$ & $0.196 \pm 0.014$ & 0.622  & 0.775 \\
$\varsigma(\Delta I_{\rm {94}})$ & $0.192 \pm 0.012$ &    0.567 & 0.717 \\ 
$d\varsigma(\Delta I_{\rm {94}})/d{\rm t}$ & $ 0.184 \pm 0.012$ & 0.577 & 0.735 \\  \hline
\multicolumn{4}{|c|}{Bottom 5 Scoring Parameters: \CCS} \\
Parameter & Brier Skill Score & \maxtss\ & $\mathcal{G}$ or \ROCSS\ \\ \hline
$d\mu(I_{\rm {193}})/d{\rm t}$ & $0.000 \pm 0.000$ &   0.027 &   0.050 \\ 
$d\mu(I_{\rm {171}})/d{\rm t}$ & $0.000 \pm 0.000$ &   0.014* &  0.002* \\ 
$d\sigma(I_{\rm {171}})/d{\rm t}$ & $0.000 \pm 0.000$ & 0.021 & 0.027 \\ 
$d\kappa(I_{\rm {211}})/d{\rm t}$ & $0.000 \pm  0.004$ & 0.217 & 0.269 \\ 
$d\kappa(I_{\rm {335}})/d{\rm t}$ & $ -0.066 \pm  0.024$ & 0.188 & 0.320 \\ \hline
\end{tabular} 
\end{center}
\flushleft{*: Top or Bottom score for \maxtss\ and for $\mathcal{G}$.}
\tablecomments{Table~\ref{tbl:resultsCCS} is published in its entirety in machine-readable format.
A portion is shown here for guidance regarding its form and content.}
\end{table}

\begin{table}[h]
\begin{center}
\caption{Results: \MMS \label{tbl:resultsMMS}}
\begin{tabular}{|llll|} \hline
\multicolumn{4}{|c|}{Top 10 Scoring Parameters: \MMS} \\
Parameter & Brier Skill Score & \maxtss\ & $\mathcal{G}$ or \ROCSS\ \\ \hline
$\kappa(\Delta I_{\rm {94}})$ & $ 0.070 \pm 0.014$ &  0.821* &  0.913* \\ 
$\sigma(I_{\rm {94}})$ & $  0.070 \pm   0.018$ & 0.707 & 0.860 \\ 
$d\varsigma(\Delta I_{\rm {131}})/d{\rm t}$ & $ 0.067 \pm 0.017$ & 0.701 & 0.846 \\ 
$\Sigma(\Delta I_{\rm {131}})$ & $  0.061 \pm 0.023$ & 0.646 & 0.806 \\ 
$\varsigma(\Delta I_{\rm {131}})$ & $  0.058 \pm 0.014$ &    0.720 & 0.819 \\ 
$d\Sigma(\Delta I_{\rm {131}})/d{\rm t}$ & $  0.056 \pm 0.011$ &  0.624 & 0.810 \\ 
$\kappa(\Delta I_{\rm {131}})$ & $  0.056 \pm 0.011$ & 0.778 & 0.886 \\ 
$\Sigma(\Delta I_{\rm {94}})$ & $  0.055 \pm 0.017$ & 0.708 & 0.836 \\ 
$\sigma(\Delta I_{\rm {131}})$ & $  0.054 \pm 0.019$ & 0.575 & 0.757 \\ 
$\varsigma(\Delta I_{\rm {94}})$ & $  0.054 \pm 0.019$ & 0.661 &  0.774 \\  \hline
\multicolumn{4}{|c|}{Bottom 5 Scoring Parameters: \MMS} \\
Parameter & Brier Skill Score & \maxtss\ & $\mathcal{G}$ or \ROCSS\ \\ \hline
$d\sigma(I_{\rm {171}})/d{\rm t}$ & $0.000 \pm 0.000$ &   0.042 &   0.036 \\ 
$\mu(I_{\rm {171}})$ & $0.001 \pm  0.002$ & 0.268 &  0.340 \\ 
$\mu(\Delta I_{\rm {1600}})$ & $0.001 \pm  0.001$ & 0.060 & 0.046 \\ 
$\mu(I_{\rm {1600}})$ & $-0.001 \pm  0.001$ &    0.141 & 0.139 \\ 
$d\mu(I_{\rm {131}})/d{\rm t}$ & $-0.003 \pm 0.006$ & 0.150 & 0.259 \\ \hline
\end{tabular} 
\flushleft{*: Top score for \maxtss\ and for $\mathcal{G}$.  In this case the worst
\maxtss= -0.104, $\mathcal{G}=-0.116$ both for $d\mu(I_{\rm {171}})/d{\rm t}$ 
which has \BSS=\,$0.000 \pm 0.000$.}
\tablecomments{Table~\ref{tbl:resultsMMS} is published in its entirety in machine-readable 
format.  A portion is shown here for guidance regarding its form and content.}
\end{center}
\end{table}

Overall, the classification results for select UV/EUV parameters
show confidence at statistically significant levels for the 
\CCL, \CCS, and \MML\ event definitions.  By this we mean that the 
sample sizes are large enough that the bootstrap-derived uncertainties
in the \BSS, plus the AARP-focused cross validation, provide good
estimates of the uncertainties and that the \BSS\ results indicate
skill above climatology (\BSS$>0.0$).  We did not perform a separate
bootstrap or sorting for the other metrics provided, but assume that
the (un)certainty levels are similar.  As has been found in other studies,
there are numerous parameters that perform similarly within the error bars.

The uncertainties related to the \BSS\ results are overall small especially compared
to the \BSS\ results for \CCL\ and \CCS.
For \MMS, while the larger error bars reflect a smaller sample of events,
the \BSS\ results barely indicate skill above the climatology.
The reliability plots (Figure~\ref{fig:reliability}) for the better
performing metrics do show a good correspondence between the predicted
probabilities and the observed frequency of occurrence, the points generally 
falling within their $1\,\sigma$ error bars of the $x=y$ line.  In other words, even
for the \MMS\ events and even with their low \BSS, the predictions
are ``reliable''.   However, the vast majority of the predictions
(especially for the \MML\ and \MMS\ events) are probabilities
close to the event rates, and this lack of sharpness is reflected in the low \BSS. 

However, the $\mathcal{G}$ results are quite high, generally, as are the 
\maxtss.  For rare events, as displayed in the ROC plots (Figure~\ref{fig:roc}),
the metrics reward a high probability of detection at the 
expense of an increased false alarm rate. Thus the predictions 
have good ability to distinguish between the event and non-event
populations, or good resolution.

Overall, the class imbalance in all event definitions, but especially
the \MML\ and \MMS\ as we define them here, is extreme.  This can lead
to impressive \maxtss\ scores.  Simultaneously, the \BSS\ is negatively
impacted by the class imbalance although it takes the climatology into
account since the climatology provides the reference prediction.

The best-performing parameters across the four event definitions are
dominated by the kurtosis of the running-difference images.  The kurtosis
detects deviation from a Gaussian distribution in terms of central peak {\it
vs.} wing relative strength.  An enhanced kurtosis or leptokurtic distribution,
which is associated with an increased probability of flaring, has an
over-population of the wings relative to a normal distribution, although it can
also indicate an under-population of the central peak (and {\it vice versa} for
a low kurtosis or platykurtic distribution). In terms of moments, the remaining
best-performing parameters are typically either the skew or the total of the
running-difference images.

There are fewer direct-image (\textit{vs.} running-difference image) and evolution
(``$d{\rm X}/d{\rm t}$'') parameters than expected in the top-10 across event
definitions (fewer than 5 of 10); evolution-based parameters in fact tend to
dominate the low-scoring \BSS\ results.  As mentioned in
Section~\ref{sec:metrics}, the ``$d{\rm X}/d{\rm t}$'' parameters may be more
susceptible to outliers, and looking beyond the top-10 their frequency becomes
higher although running-difference images still dominate over direct 
images.  The $\cos(\theta)$ location (observing angle) parameter shows
minimal but not zero classification power.  This result is due to the
HARP selection criteria that includes numerous small plage regions 
at greater absolute latitudes than spot-containing active regions.
These plage regions generally belong to the ``no-event'' population, 
providing a small discriminating advantage to the middle latitudes
and the corresponding $\cos(\theta)$ ranges.

\subsubsection{Wavelength-compared Classification Performance}
\label{sec:wave}

The different filters of AIA are sensitive to plasma
at different temperatures, and often sensitive to more
than one temperature \citep{aia_Lemen}.  The behavior of the
plasma in the corresponding physical regimes may reflect different
thermal or density responses to energy build up, or 
different kinematic responses to photospheric driving
motions, for example.  To address these questions, we first simply 
evaluate the parameters' performance as grouped by wavelength; in 
Section~\ref{sec:interp_wave} we discuss more the physical implications
of the results.

A cursory look at Tables~\ref{tbl:resultsCCL}--\ref{tbl:resultsMMS} gives
the impression that filters which detect hotter plasma more frequently
appear in the ``Top-10'', across event definitions.  The \ion{C}{4}
1600\AA-based parameters are never in the ``top-10'', the \ion{He}{2}
304\AA- and \ion{Fe}{9} 171\AA-based parameters do make the top tiers
in \BSS\ but rarely.  The top parameters are dominated by parameters
built from the \ion{Fe}{18} 94\AA\ filter and the other filters
sensitive to hotter plasma, for example the \ion{Fe}{21}-sensitive
131\AA\ filter.  We note that the top-performing parameters for
the \CC\ event definitions include parameters across all analyzed
EUV filters, while for the \MM\ event definitions the top-ranked
parameters are predominantly those derived from 94 and 131 \AA\ filters
(Tables~\ref{tbl:resultsCCL}--\ref{tbl:resultsMMS}).

\begin{figure}
\centerline{
\includegraphics[width=0.250\textwidth,clip, trim = 0mm 0mm 0mm 0mm, angle=0]{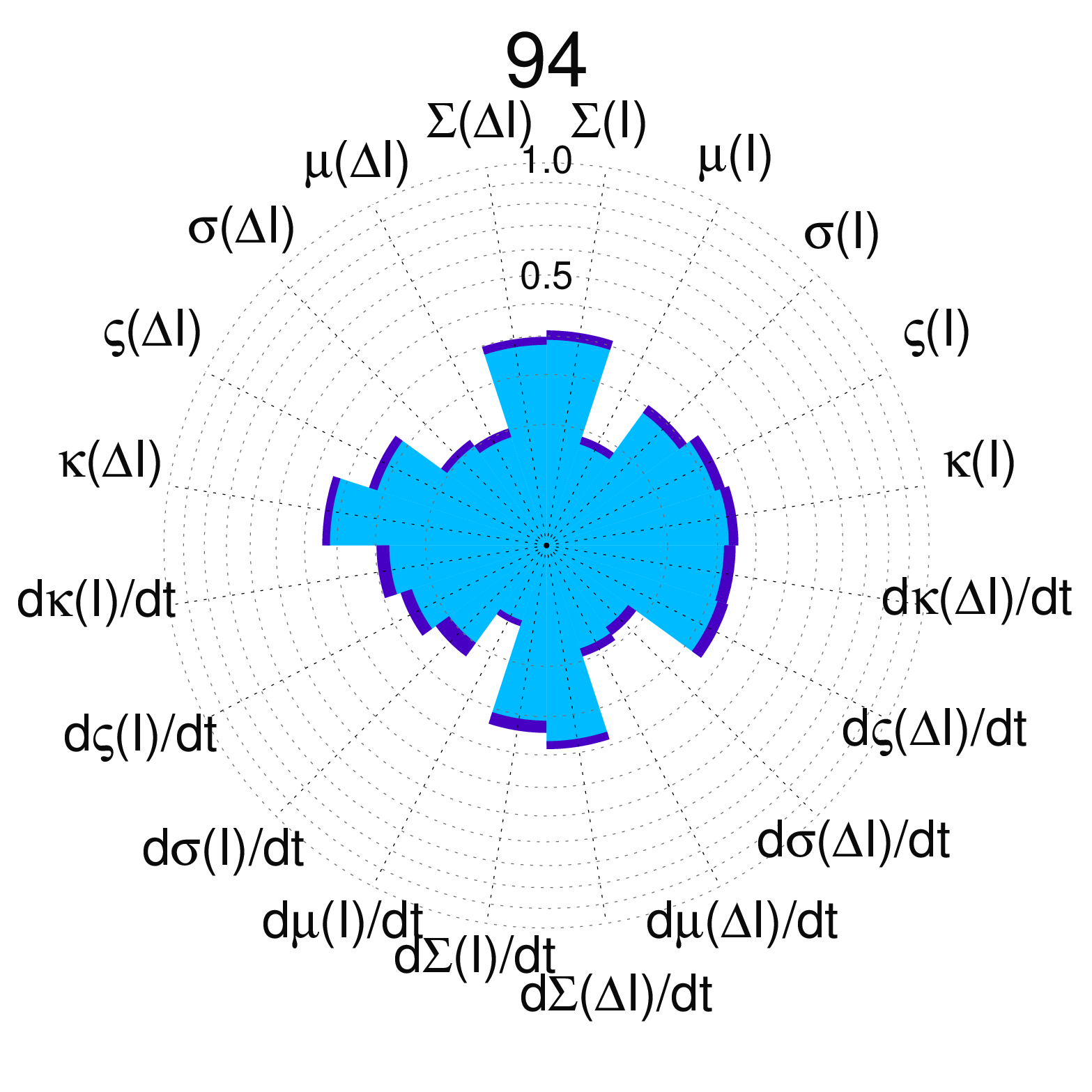}
\includegraphics[width=0.250\textwidth,clip, trim = 0mm 0mm 0mm 0mm, angle=0]{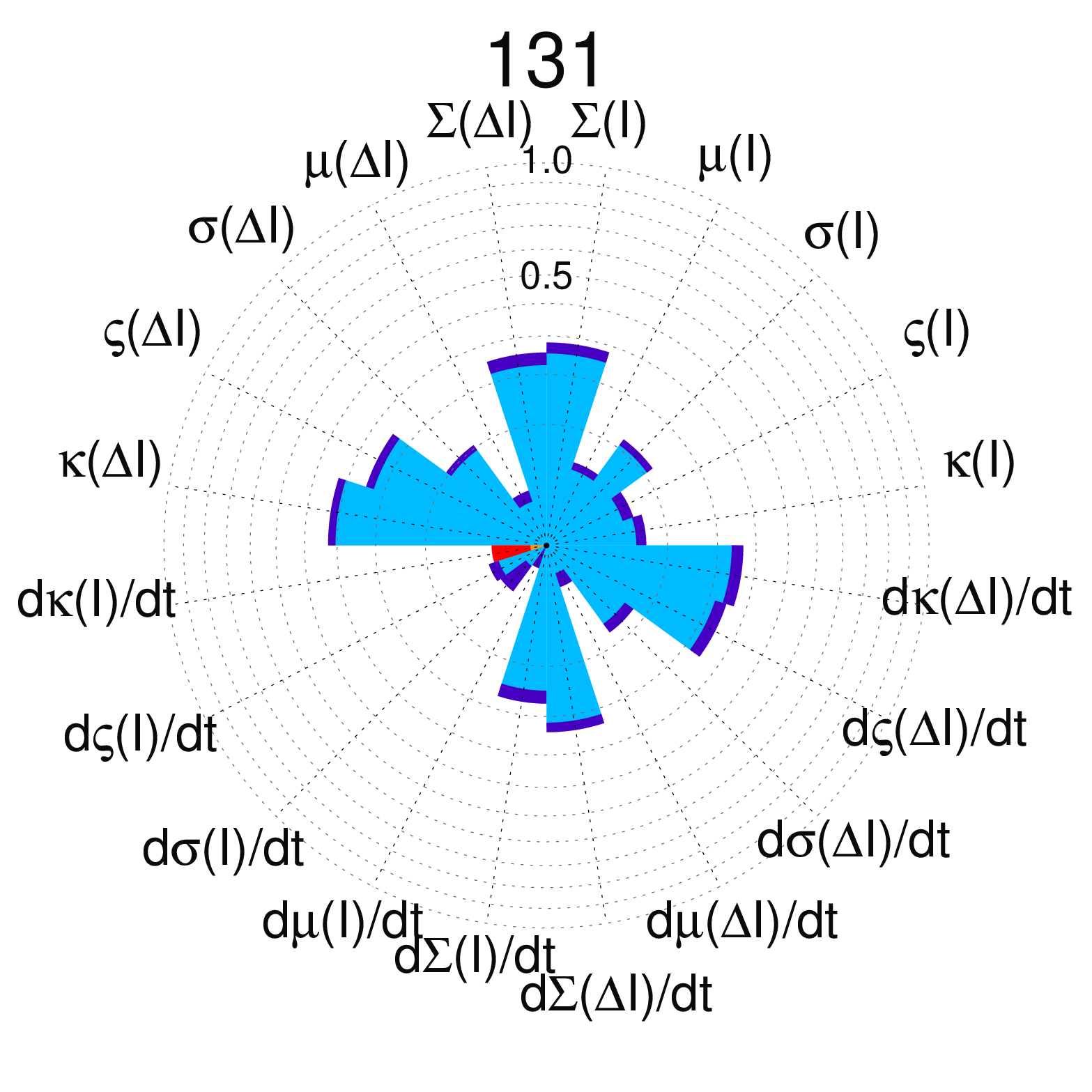}
\includegraphics[width=0.250\textwidth,clip, trim = 0mm 0mm 0mm 0mm, angle=0]{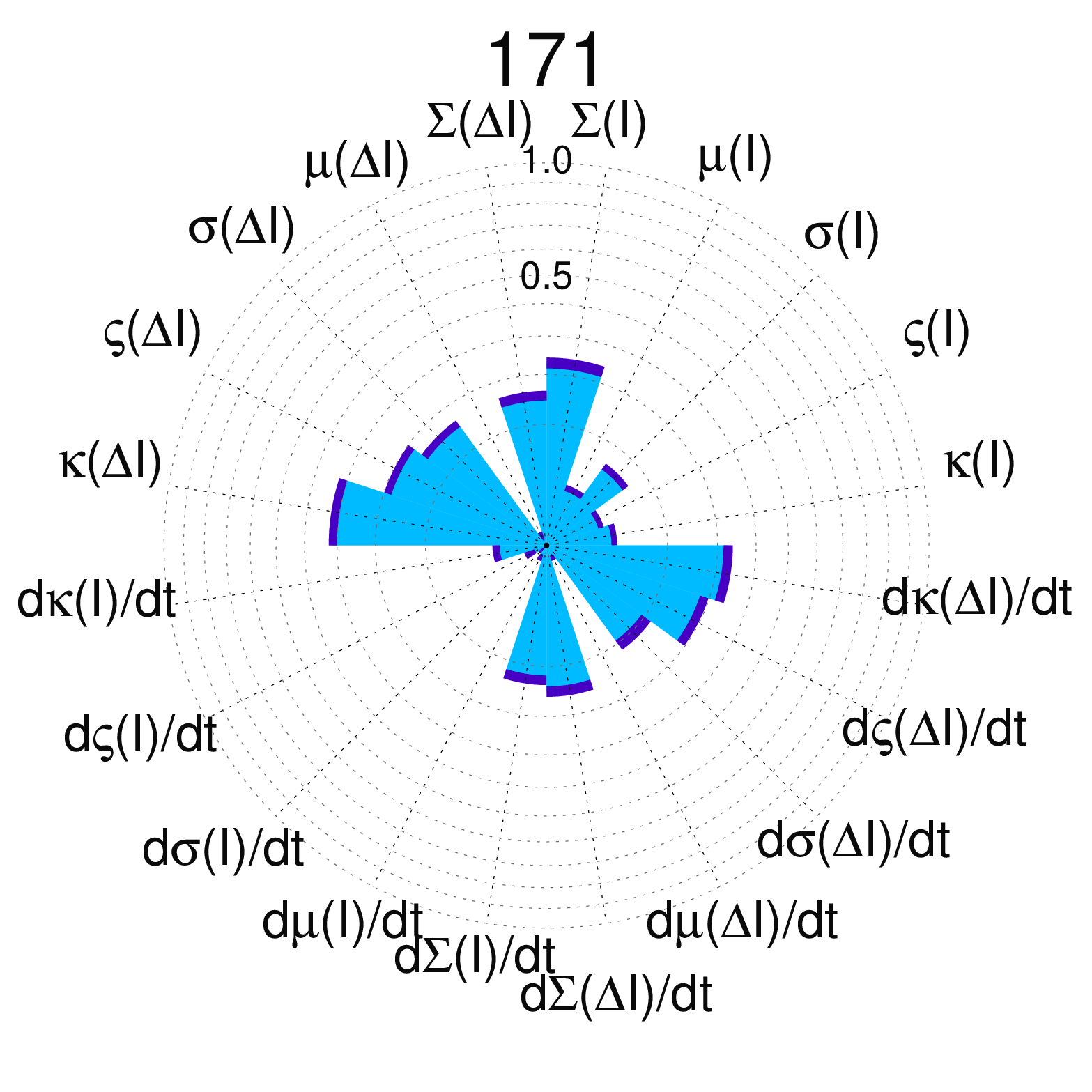}
\includegraphics[width=0.250\textwidth,clip, trim = 0mm 0mm 0mm 0mm, angle=0]{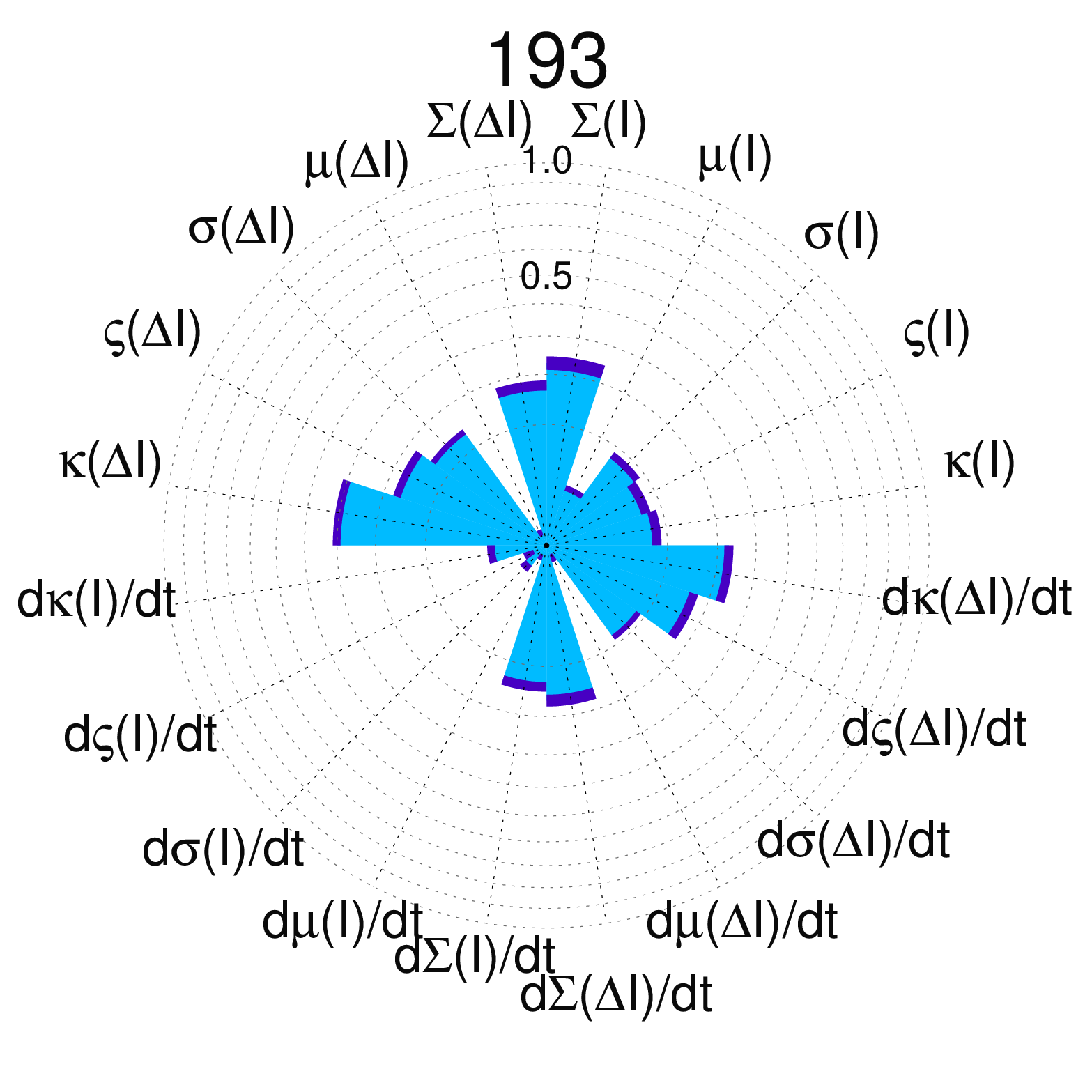}}
\centerline{
\includegraphics[width=0.250\textwidth,clip, trim = 0mm 0mm 0mm 0mm, angle=0]{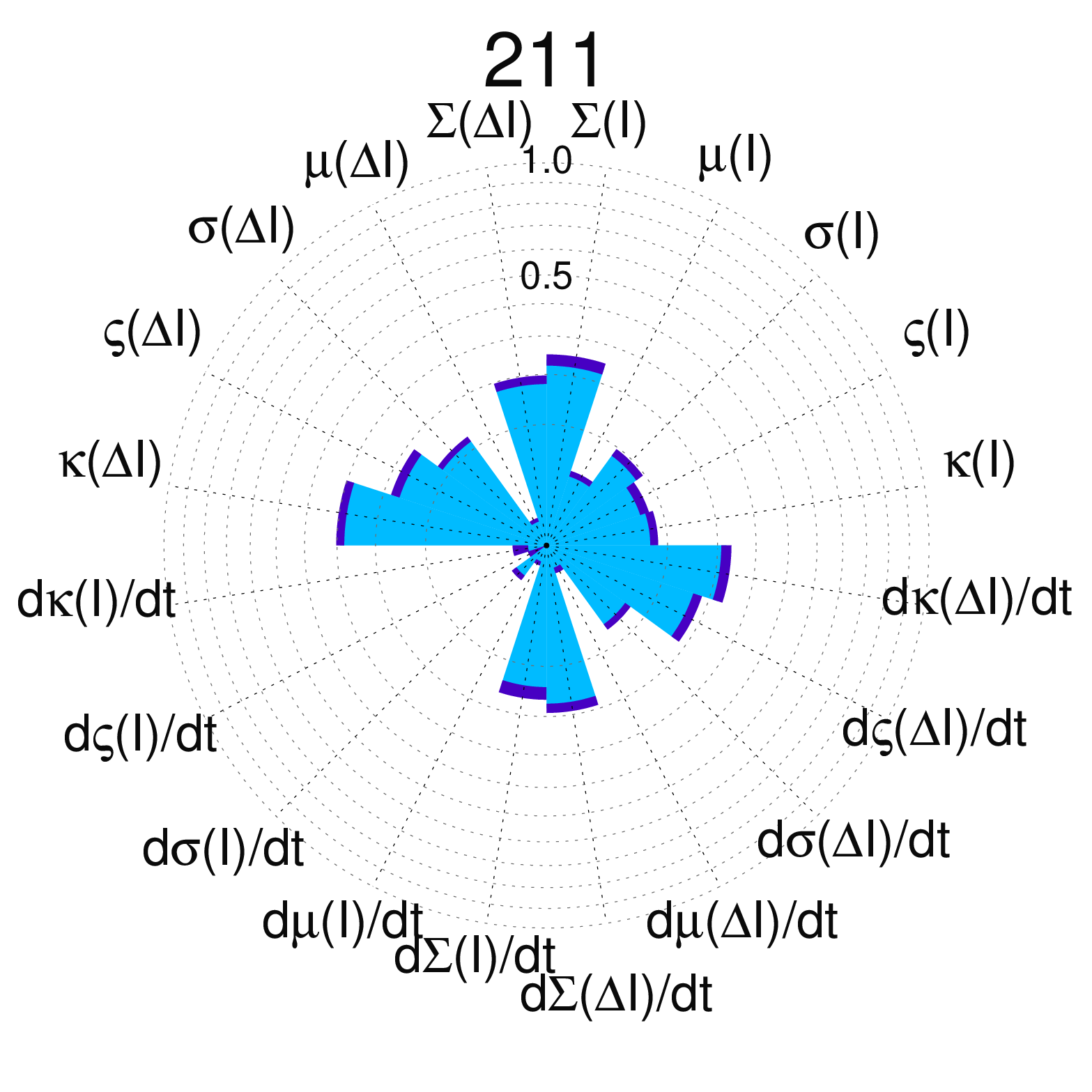}
\includegraphics[width=0.250\textwidth,clip, trim = 0mm 0mm 0mm 0mm, angle=0]{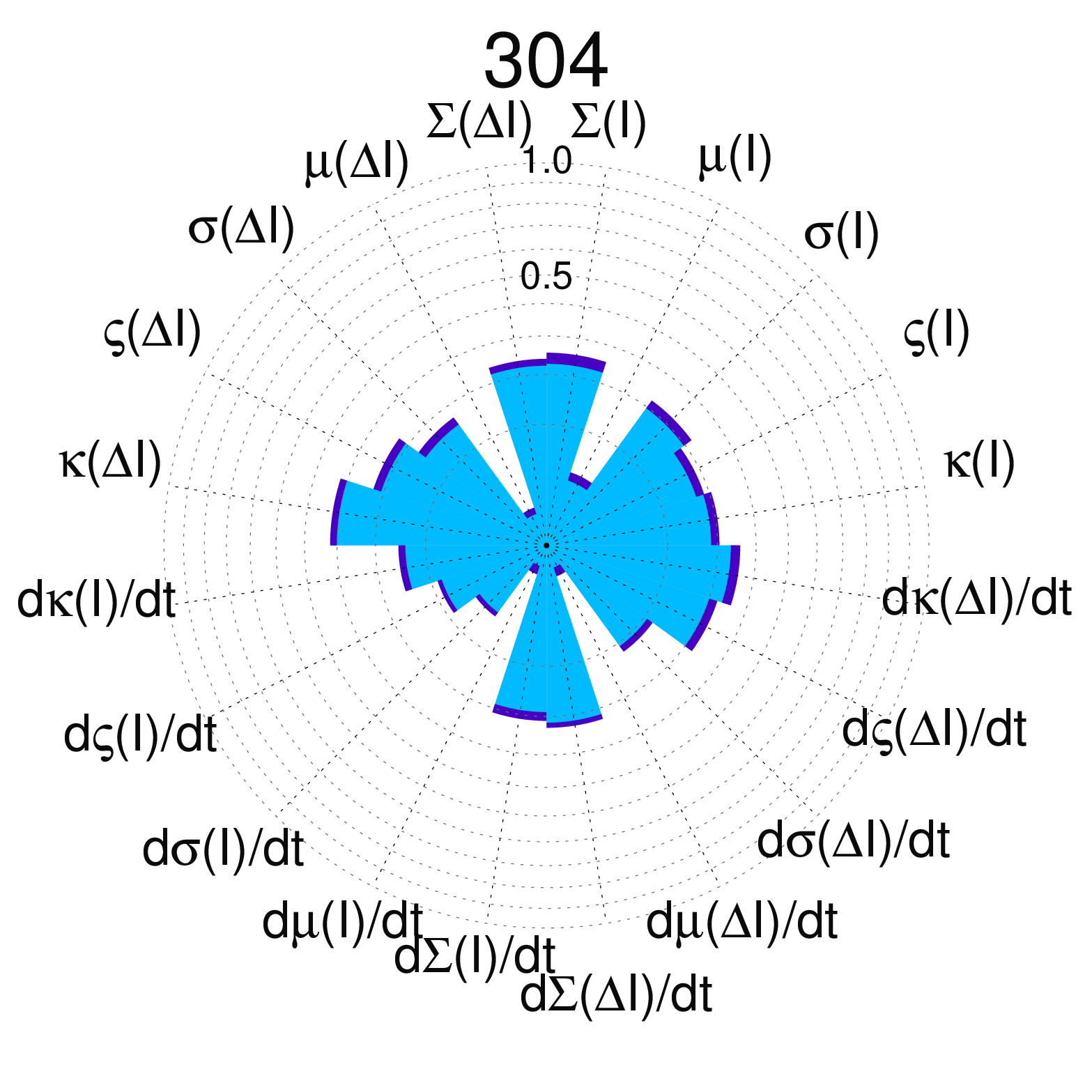}
\includegraphics[width=0.250\textwidth,clip, trim = 0mm 0mm 0mm 0mm, angle=0]{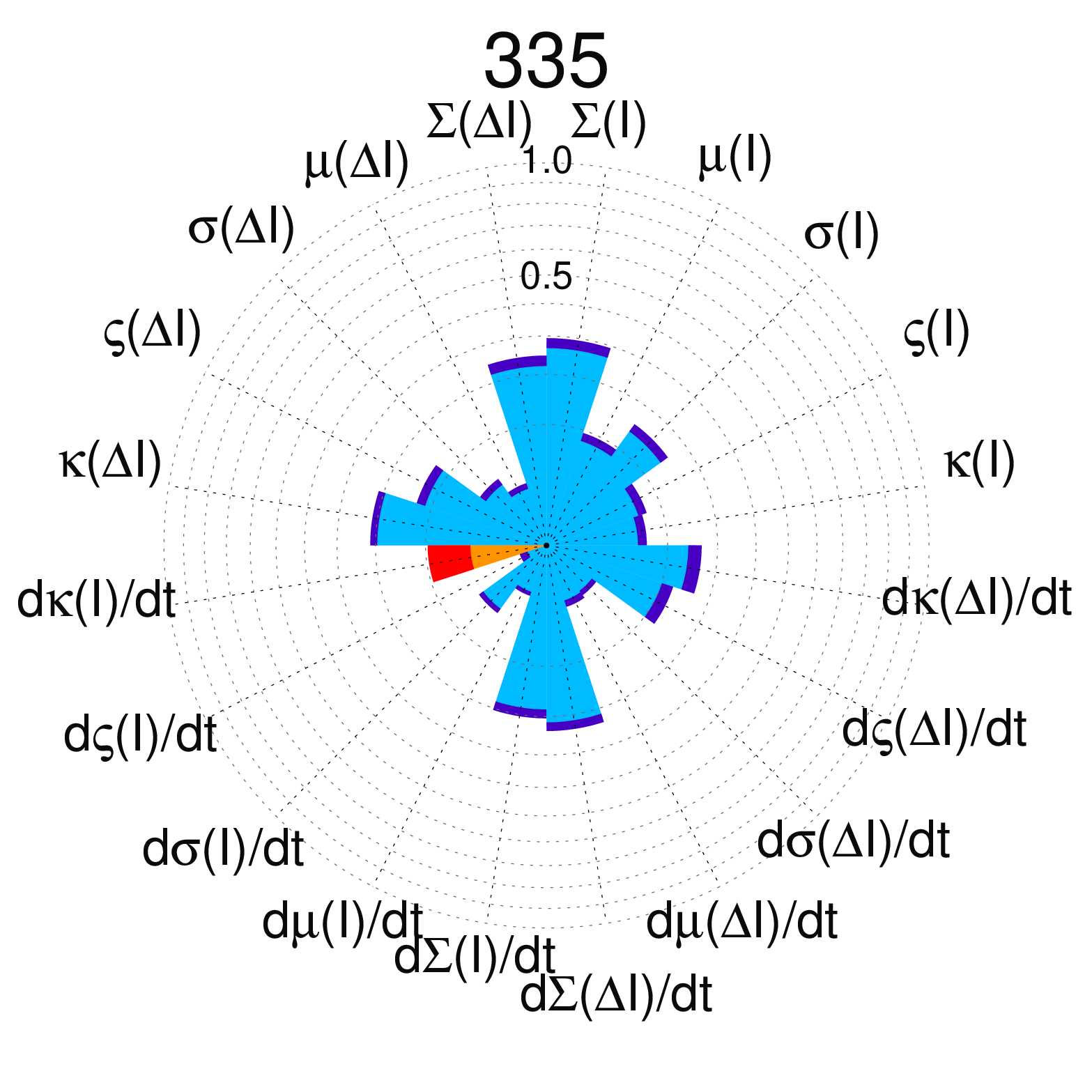}
\includegraphics[width=0.250\textwidth,clip, trim = 0mm 0mm 0mm 0mm, angle=0]{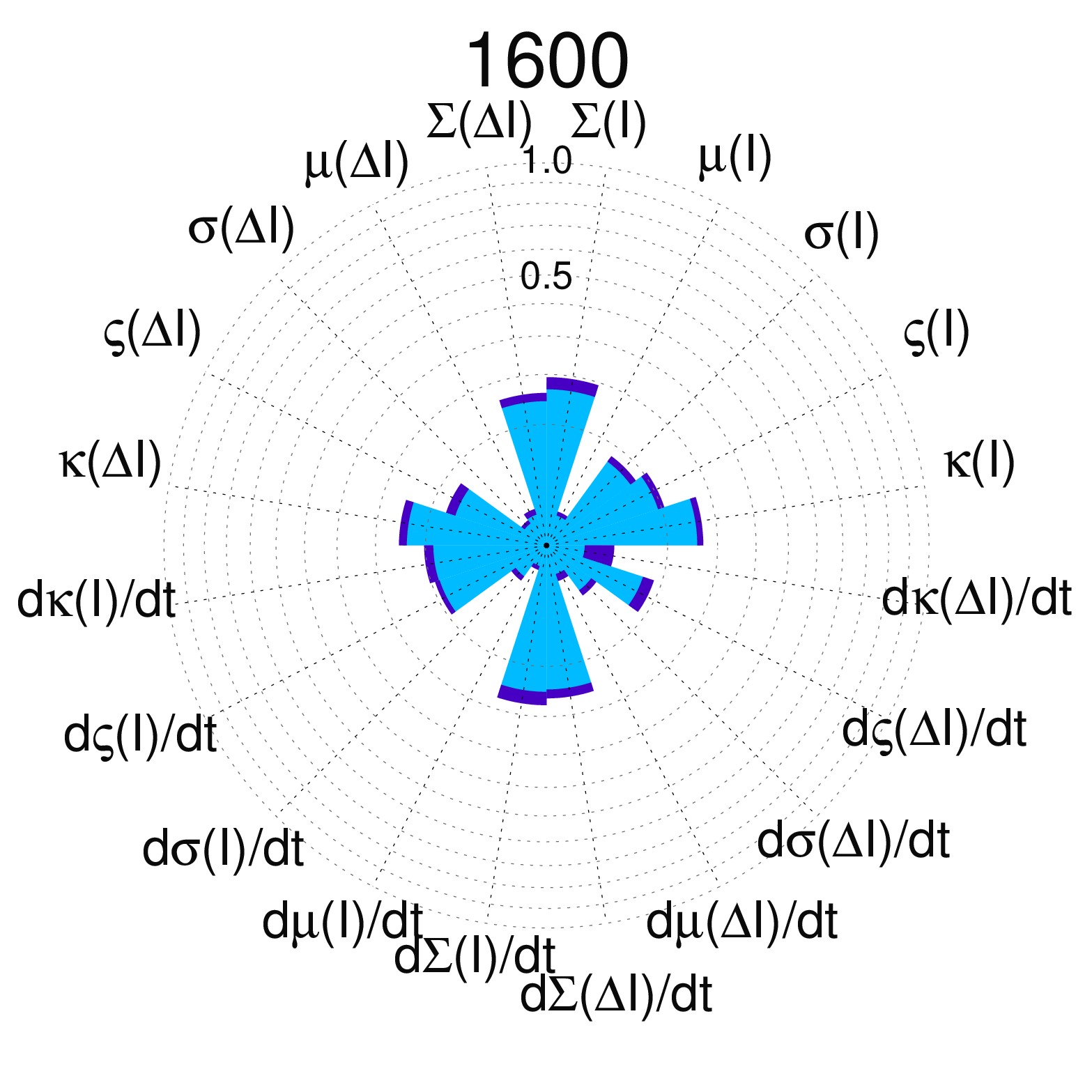}}
\caption{Radar plots showing Brier Skill Score for all parameters, grouped by filter, as labeled.
Arcs indicate the range of the \BSS$\pm \sigma_{\rm BSS}$, \BSS$>0$ (blue) and $|\BSS|$ for \BSS$<0$
(orange), with darker hues indicating the uncertainty ranges. Shown: \CCL\ event definition results.}
\label{fig:radar_C1_24}
\end{figure}

\begin{figure}
\centerline{
\includegraphics[width=0.250\textwidth,clip, trim = 0mm 0mm 0mm 0mm, angle=0]{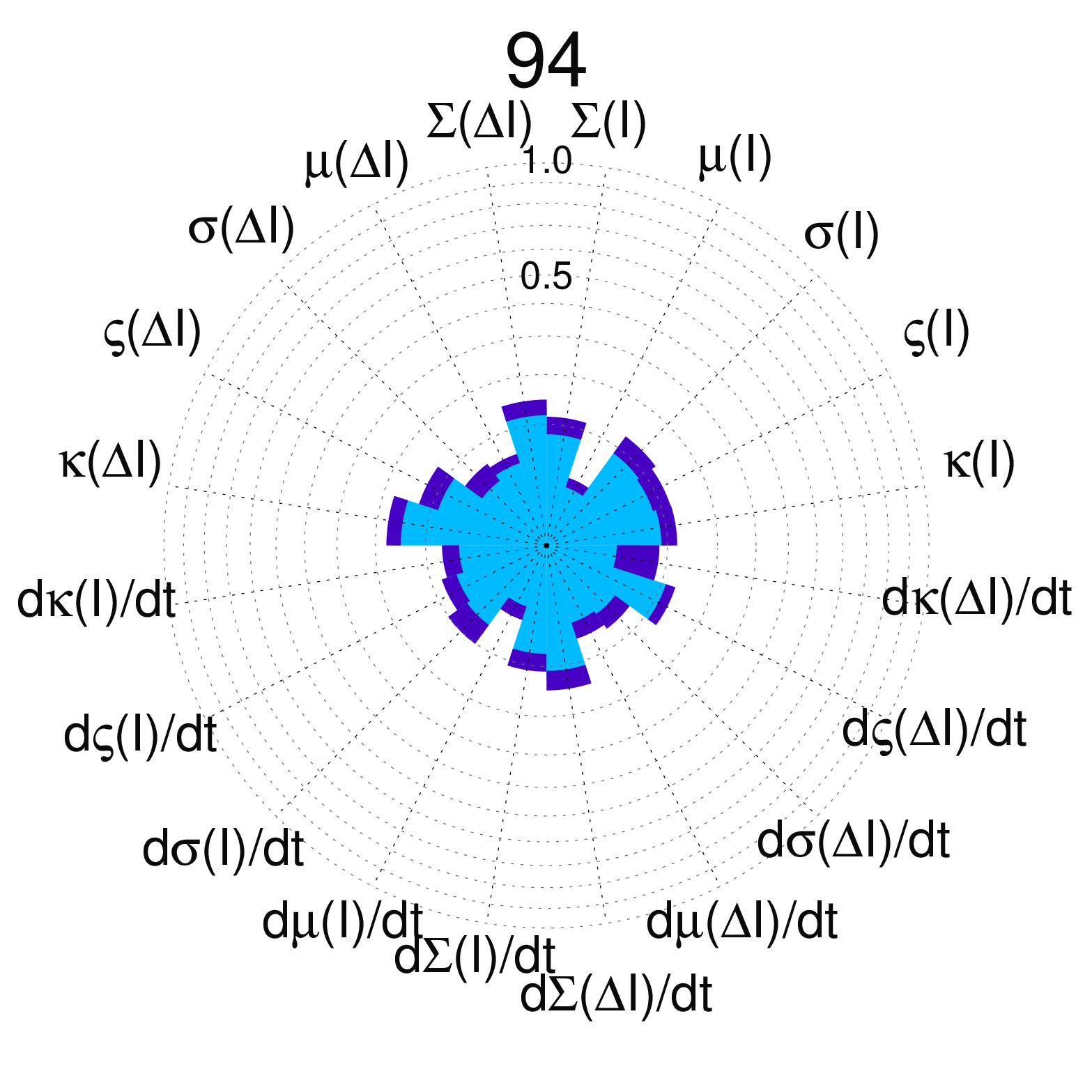}
\includegraphics[width=0.250\textwidth,clip, trim = 0mm 0mm 0mm 0mm, angle=0]{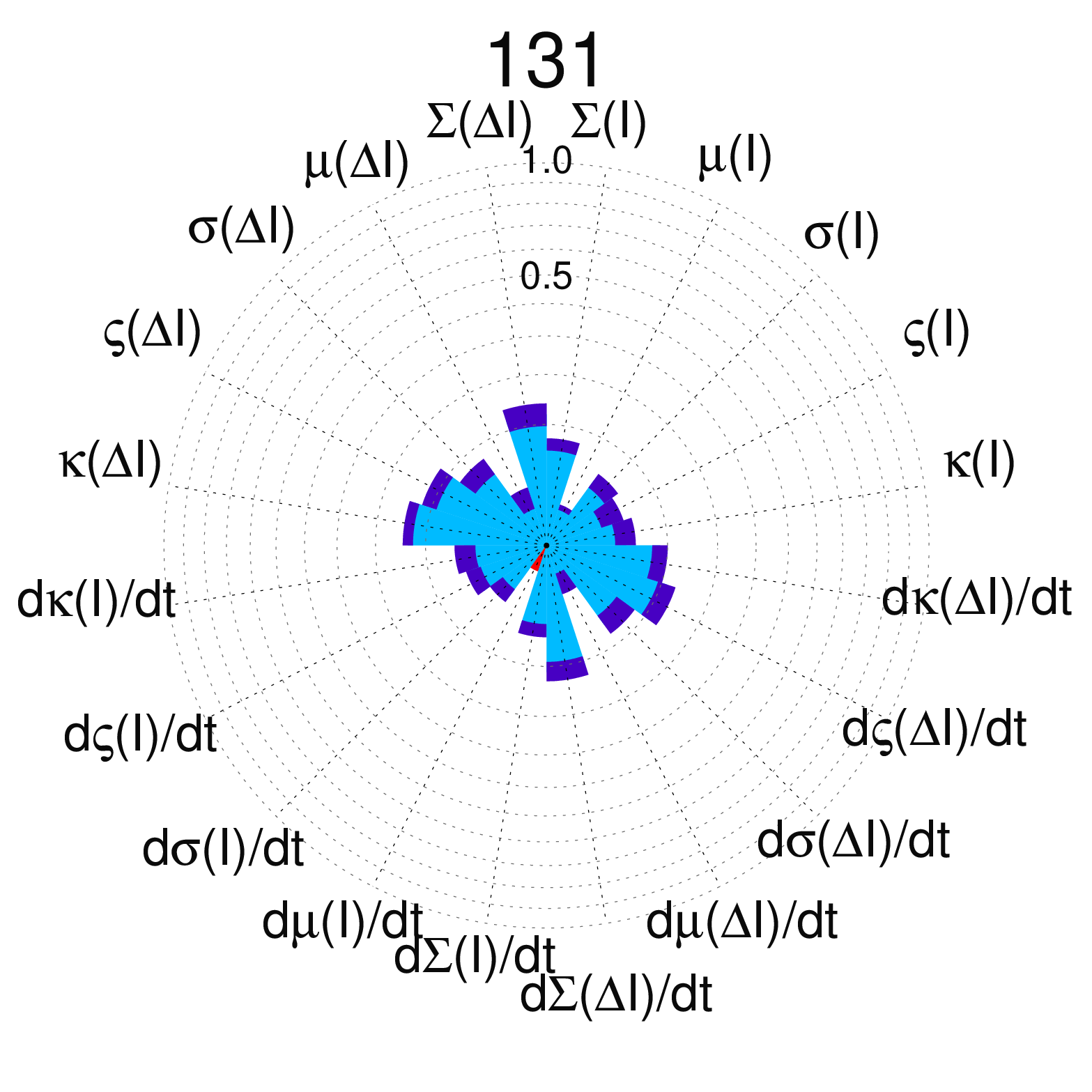}
\includegraphics[width=0.250\textwidth,clip, trim = 0mm 0mm 0mm 0mm, angle=0]{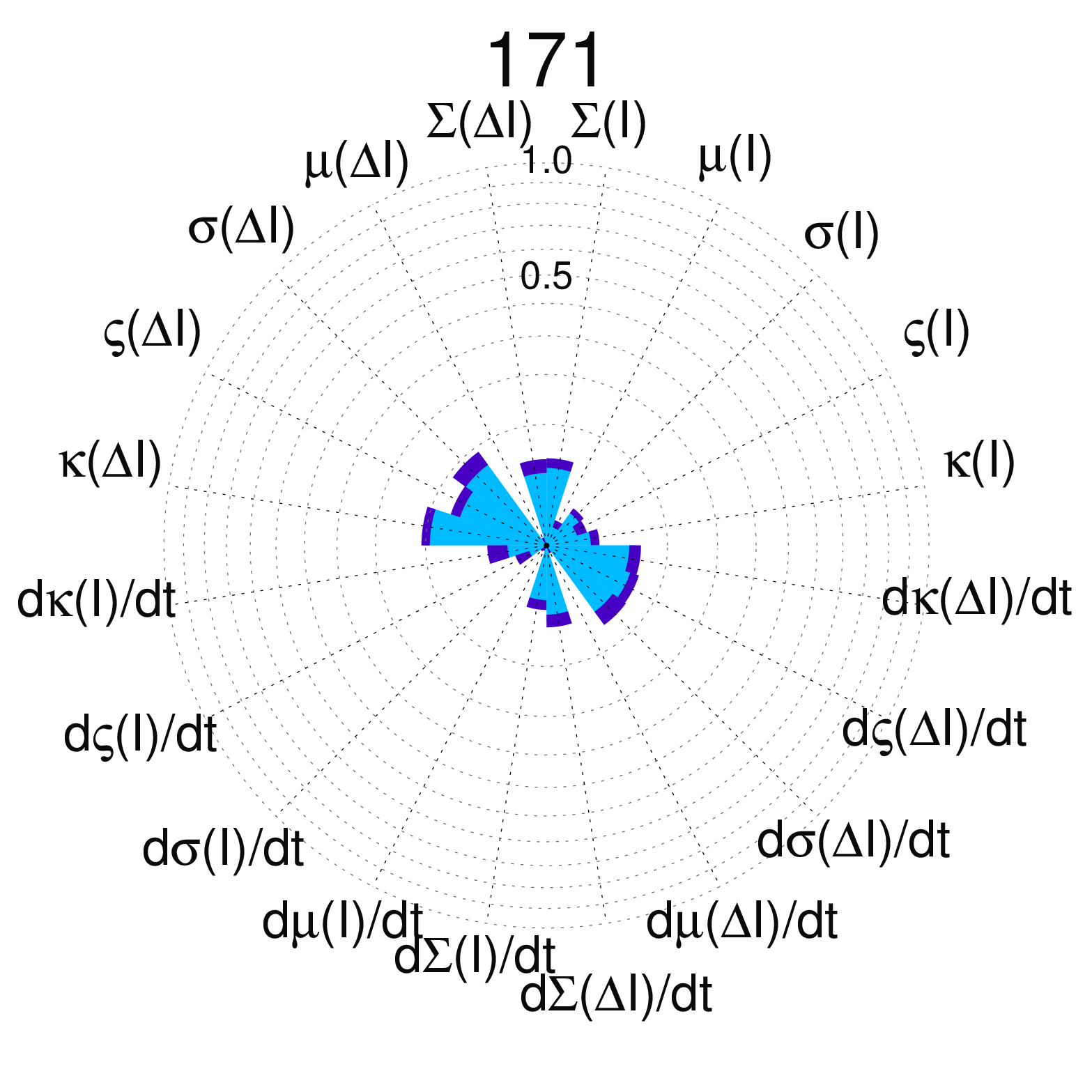}
\includegraphics[width=0.250\textwidth,clip, trim = 0mm 0mm 0mm 0mm, angle=0]{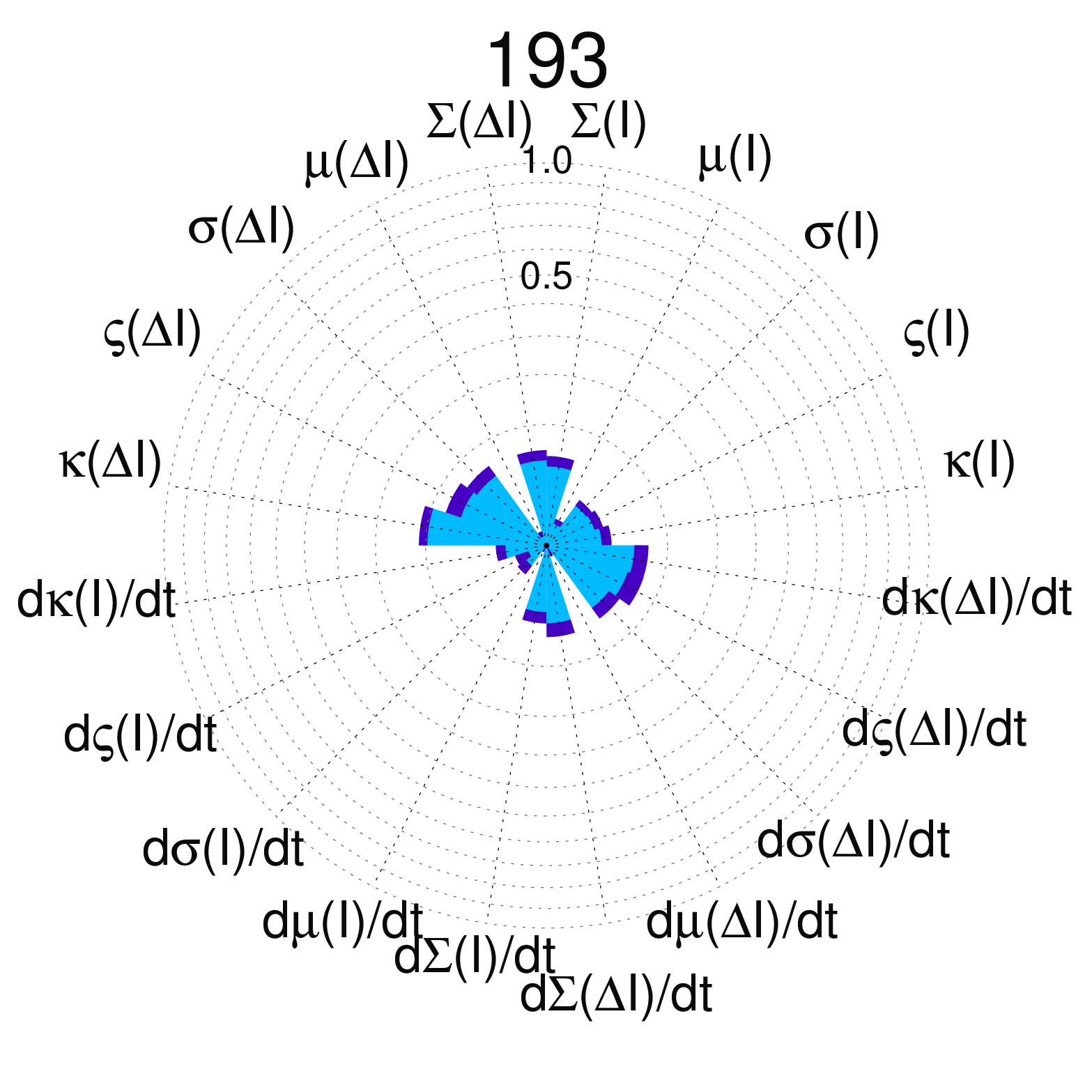}}
\centerline{
\includegraphics[width=0.250\textwidth,clip, trim = 0mm 0mm 0mm 0mm, angle=0]{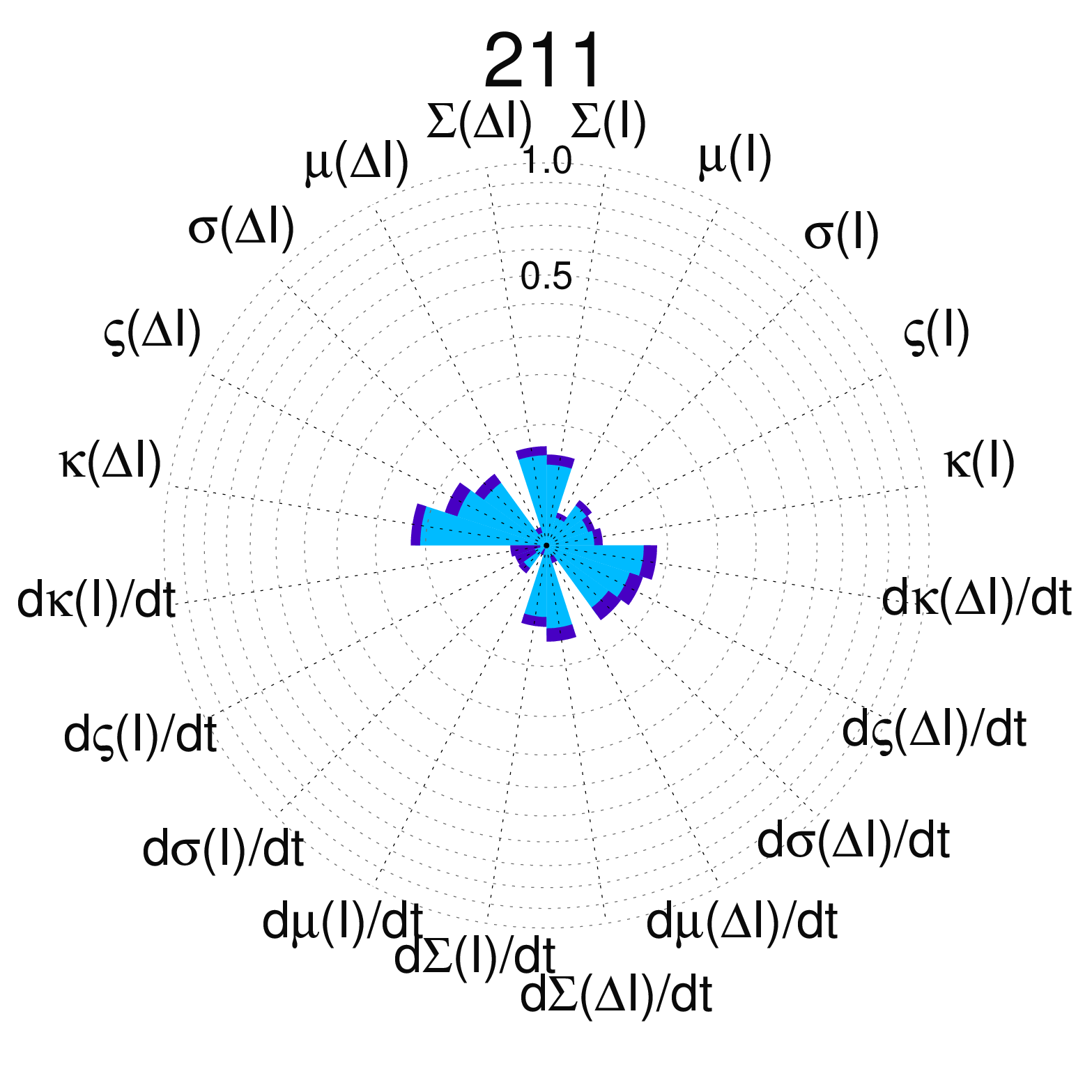}
\includegraphics[width=0.250\textwidth,clip, trim = 0mm 0mm 0mm 0mm, angle=0]{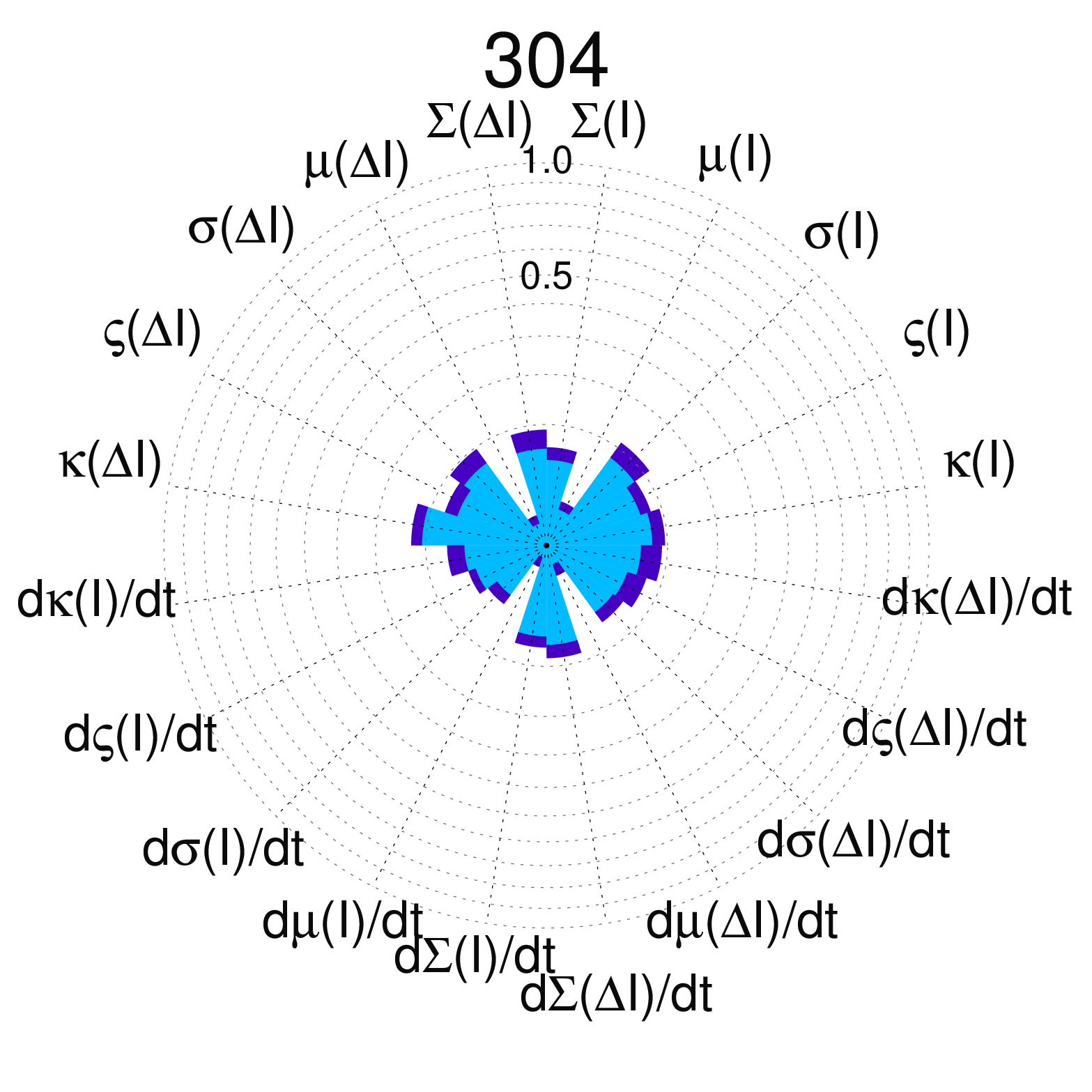}
\includegraphics[width=0.250\textwidth,clip, trim = 0mm 0mm 0mm 0mm, angle=0]{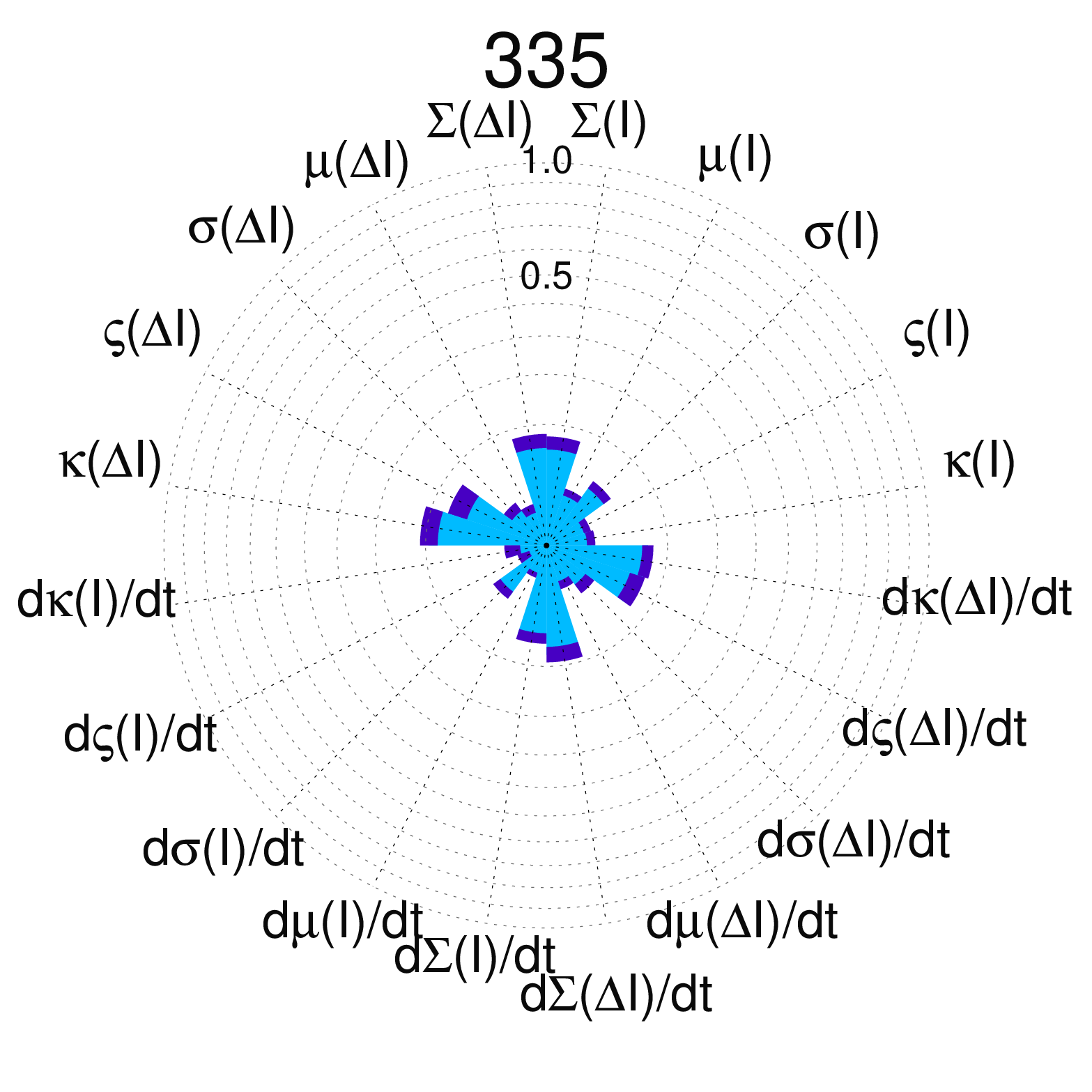}
\includegraphics[width=0.250\textwidth,clip, trim = 0mm 0mm 0mm 0mm, angle=0]{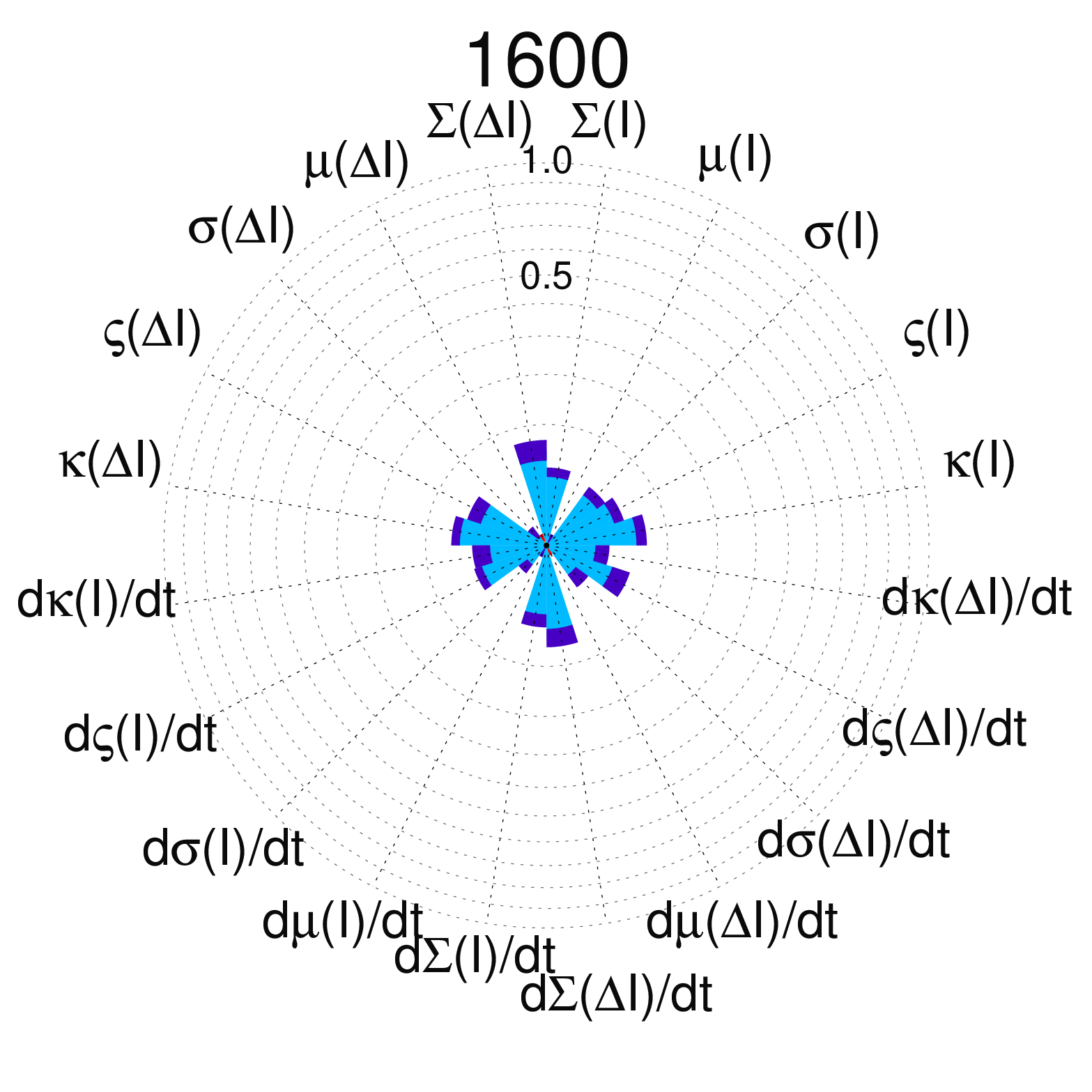}}
\caption{Same as Figure~\ref{fig:radar_C1_24} for the \MML\ event definition results.}
\label{fig:radar_M1_24}
\end{figure}

\begin{figure}
\centerline{
\includegraphics[width=0.250\textwidth,clip, trim = 0mm 0mm 0mm 0mm, angle=0]{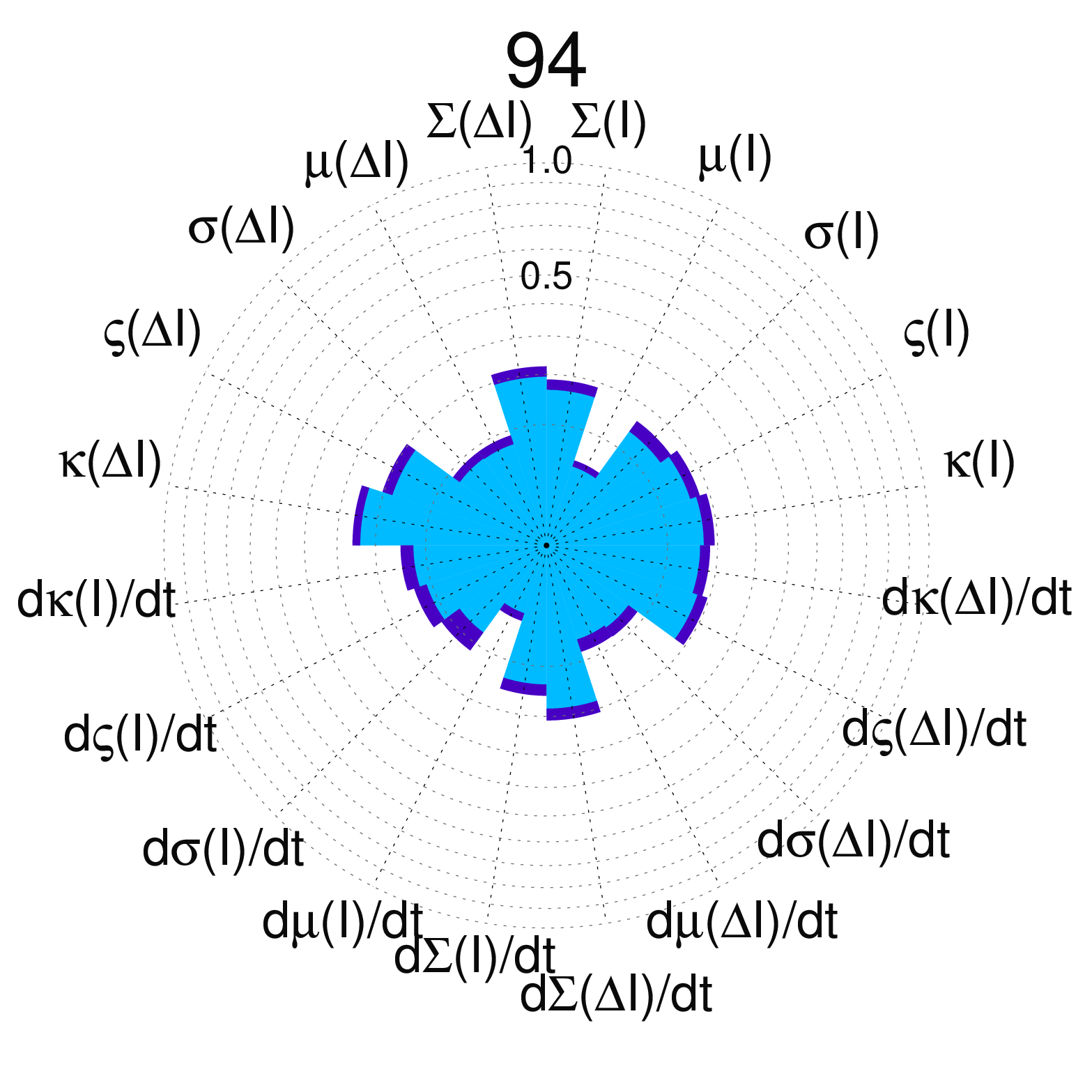}
\includegraphics[width=0.250\textwidth,clip, trim = 0mm 0mm 0mm 0mm, angle=0]{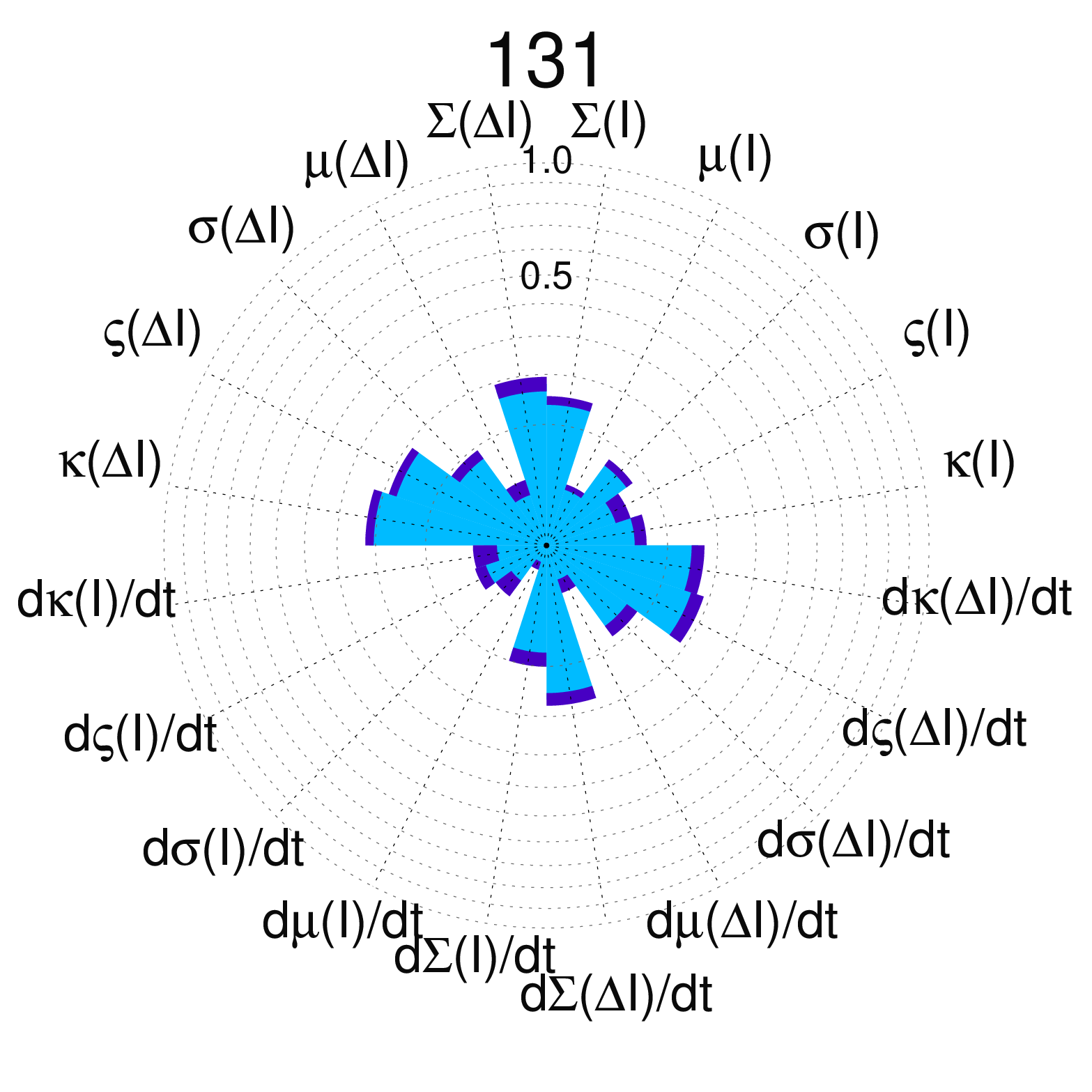}
\includegraphics[width=0.250\textwidth,clip, trim = 0mm 0mm 0mm 0mm, angle=0]{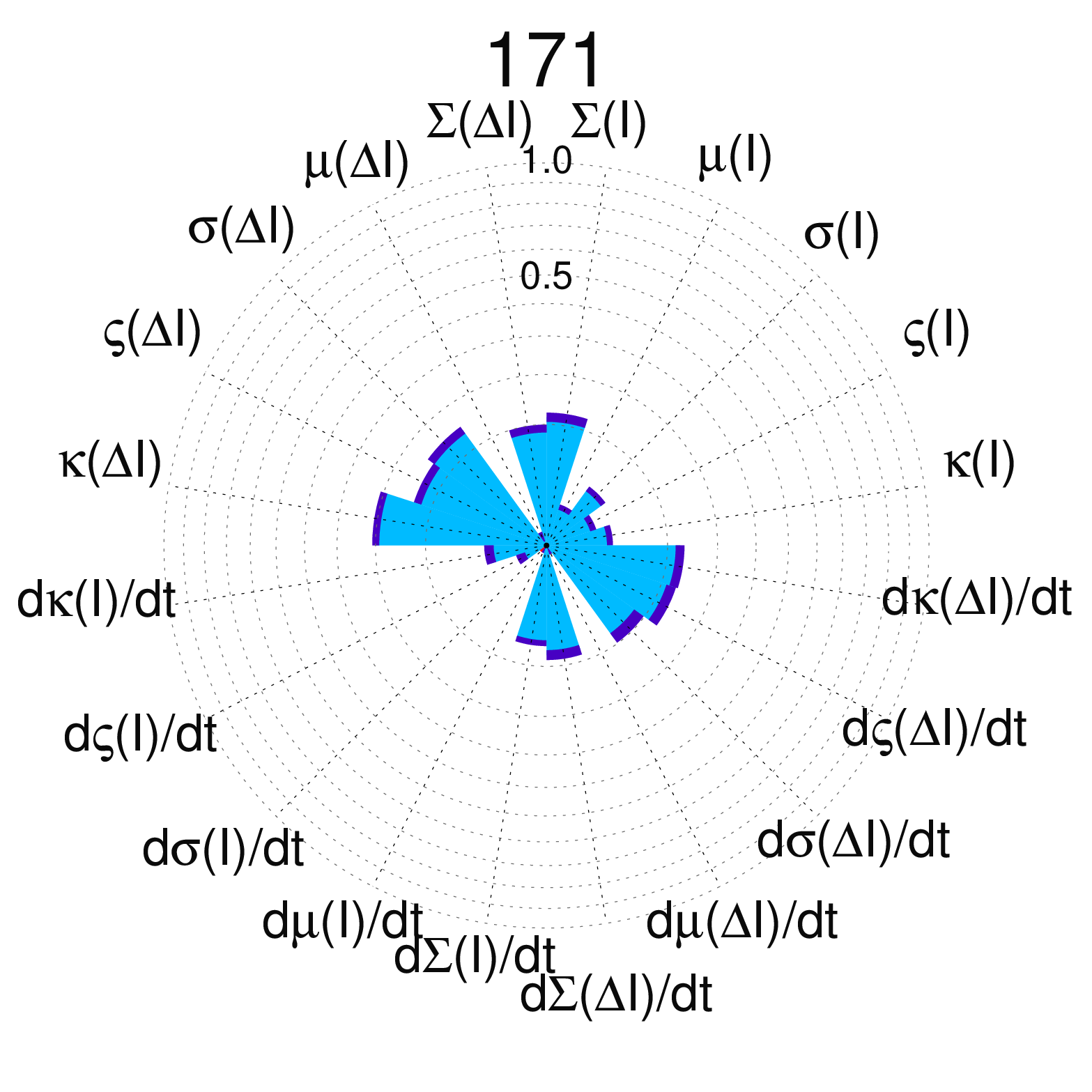}
\includegraphics[width=0.250\textwidth,clip, trim = 0mm 0mm 0mm 0mm, angle=0]{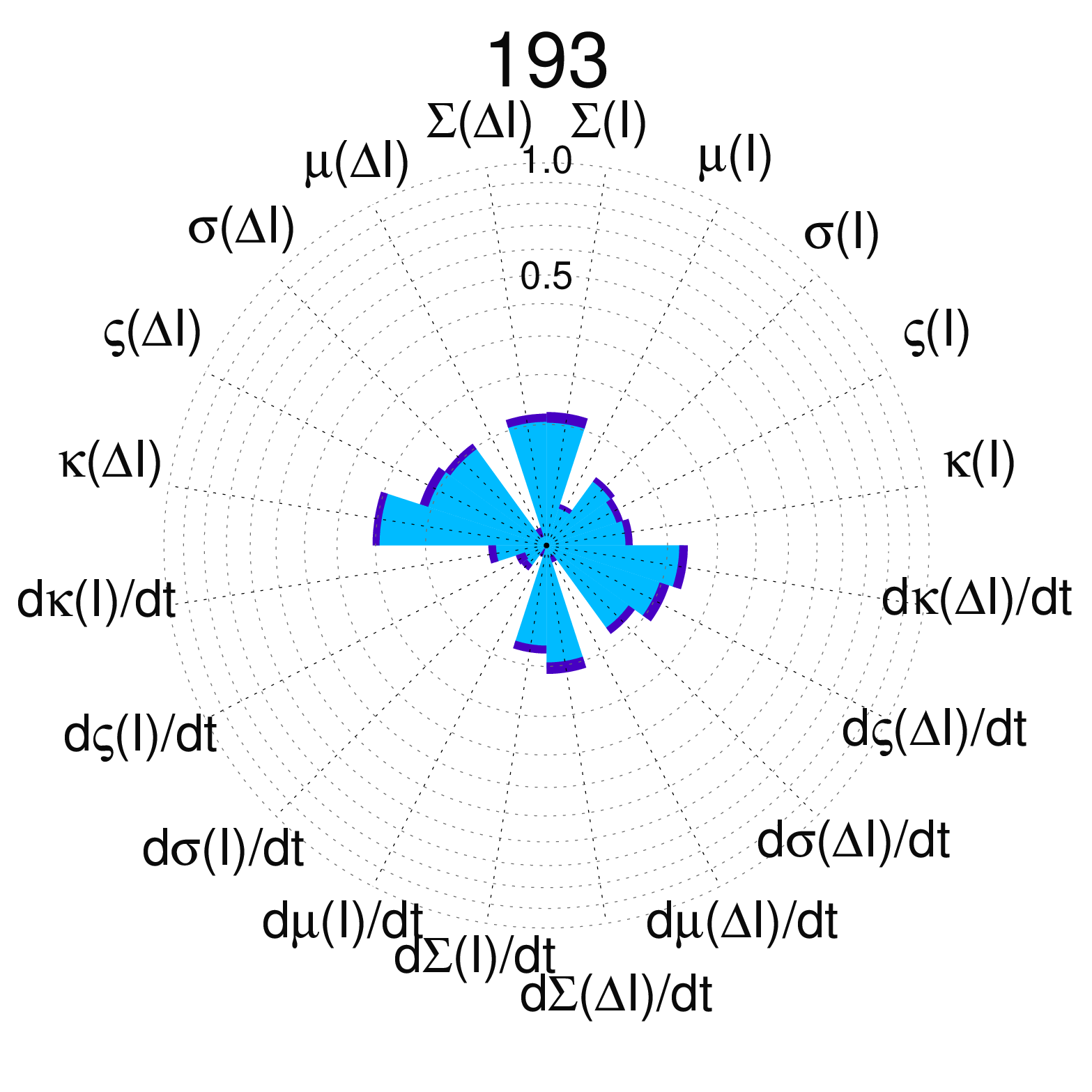}}
\centerline{
\includegraphics[width=0.250\textwidth,clip, trim = 0mm 0mm 0mm 0mm, angle=0]{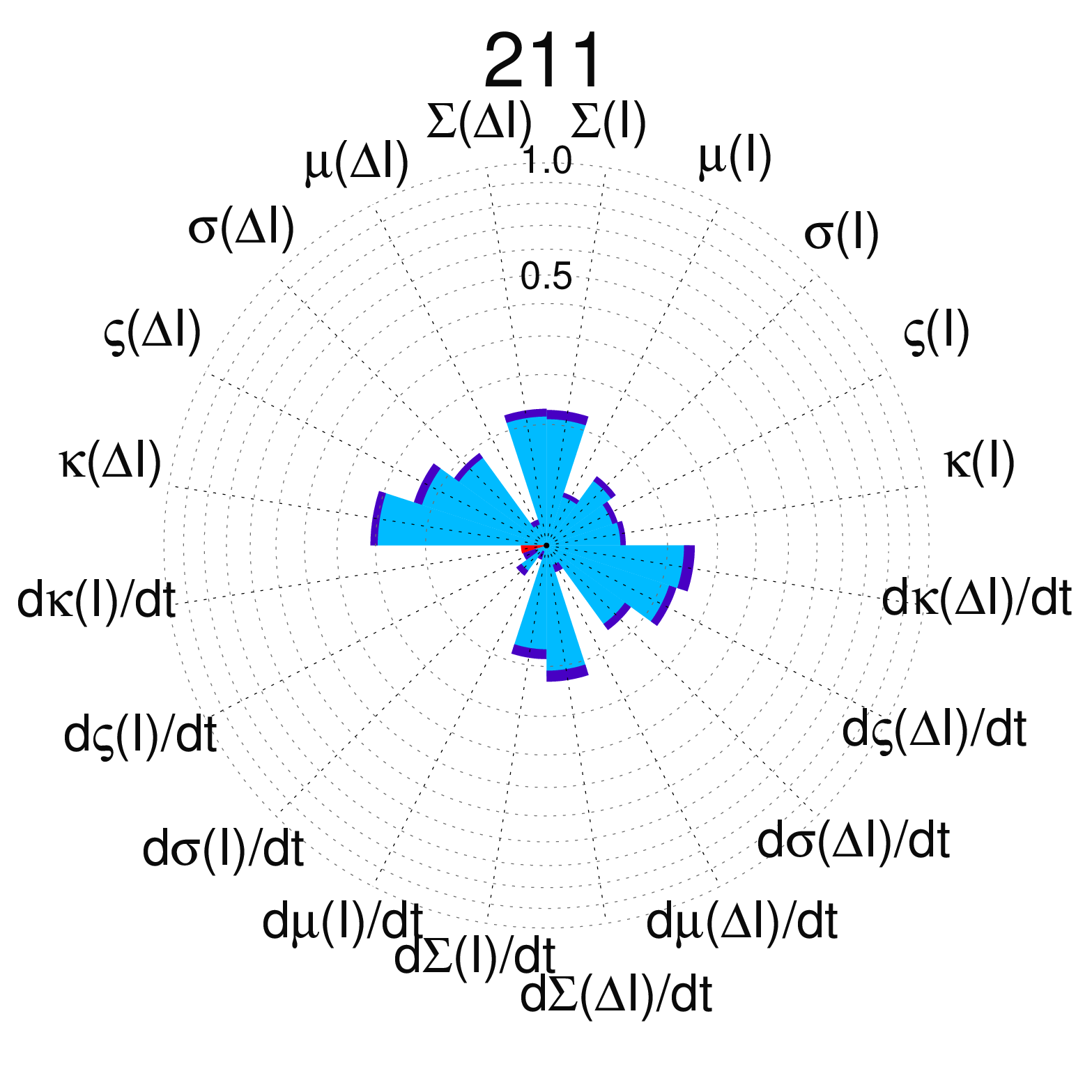}
\includegraphics[width=0.250\textwidth,clip, trim = 0mm 0mm 0mm 0mm, angle=0]{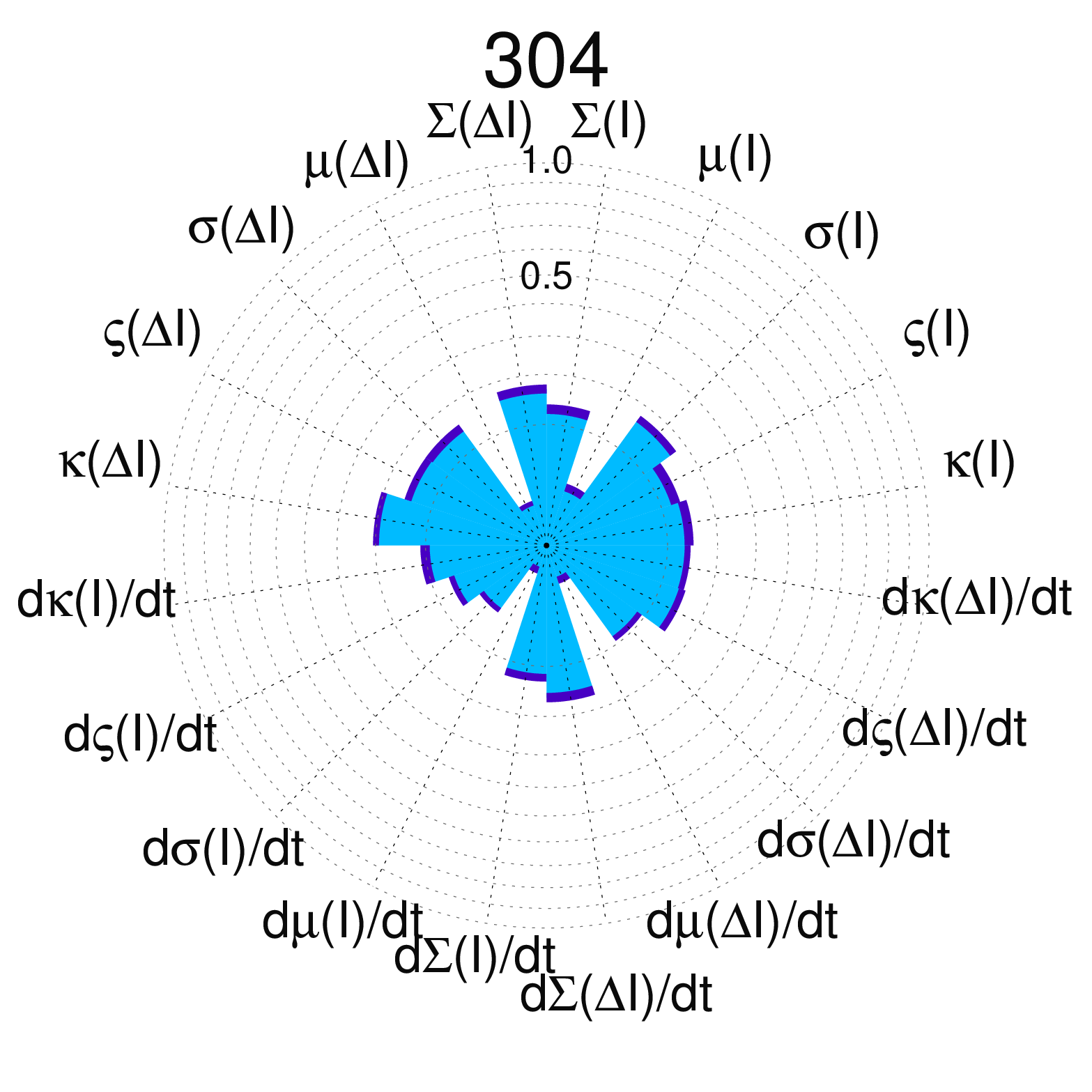}
\includegraphics[width=0.250\textwidth,clip, trim = 0mm 0mm 0mm 0mm, angle=0]{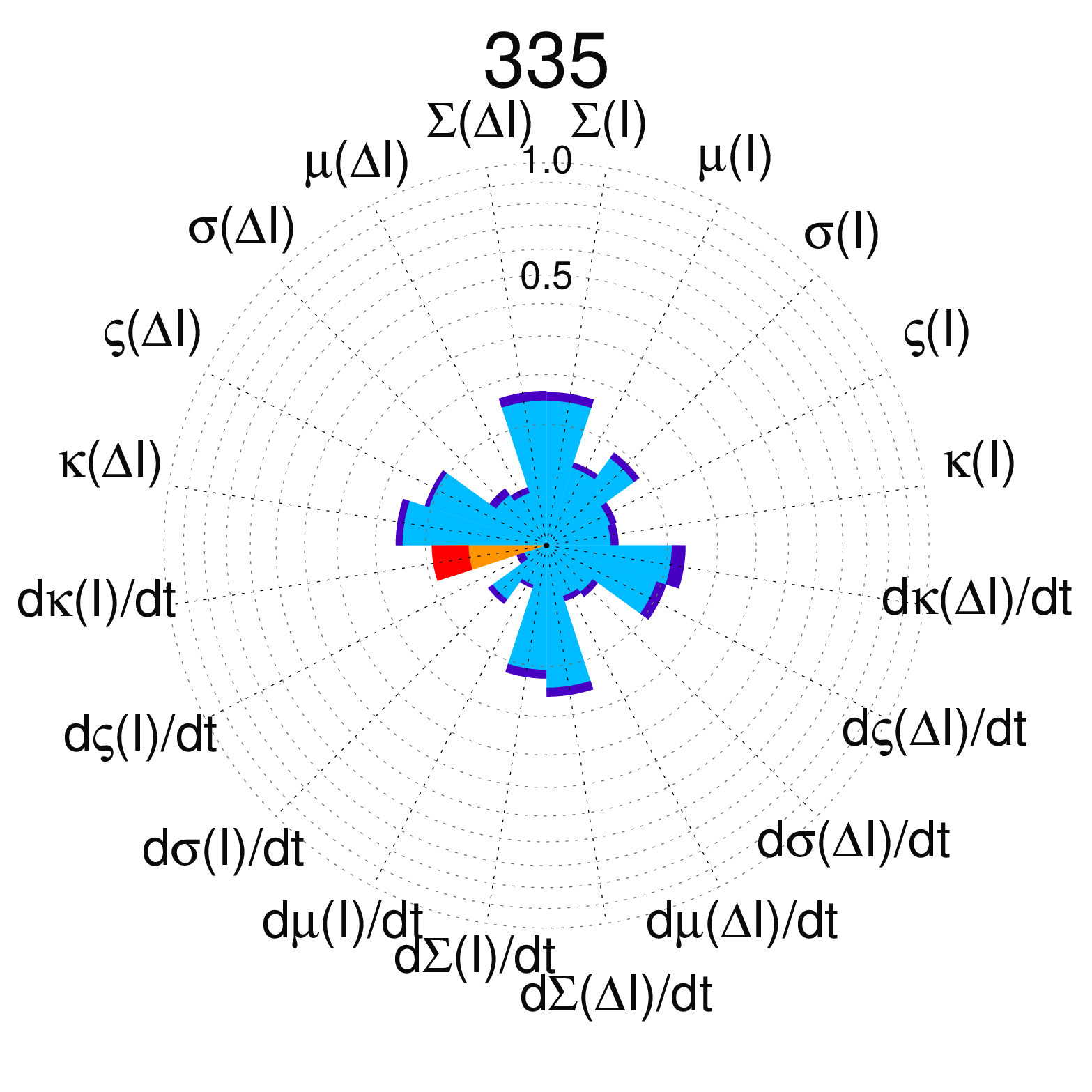}
\includegraphics[width=0.250\textwidth,clip, trim = 0mm 0mm 0mm 0mm, angle=0]{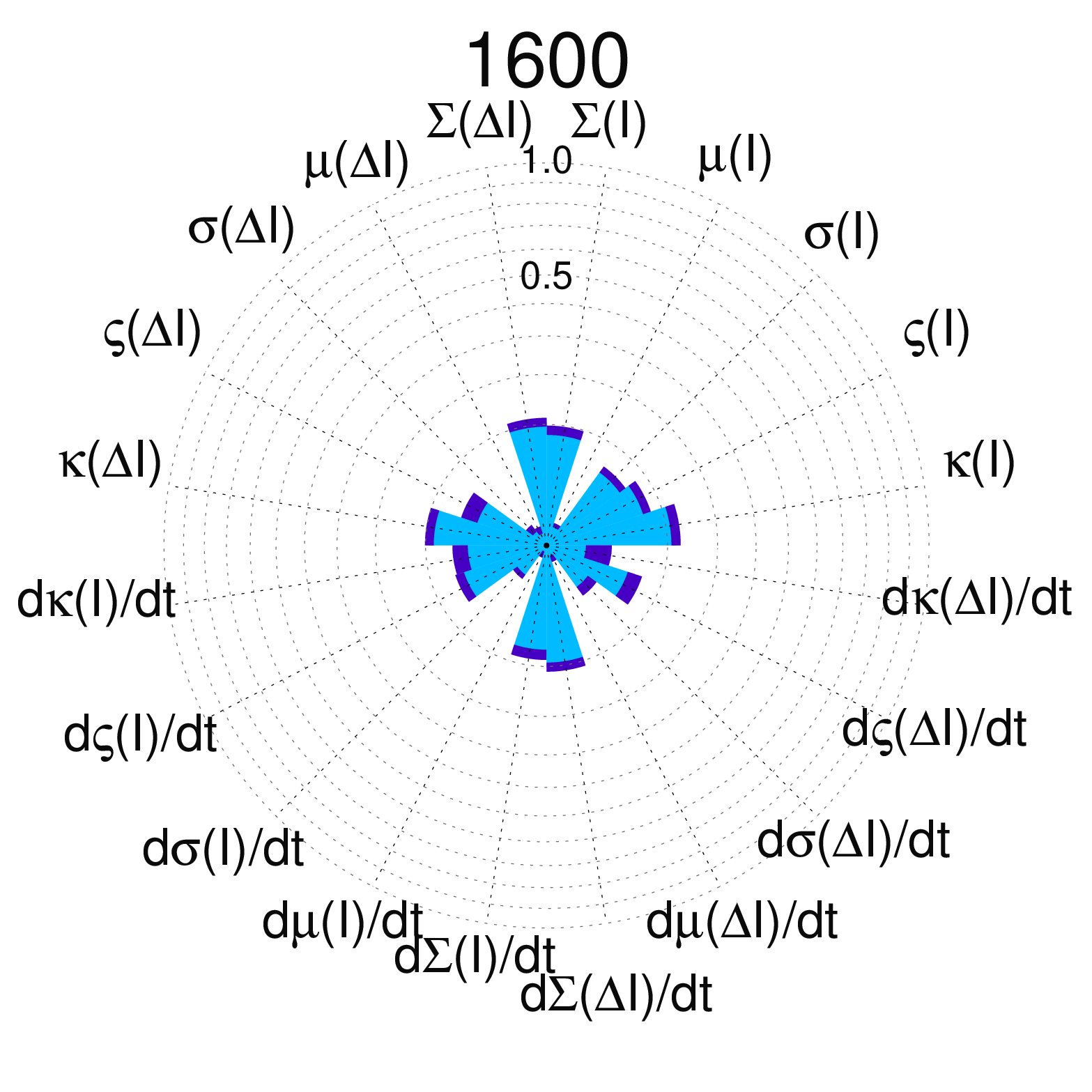}}
\caption{Same as Figure~\ref{fig:radar_C1_24} for the  \CCS\ event definition results.}
\label{fig:radar_C1_6}
\end{figure}

\begin{figure}
\centerline{
\includegraphics[width=0.250\textwidth,clip, trim = 0mm 0mm 0mm 0mm, angle=0]{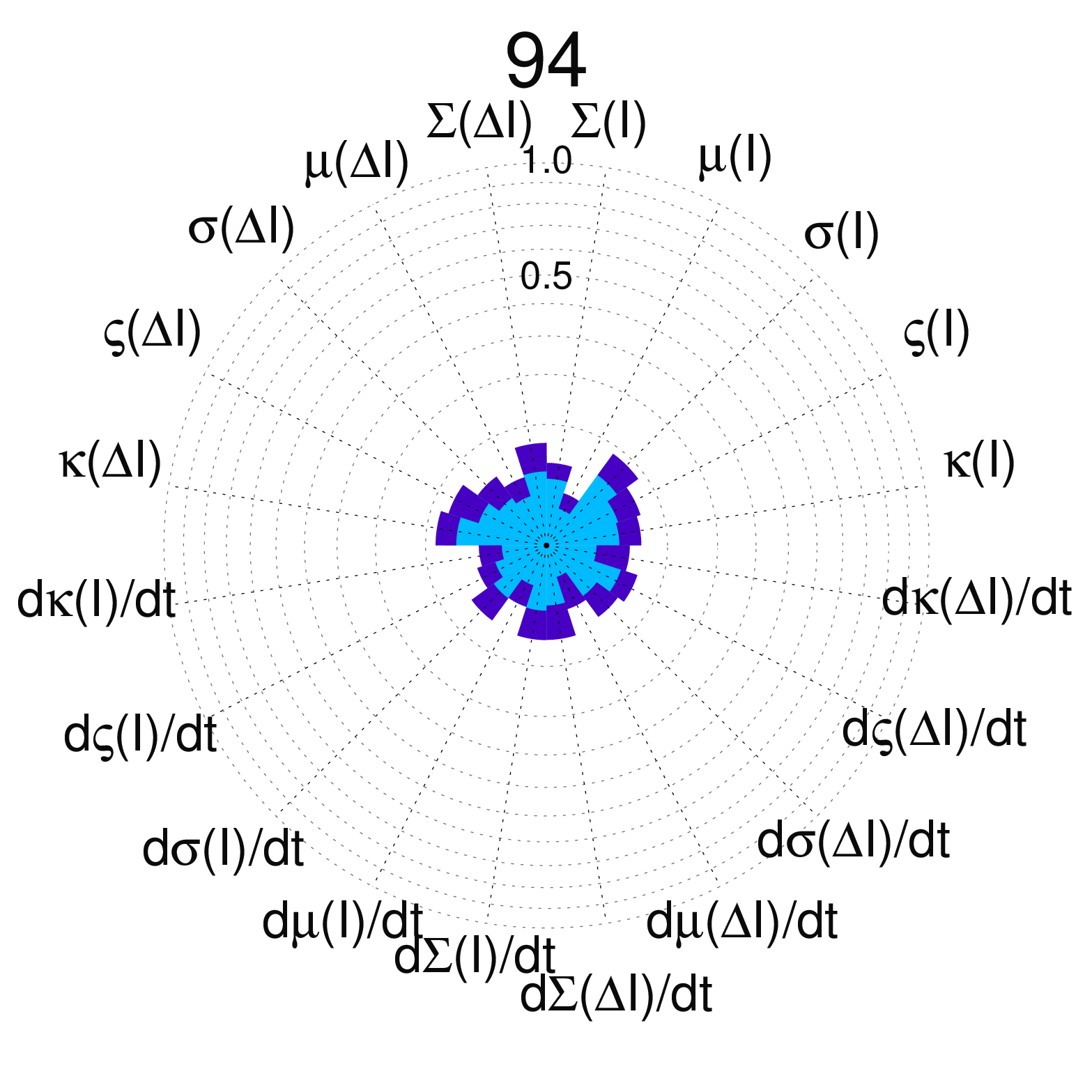}
\includegraphics[width=0.250\textwidth,clip, trim = 0mm 0mm 0mm 0mm, angle=0]{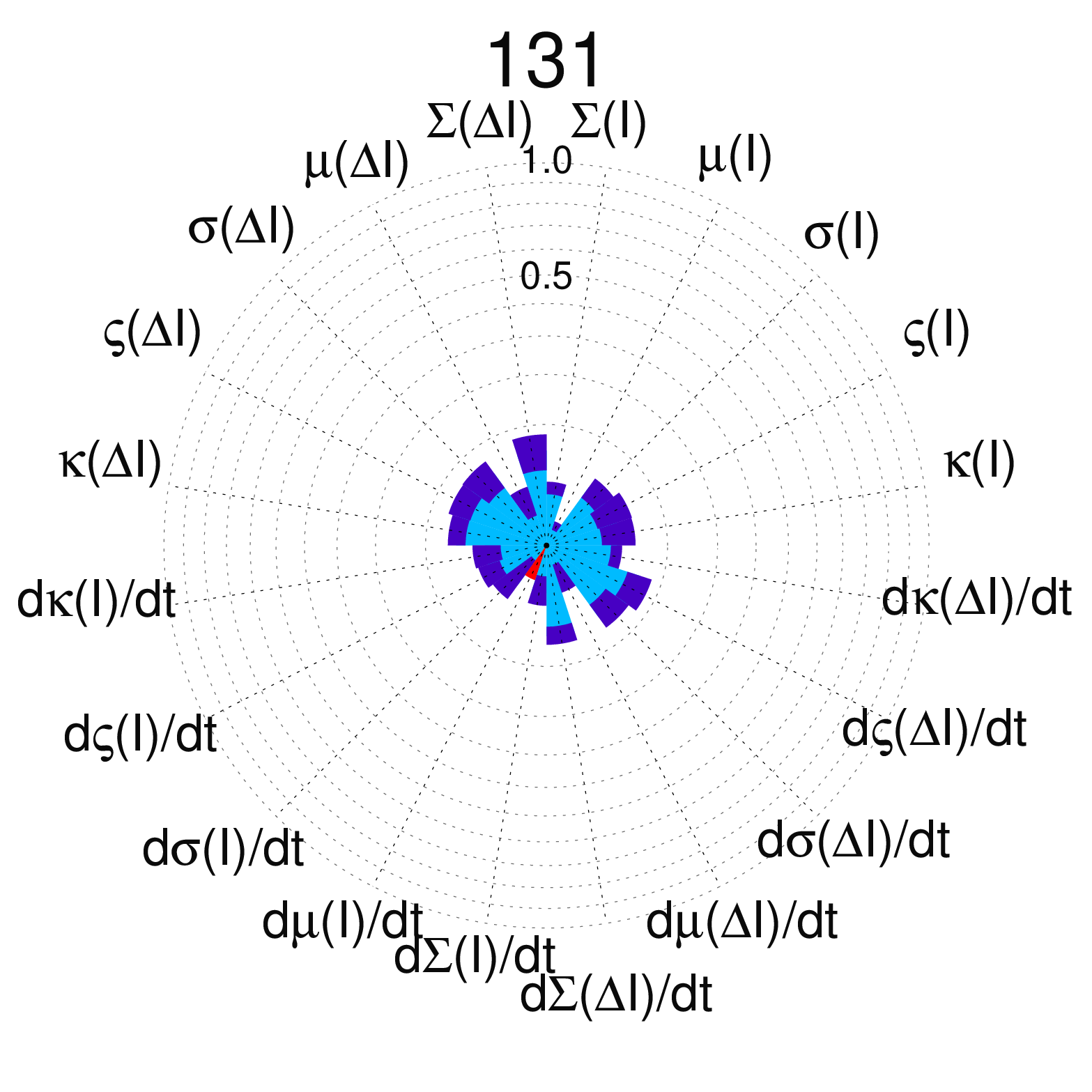}
\includegraphics[width=0.250\textwidth,clip, trim = 0mm 0mm 0mm 0mm, angle=0]{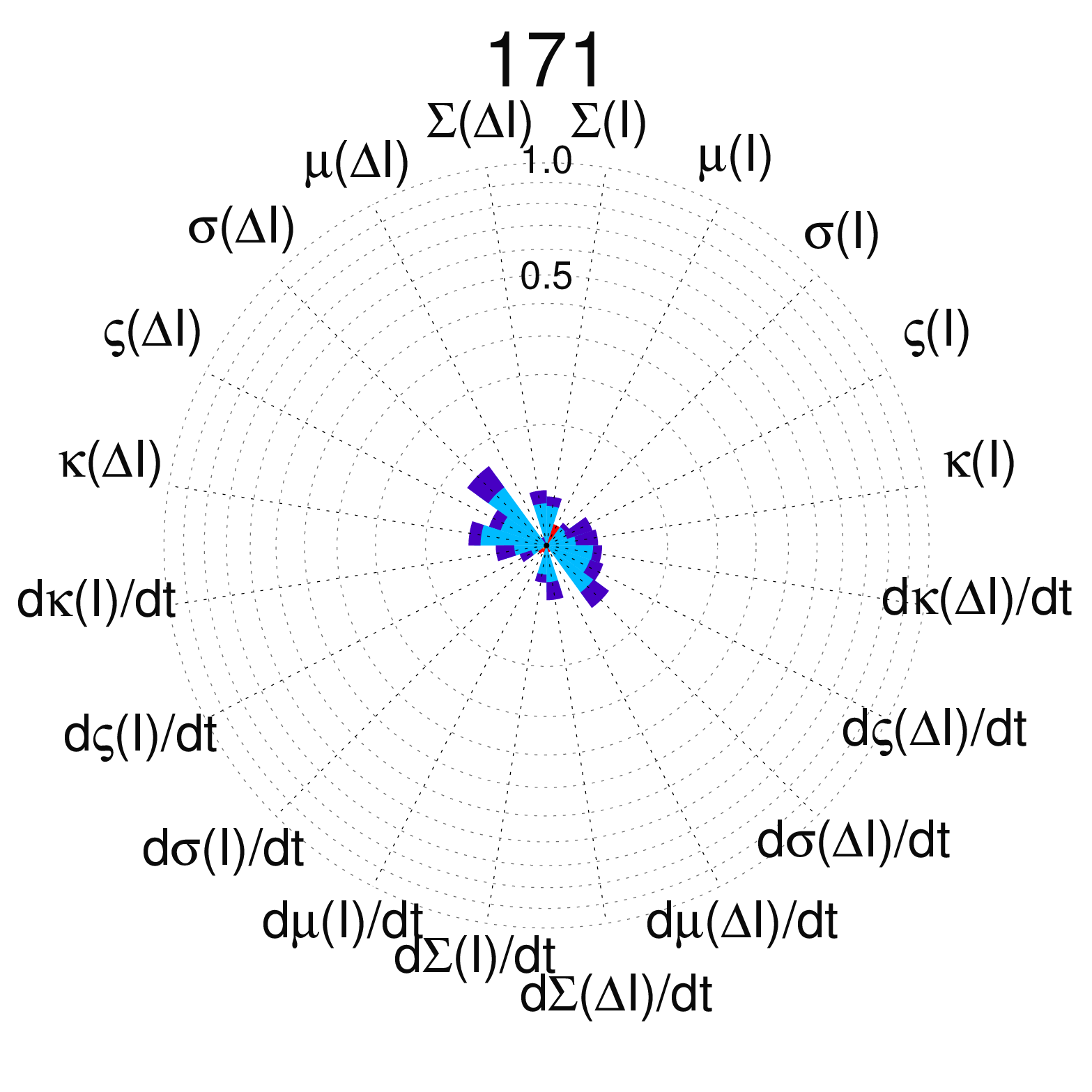}
\includegraphics[width=0.250\textwidth,clip, trim = 0mm 0mm 0mm 0mm, angle=0]{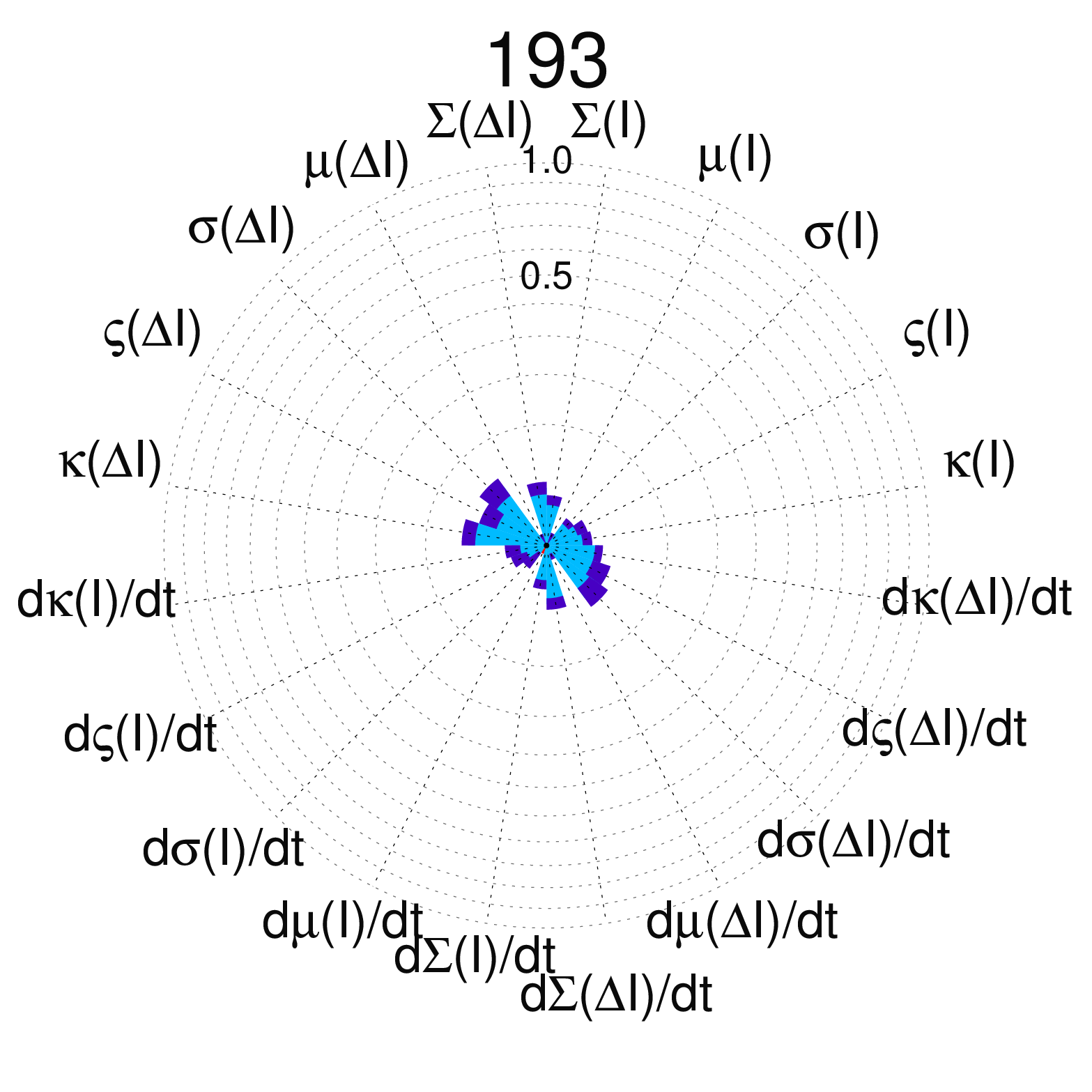}}
\centerline{
\includegraphics[width=0.250\textwidth,clip, trim = 0mm 0mm 0mm 0mm, angle=0]{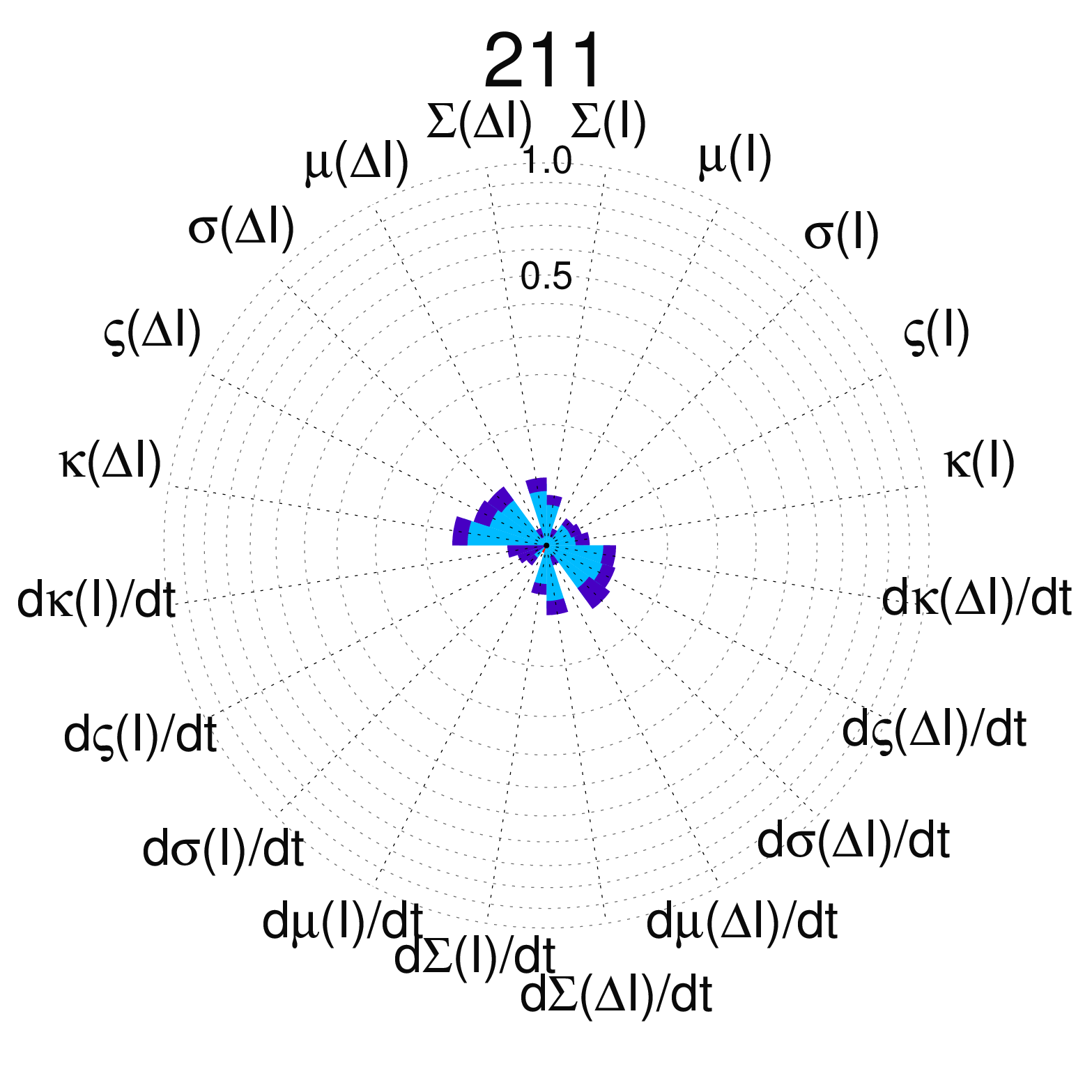}
\includegraphics[width=0.250\textwidth,clip, trim = 0mm 0mm 0mm 0mm, angle=0]{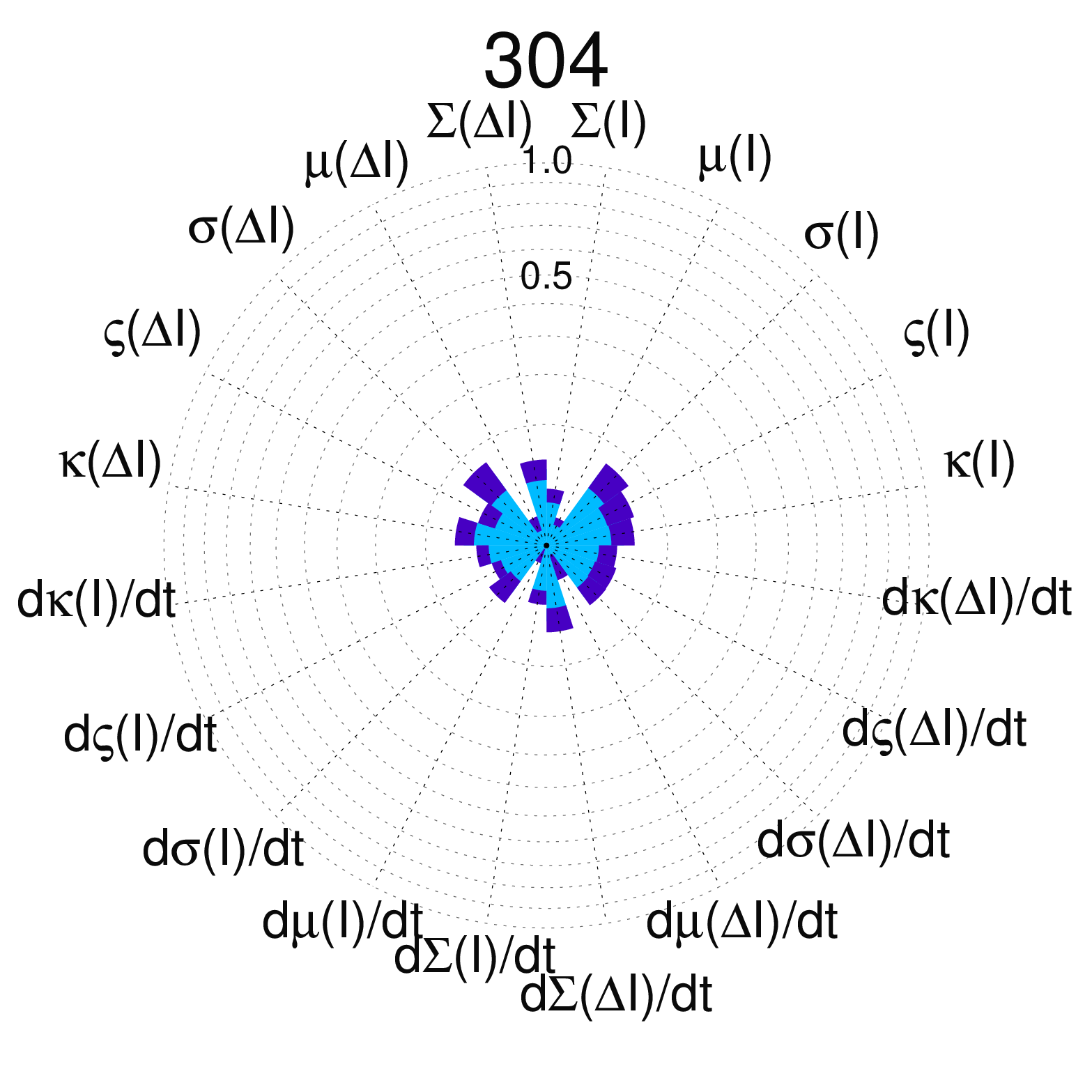}
\includegraphics[width=0.250\textwidth,clip, trim = 0mm 0mm 0mm 0mm, angle=0]{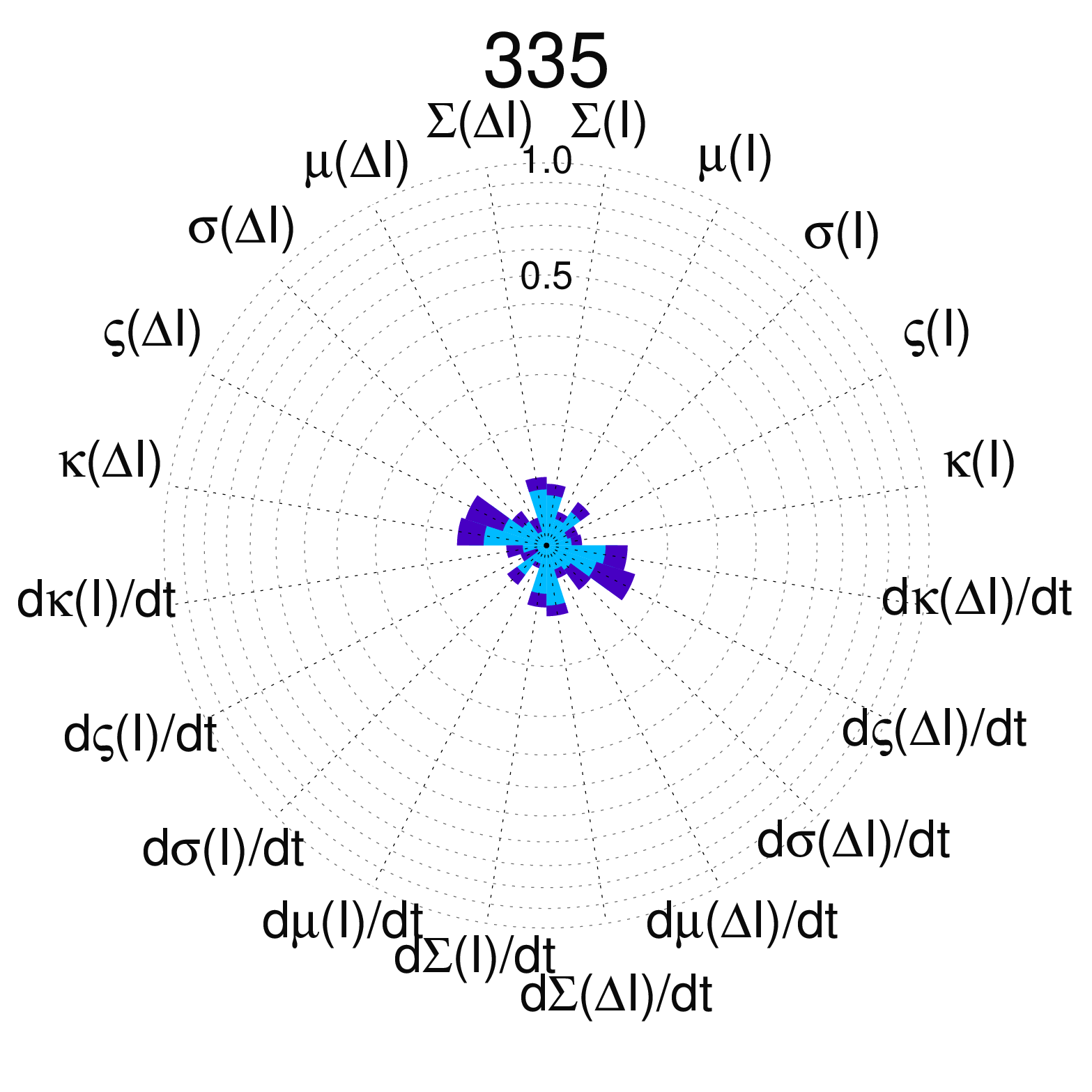}
\includegraphics[width=0.250\textwidth,clip, trim = 0mm 0mm 0mm 0mm, angle=0]{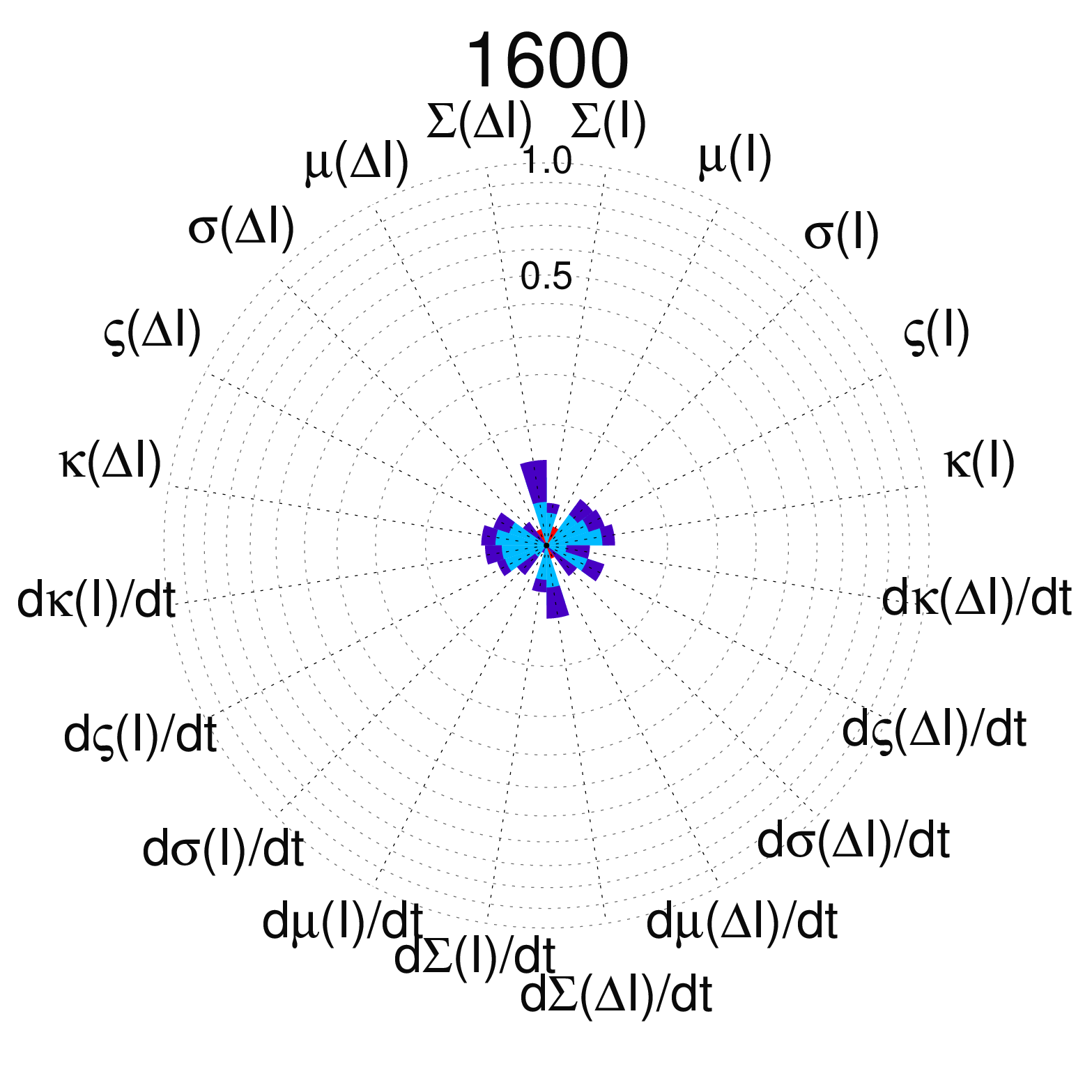}}
\caption{Same as Figure~\ref{fig:radar_C1_24} for the  \MMS\ event definition results.}
\label{fig:radar_M1_6}
\end{figure}

In Figures~\ref{fig:radar_C1_24}, ~\ref{fig:radar_M1_24},
~\ref{fig:radar_C1_6}, ~\ref{fig:radar_M1_6} we group the \BSS\ results
by wavelength.  What is striking in these plots with regards to the
performance by different AIA filters is that the 94\AA\ parameters by and
large perform consistently well (comparatively speaking), with all ``radar
sectors'' filled in at least somewhat.  In contrast, the radar plots
for 211\AA, for example, have definite gaps; for example, while 
the $\kappa(\Delta \rm{I_{211}})$ scores well, 
the $d\kappa({\rm I_{211}})/d{\rm t}$ parameter does not.

Overall, this presentation confirms the highlights of
Tables~\ref{tbl:resultsCCL}--\ref{tbl:resultsMMS}: the performance is
overall lower for the shorter-validity definitions, and uncertainties
are larger for the event definitions that have smaller event-population
sample sizes (higher class imbalance).  There are more parameters that
perform with higher classification success for the 94\,\AA\ filter than
most of the others, but then the 304\,\AA\ parameters also have a fairly
high frequency of similar performance (albeit not ``high performing''
by this metric {\it per se}).  The other AIA filters show a more mixed
performance, with the 1600\,\AA\ arguably the lowest overall.  Notably,
for all wavelengths, the kurtosis- and skew-\ and total-based evaluation
of running-difference images are often the highest performing parameters
of any particular wavelength.


\subsubsection{Performance Changes between Event Definitions}

Generally speaking, the \BSS\ scores decrease while \maxtss\ and
$\mathcal{G}$ stay the same or increase between \CC\ and \MM\ definitions,
and between, for example, the 24\,hr and 6\,hr validity times.  This is
fairly evident as a general rule from the discussion thus far and
is not unexpected given the sensitivity of \BSS\ to event rates
and relative insensitivity of \maxtss\ to the same \citep{Bloomfield_etal_2012,allclear}.

\begin{figure}
\centerline{
\includegraphics[width=0.30\textwidth, clip, trim = 6mm 0mm 0mm 0mm]{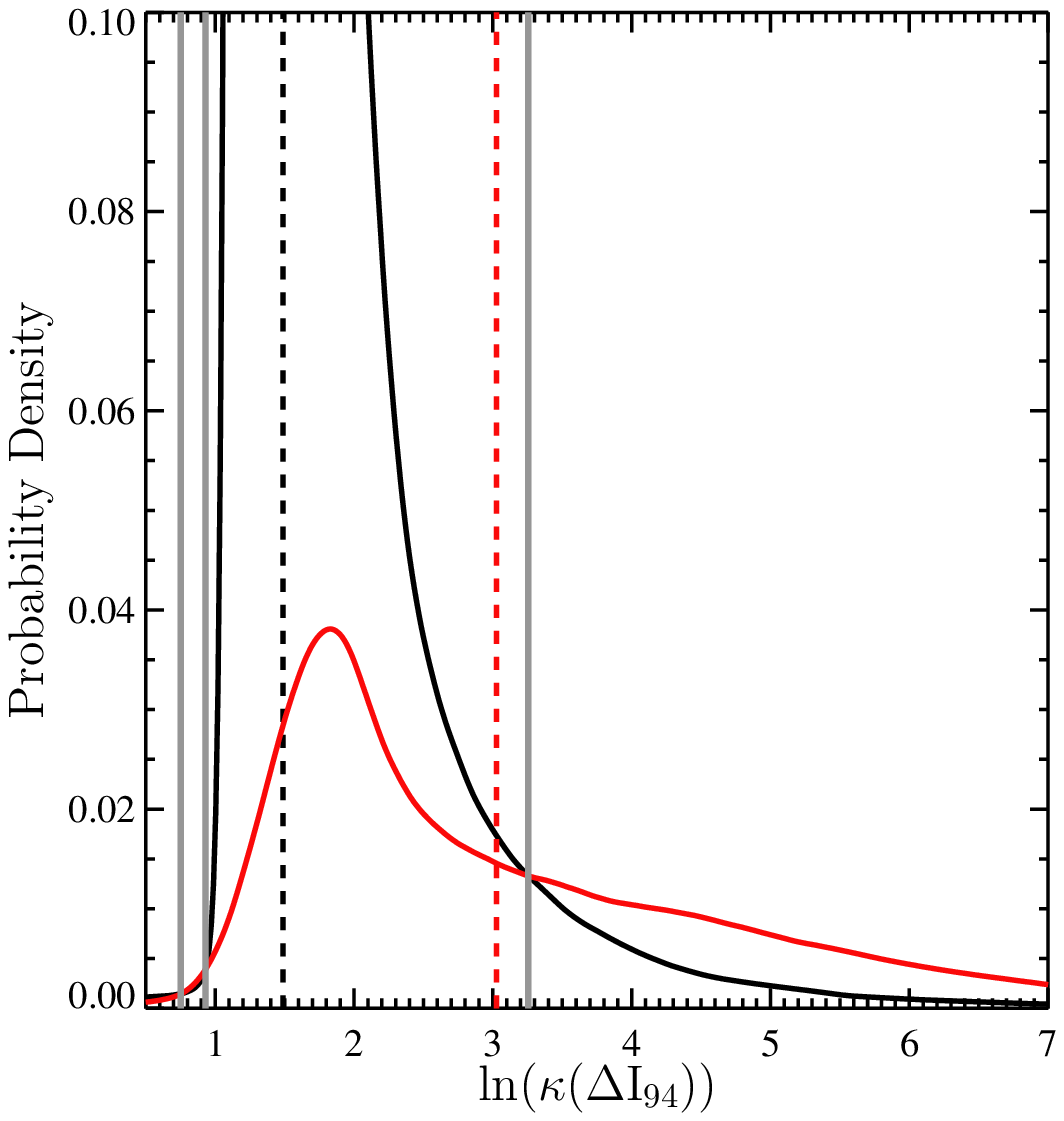} 
\includegraphics[width=0.30\textwidth, clip, trim = 6mm 0mm 0mm 0mm]{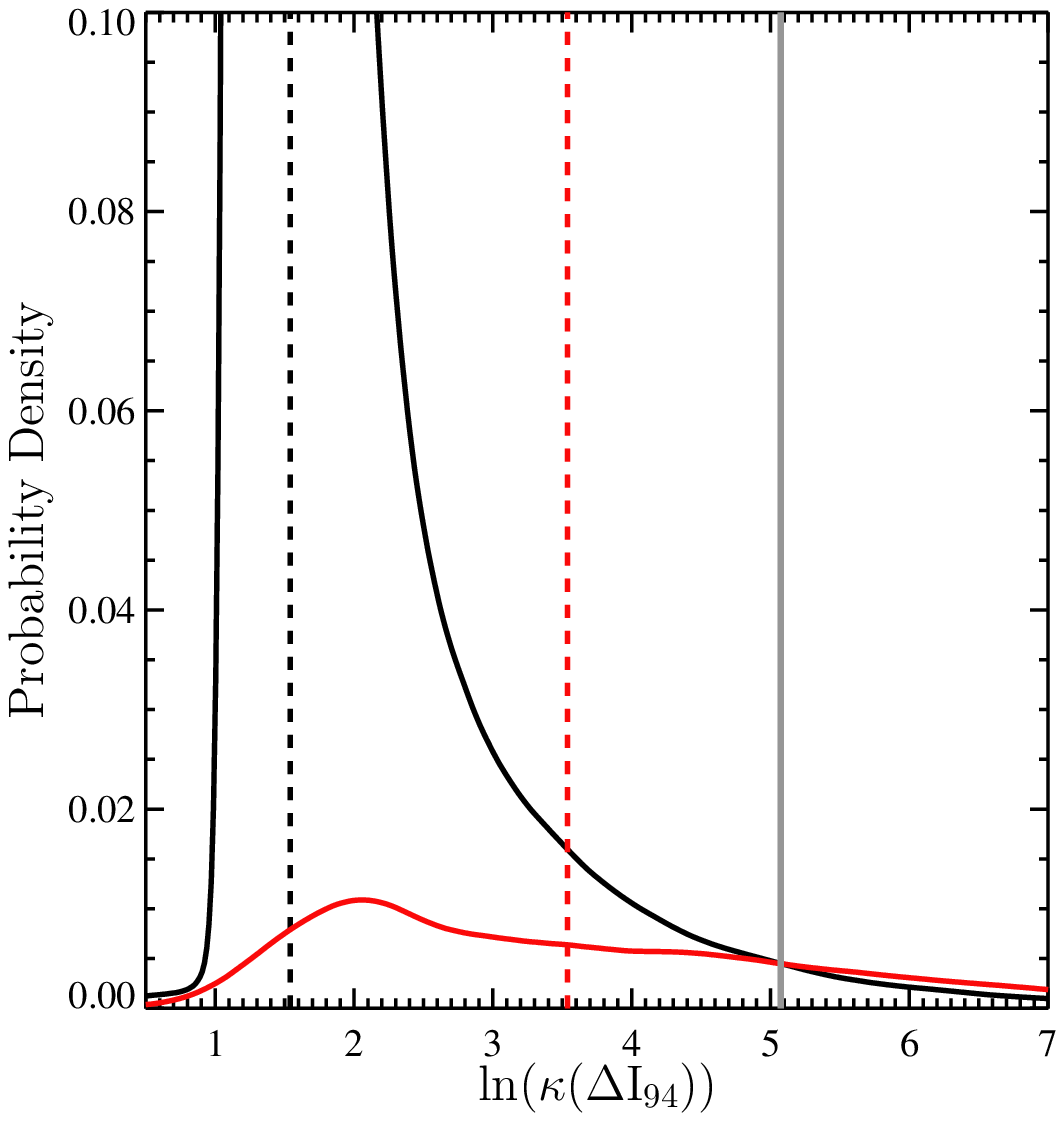} 
\includegraphics[width=0.30\textwidth, clip, trim = 6mm 0mm 0mm 0mm]{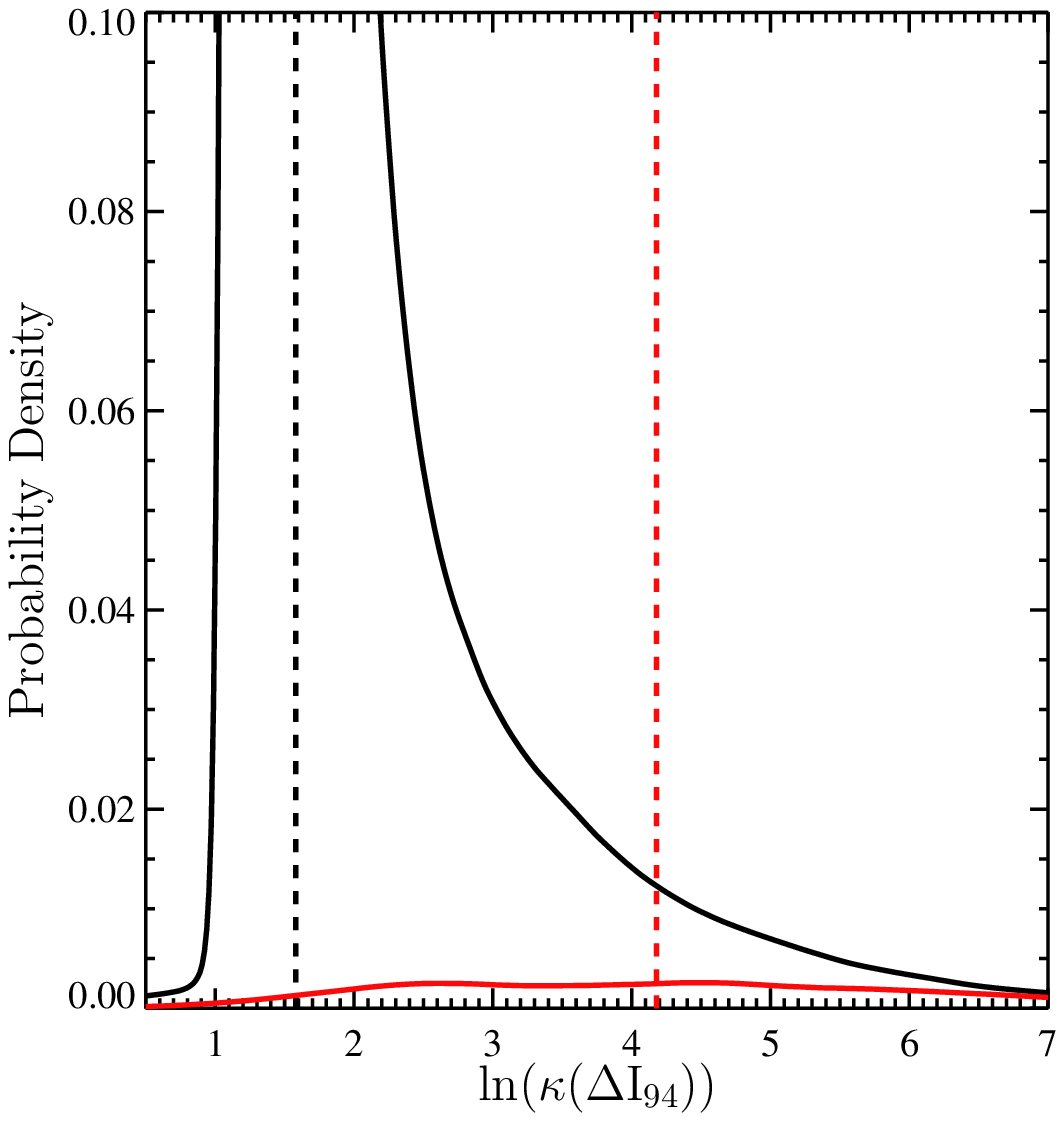}}
\centerline{
\includegraphics[width=0.30\textwidth, clip, trim = 0mm 0mm 0mm 0mm]{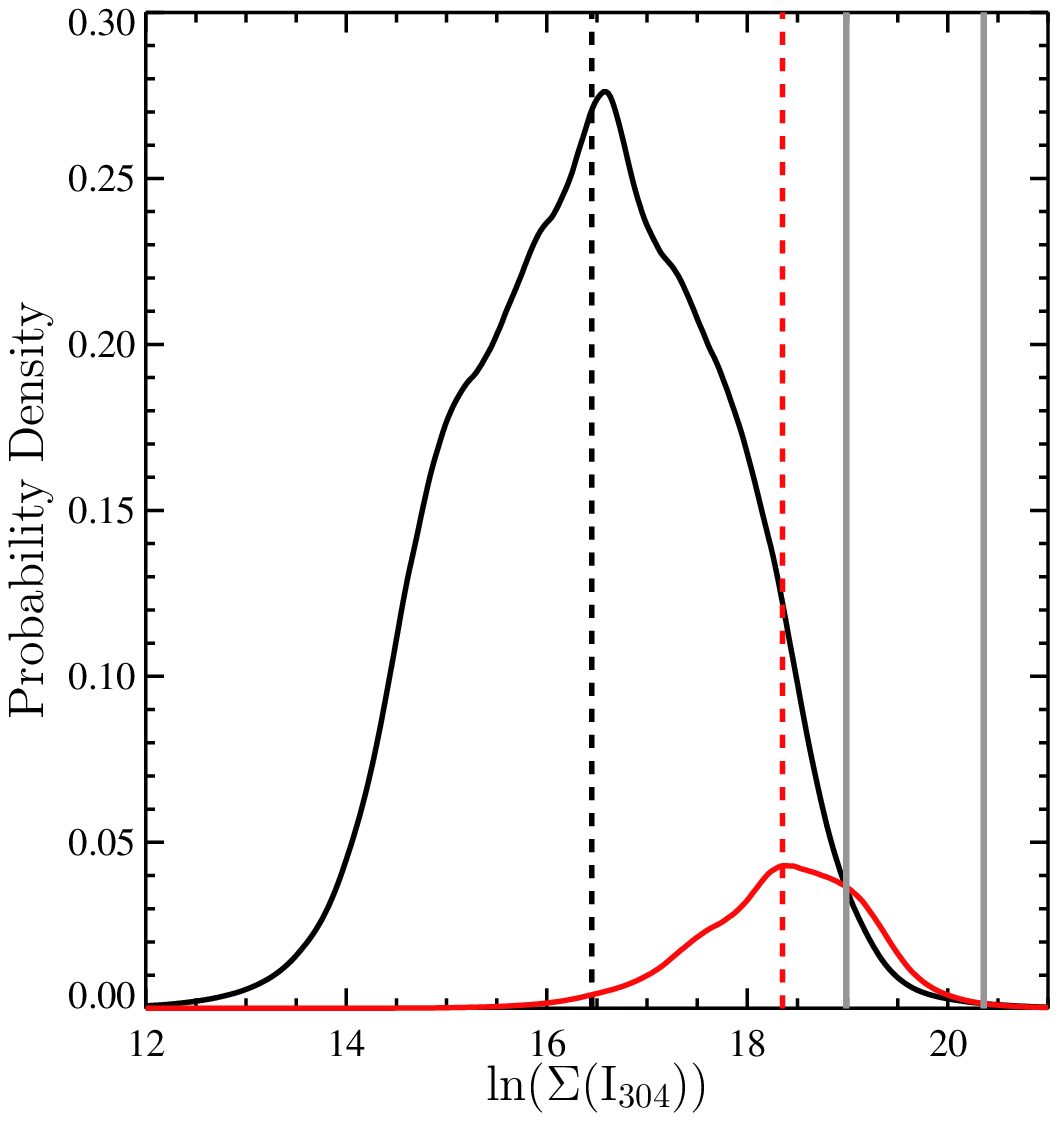}
\includegraphics[width=0.30\textwidth, clip, trim = 0mm 0mm 0mm 0mm]{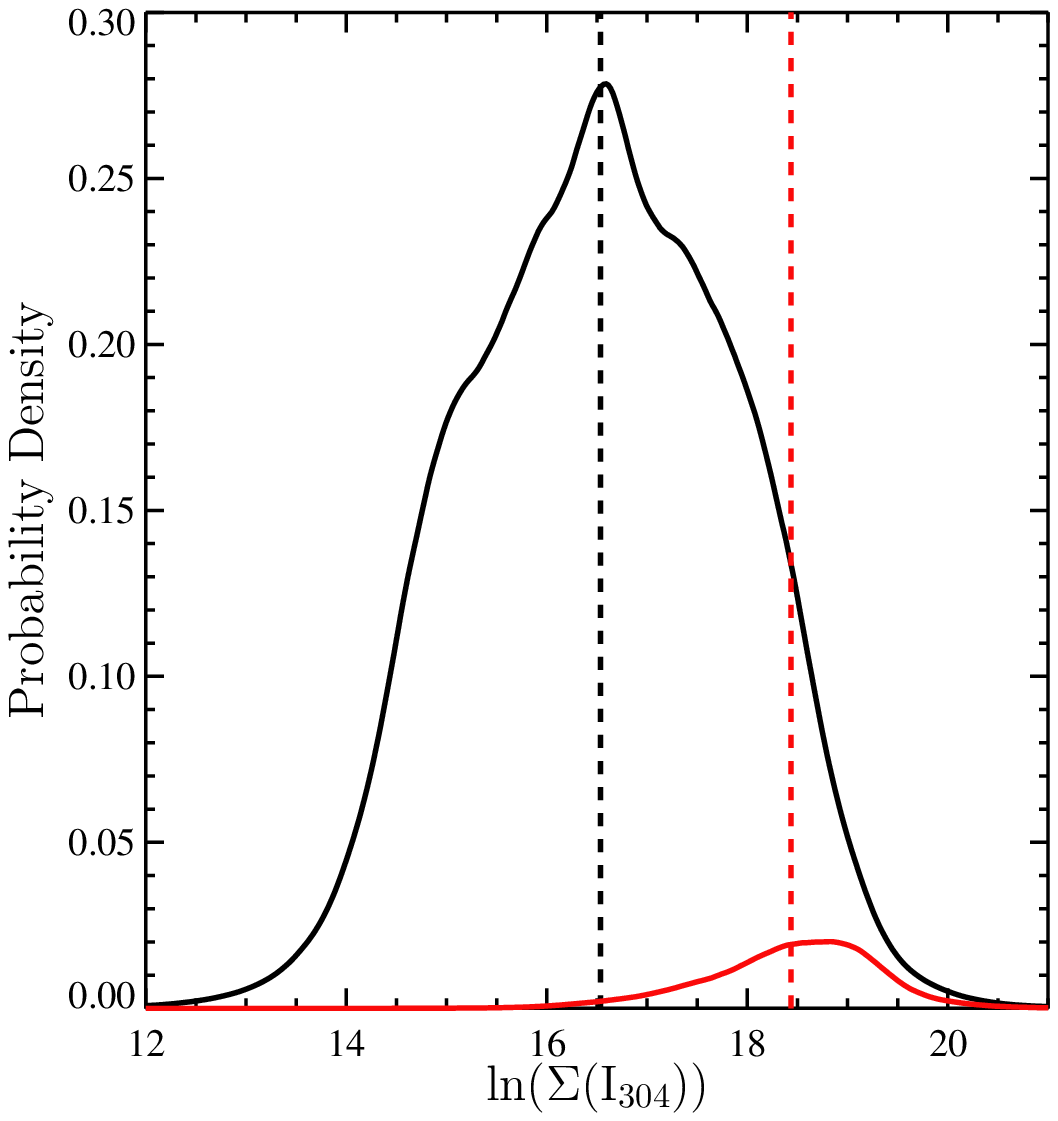}
\includegraphics[width=0.30\textwidth, clip, trim = 0mm 0mm 0mm 0mm]{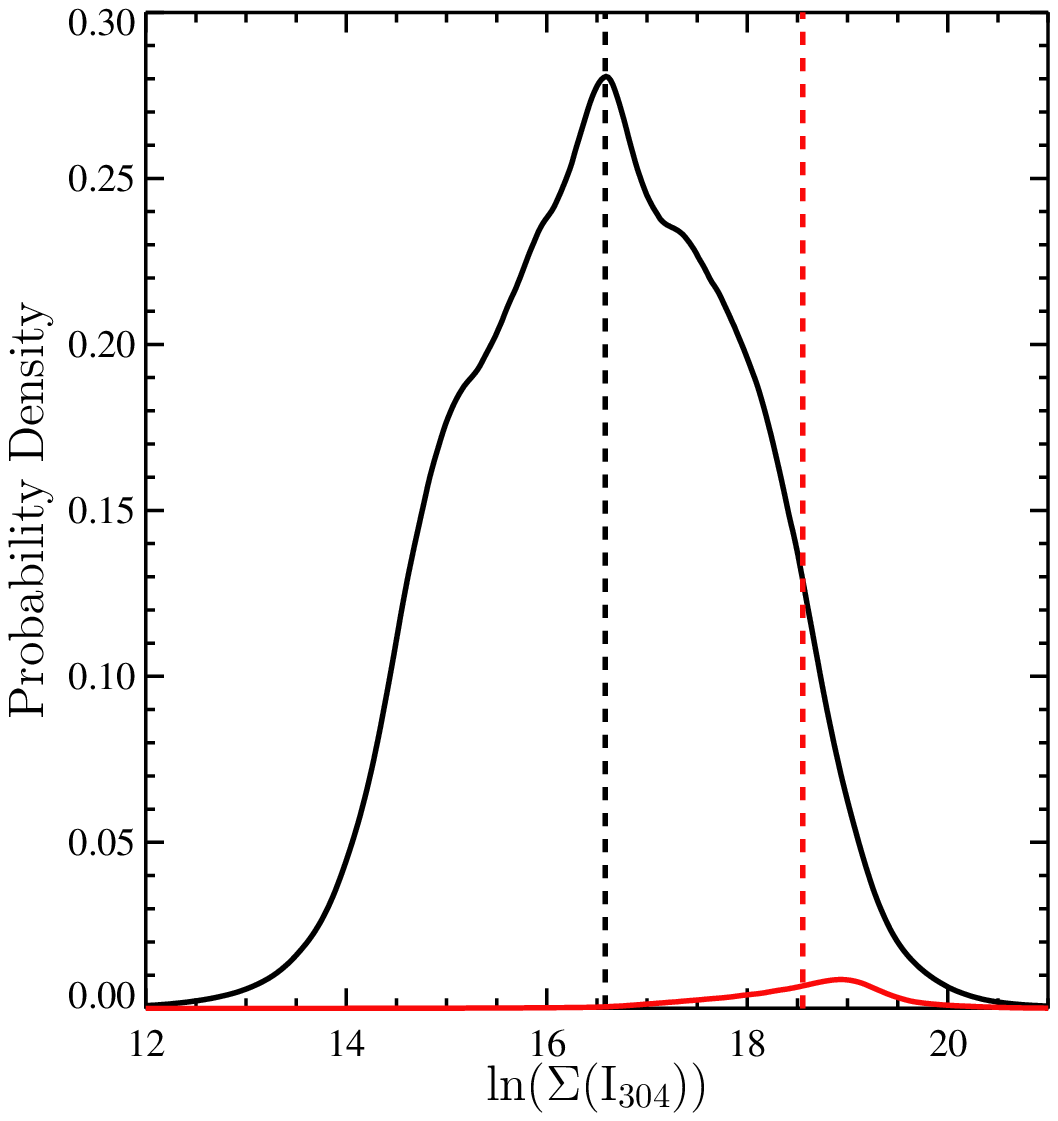}}
\begin{tabular}{|lllll|} \hline
Parameter & Event Definition & \BSS\ & \maxtss\ & $\mathcal{G}$ or \ROCSS\ \\ \hline
$\kappa(\Delta{\rm I_{94}})$ & \CCL\ &  $0.332 \pm 0.011$ & 0.650  & 0.816    \\
  & \CCS\ & $0.247 \pm 0.010$ & 0.703 & 0.853 \\
  & \MML\ & $0.160  \pm 0.015$ & 0.794  & 0.909 \\ \hline
$\Sigma({\rm I_{304}})$ &  \CCL\ & $0.240 \pm 0.014$ & 0.613 & 0.775 \\
  &  \CCS\ & $ 0.167 \pm  0.008$ &  0.589 & 0.747 \\
  & \MML\ & $0.095 \pm 0.015$ & 0.616 & 0.766 \\ \hline
\end{tabular}
\caption{NonParametric Probability Density Functions of two parameters (top) $\kappa(\Delta {\rm I_{94}})$
and (bottom) $\Sigma({\rm I_{304}})$ across three event definitions: \CCL, \CCS, \MML.  Presentation
is the same as Figure~\ref{fig:npda_examples}.  Also shown are the relevant entries
for the performance metrics.  The distribution of the  {\bf {\color{red}{event}}} and 
{\bf non-event} density estimations vary significantly for $\kappa(\Delta {\rm I_{94}})$ 
across event definitions, 
most easily seen by the increase in the mean for the event population, and is 
reflected in changes in their relative \BSS, but for $\Sigma({\rm I_{304}})$ 
the distributions change primarily in amplitude, due to the different
prior probabilities from the different event rates, so the differences in performance in 
particular for \maxtss\ are much less.}
\label{fig:npda_examples2}
\end{figure}

However, there are some variations in this behavior.  There are some
parameters for which the relative distributions (event- {\it vs.}
non-event) vary noticeably with an expected increase in \maxtss\
between, for example, \CC\ and \MM\ definitions - reflecting a shift
to higher parameter values for the event population, for example
(Figure~\ref{fig:npda_examples2}, top), and a relatively smaller
decrease in the \BSS.  For other parameters, the distributions vary
in relative magnitude reflecting the different relative sample sizes,
but the distribution means, for example, do not significantly change
(Figure~\ref{fig:npda_examples2}, bottom).  In this case, the \maxtss\
does not appreciably change because the change in magnitude is offset
by the change in the value of $\mathcal{R}$, and the value of the \BSS\
decreases more substantially.  We found no obvious or systematic behavior
in this regard between parameter ``classes'' (those based on direct {\it
vs.} running-difference images, or static  {\it vs.} $d{\rm X}/d{\rm
t}$ parameters) except that similar parameters often (but not always)
behave the same across wavelengths.

\subsection{Performance Changes with Solar Cycle}
\label{sec:solarcycle}

Solar-cycle-related variations may impact the ability of the parameters
generated here to classify flare-imminent active regions.  The background
UV- and EUV emission \citep{Argiroffi_etal_2008,Schonfeld_etal_2017}
may add a constant to the mean or summation-based parameters,
and varying event rates can change the prior probabilities
\citep{McCloskey_etal_2018,ffc3_1}.  Even running-difference images
may be subject to subtle changes in signal-to-noise ratios due to high
background contamination, potentially impacting
their ability to detect changes in active-region structure.

To examine the behavior of these parameters against cycle-related
influences, we break the data set into two subsets, first with years
that were ``active'' parts of the cycle (2011-2015 inclusive, plus 2017) and ``quiet''
(the rest), based partly on the start of the high-activity time as defined by 
coronal temperature \citep{Schonfeld_etal_2017}, and partly due to flaring rates.
This partitioning provided total sample size of 4898 AARPS (quiet) and 27169 AARPS (active).
We run the full analysis, then look in detail for two very different but originally 
high-scoring parameters, $\Sigma(I_{\rm 94})$ and $\kappa(\Delta I_{\rm 94})$.

\begin{figure}
\centerline{
\includegraphics[width=0.36\textwidth,height=0.30\textwidth, clip, trim = 5mm 5mm 0mm 0mm]{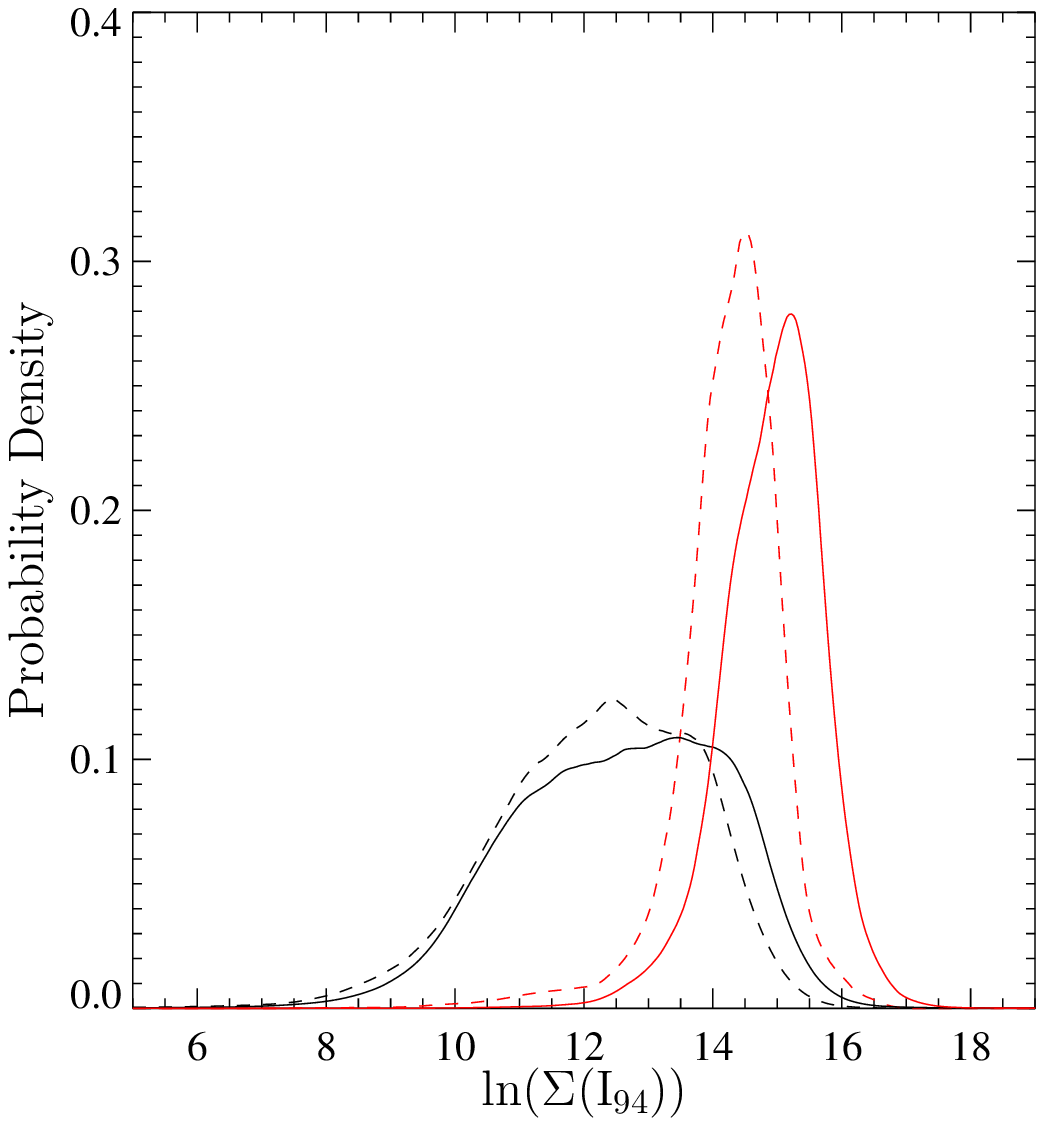}
\includegraphics[width=0.36\textwidth,height=0.30\textwidth, clip, trim = 5mm 5mm 0mm 0mm]{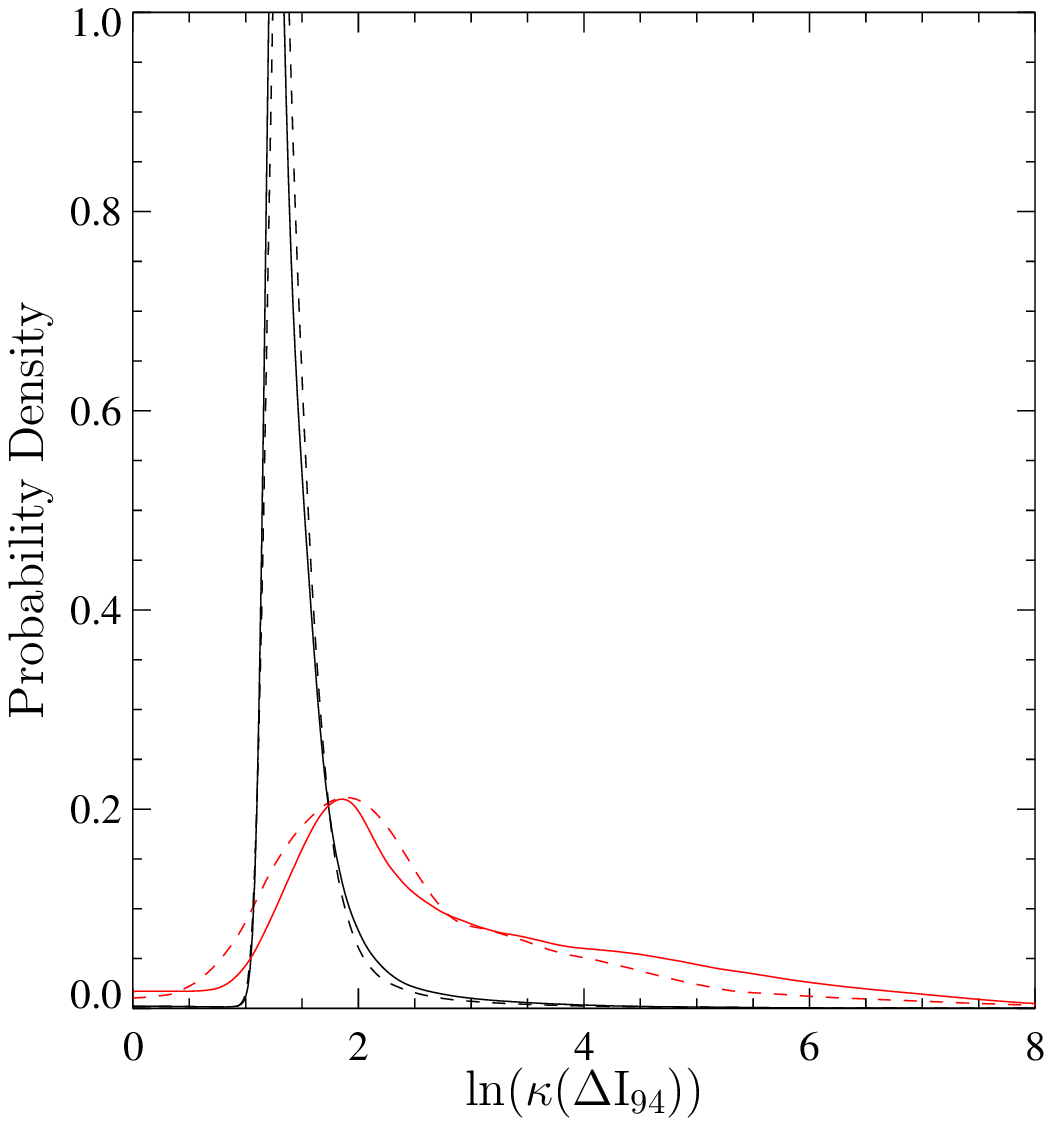}}
\centerline{
\includegraphics[width=0.38\textwidth, clip, trim = 6mm 0mm 0mm 5mm]{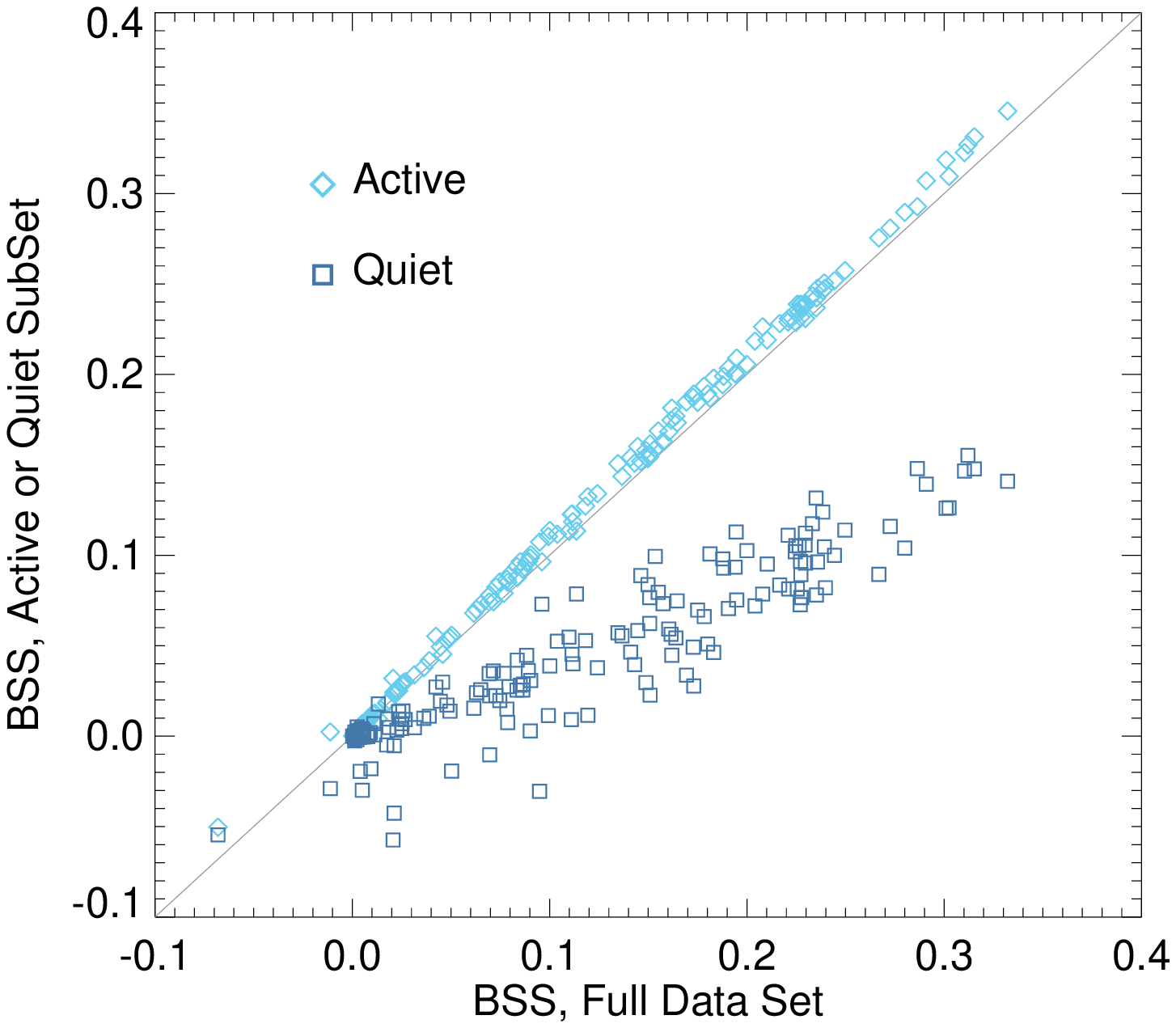}
\includegraphics[width=0.38\textwidth, clip, trim = 6mm 0mm 0mm 5mm]{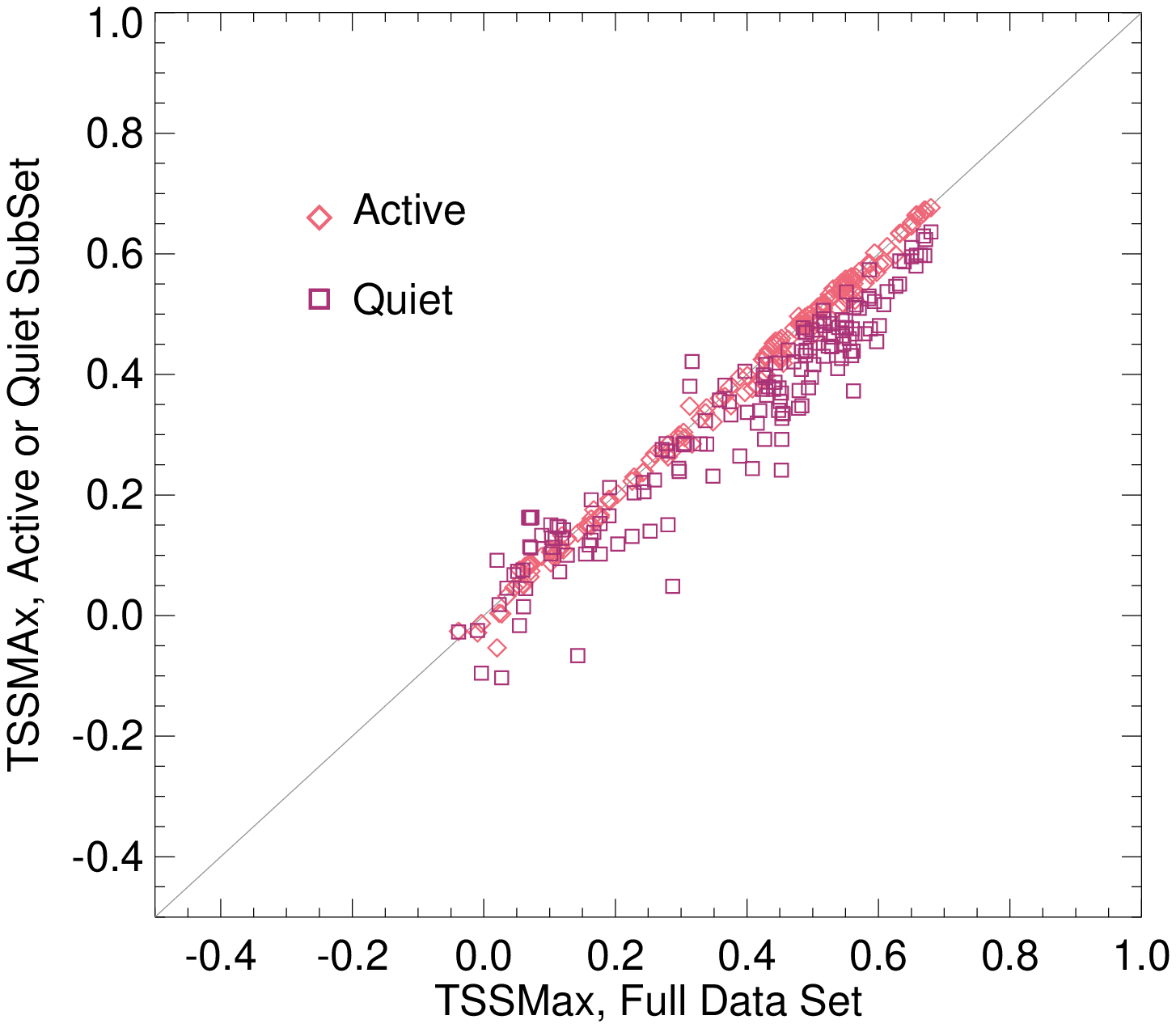}}
\hspace{-1cm}
\begin{tabular}{|llccc|} \hline
Event & Subset & \# Events, & median \%change, & median \%change, \\ 
Definition &  & (Events Rate $\mathcal{R}$) &  vs. full set, \BSS\ & vs. full set, \maxtss\ \\ \hline
\CCL\ & Quiet & 178 (0.036) & -61 & -11 \\
      & Active & 2574 (0.095)  & +6.1 & -0.02 \\ \hline
\end{tabular}
\caption{[Top]: Probability density functions
for \CCL\ {\bf \color{red}{event-}} and {\bf non-event} distributions
comparing the ``quiet'' part of the solar cycle (dashed) with the
``active'' part of the solar cycle (solid).  Equal prior probabilities are
used to highlight differences in the shape of the PDEs, separate from the
changes due to the different event rates.  (Left): $\Sigma(I_{\rm 94})$,
(Right): $\kappa(\Delta I_{\rm 94})$.  [Bottom]:
Comparisons of \BSS\ (left) and \maxtss\ (right) for all parameters,
showing results for the quiet- and active- parts of the cycle (as indicated)
vs. the metrics for the full data set.  
The table summarizes the subset characteristics and the resulting differences 
for two metrics for the \CCL\ event definition.}
\label{fig:cyclediffs}
\end{figure}

The resulting probability density functions for the quiet and active periods for the \CCL\ 
event definition are shown in Figure~\ref{fig:cyclediffs}, using equal prior probabilities 
for clarity.  Overall, we find very little difference in the
distributions between the subsets.  There is a very small shift toward 
higher values for the  $\Sigma(I_{\rm 94})$ parameter during the ``active''
years, but it shifts for both event- and non-event distributions.
There is almost no discernible difference in the distributions for 
$\kappa(\Delta I_{\rm 94})$.

The event rates differ significantly between the active and quiet periods, as designed.
The sample sizes under this division are very small in most cases, leading to
the situation that the adaptive-kernal NPDA is no longer an appropriate model to use.
The quiet period is most susceptible, with the number of events for these years
being: \MML: 15; \CCS: 65; \MMS: 5. These small
numbers mean that for those event definitions, we cannot compare the results for active 
years to those for quiet years with confidence;
hence we concentrate on \CCL\ for the statistical analysis.

In Figure~\ref{fig:cyclediffs} we show scatter plots of the \BSS\
and \maxtss\ for \CCL\ for the quiet and
active periods separately against those scores resulting from the
full dataset.  For the active subset, the difference against
the full dataset is minimal for both metrics.  For the quiet subset however, the \BSS\
shows a strong systematic decrease whereas the \maxtss\ shows scatter
that is, within the expected uncertainties, without significant trend.
Recalling that \BSS\ is sensitive to climatological event rates whereas \maxtss\ is not
\citep{JolliffeStephenson2012}, we demonstrate that the varying event rates 
have a measureable impact on some evaluation metrics.  

Combining this result with the minimal differences in the probability
densities between quiet and active parts of the solar cycle, we conclude
that cycle-related event-rate variations have a much larger impact on the
ability to classify our parametrizations, as measured by some metrics,
than the impact of variation in background emission.

\section{Interpretation}
\label{sec:interp}

Because we construct the parametrizations ourselves, they enable
physical interpretation to the extent allowed by analysis of just the 
images themselves.  The span of regimes sampled, in temperature/density,
height, and temporal dimensions, provides the opportunity to understand
the causes and effects of upper-atmosphere behavior in this context.

\subsection{Temporal Variability}
\label{sec:interp_time}

The parameterizations examine the variability of the corona on two
different time scales.  All of the $\Delta I_*$ parameters look at the
variation in intensity on 72\,s cadence which tracks both small-scale
short-lived brightening events and (dis)appearances and kinematics of
structures including coronal loops.  The moments of the running-difference
images \moments{$\Delta I_*$} further quantify the behavior: increased
or decreased mean indicates a preferential brightening or dimming on
these timescales, or the appearance / disappearance of structures.
The standard deviation indicates the spatial (lack of) quietness.
the skew and kurtosis provide sensitivity to
the far wings of the distributions indicating small-scale dynamics related to 
temperature changes or to kinematic variations.

The \moments{$\Delta I_*$} overwhelmingly dominate the top-10
performing parameters across all event definitions, and in particular
the higher-order moments $\varsigma(\Delta I_*), \kappa(\Delta I_*)$.
The density estimates (see example in Figure~\ref{fig:npda_examples})
show enhanced kurtoses for the event populations relative to the non-event
populations, indicating wing enhancements rather than degradation of the
distribution peaks.  Consistently high kurtosis over the 13\,min indicates
continual presence of rapidly-changing but large-amplitude brightness
fluctuations (see Figure~\ref{fig:aia_params_images}).  In contrast,
the $\mu(\Delta I_*), \sigma(\Delta I_*)$ parameters that should be
sensitive to more subtle variations such as expected from gradual loop
motion, do not generally perform well in \BSS\ although some have notable
$\mathcal{G}$.  Non-activity-related intensity changes as due to gradual
loop motion or gradual loop heating / cooling generally proceed slower
than the cadence here, and additionally involve preferentially larger
(full-loop) structures \citep{ViallKlimchuk2012}.  Hence, there is strong
indication, from multiple parameter results, that enhanced variability
in brightness or enhanced kinematic activity, on short timescales and
small spatial scales, is a discriminating feature of flare-imminent
active regions.

Longer-term evolution is reflected in the slope of the linear fit to the 7
hourly samples (Figure~\ref{fig:aia_params_plots}).  
We find that, for example, for \CCL\
the $d\Sigma(\Delta I_{\rm {94}})/d{\rm t}$ parameter performs well,
and has discriminant boundaries in the wings of the distribution
(Figure~\ref{fig:npda_examples}).  This means that impending activity
is indicated by either a rapid increase or a rapid decrease in the level of
rapid intensity fluctuations in the 94\AA\ filter.  A similar scenario
is found for \MML\ for $d\varsigma(\Delta I_{\rm 94, 131})/d{\rm t}$,
the temporal variation of the skew of the running-difference analysis
from 94\AA\ and 131\AA\ filter images (Table~\ref{tbl:resultsMML}):
on longer timescales, either increasing {\it or decreasing} levels
of short-term brightness variability can indicate upcoming activity
(Table~\ref{tbl:resultsMML}).  In contrast, the 
$d\varsigma(I_{\rm {211}})/d{\rm t}$ parameter for the \CCL\ definition performs very
poorly and in Figure~\ref{fig:npda_examples} it is easy to see why:
the event and non-event density estimates are essentially identical,
apart from the different prior probabilities.

Across the event definitions, parameters describing the evolution on
hours-long timescales (the $d{\rm X} / d{\rm t}$ parameters) are not
generally overall better- or worse- performing than the static parameters.
They do not appear as frequently as would be expected by even chance in
the top-most tiers of performance, but have \BSS\ that are within the
uncertainties of many static parameters, and {\it vice versa}.  In other
words, while in certain cases for certain parameters and certain event
definitions there may be a $d{\rm X} / d{\rm t}$ parameter that shows
promise for relating coronal evolution to imminent flare activity,
there will be at least a few other parameters that do {\it not} track
the evolution but which perform as well.  The results here show a small
preference for static parameters, but we note this may be a result of
outliers rather than a true property of the Sun.

\subsection{The Totals, The Moments}
\label{sec:interp_moments}

The extensive $\Sigma(I_*)$ parameters 
scale with the size of the AARP, whereas the intensive parameters (the moments
\moments{$I_*$}) do not.  We see here that extensive parameters
can perform at least as well as some of the intensive parameters.
In addition to $\Sigma({\rm I_{94}})$ 
being a ``top-10'' discriminator for \CCL\ (see also Figure~\ref{fig:npda_examples}), 
most $\Sigma({\rm I_{\lambda}})$ parameters for EUV wavelengths (meaning, all but 1600\AA)
have high ranking across the event definitions.
The general ability of extensive AIA-based parameters to differentiate
between flare-imminent and flare-quiet groups is consistent with the results of 
numerous prior studies, in particular those based on the 
photospheric magnetic flux (reflecting long-held observers' wisdom)
that, simply put, ``size matters'' \citep[see discussions in][]{flareprediction,dfa,dfa3}.
Larger active regions have
more total emission in the corona and chromosphere (as heating
functions are believed to scale with magnetic flux, \citep[e.g.][]{Warren_etal_2012}), 
and are also the more flare-productive, so this is an example of ``large active region bias''.

However, the $\mu(I_*)$ parameters perform poorly across wavelengths
and event definitions: see for example $\mu({\rm I_{94}})$ for \CCL\ in
Figure~\ref{fig:npda_examples}.  The distributions are distinguishable
(the means are separated), and the ``event'' distribution tends toward
higher values, but there is no discriminant boundary within the bulk of
the data.  Pairing of these results (the $\Sigma(I_*)$ and $\mu(I_*)$
parameter performances), and looking in detail at the distributions,
confirms that while in fact the $\mu(I_*)$ values are higher for the
event populations, it is by not enough so as to provide good predictive
power due to the class imbalance.

In other words, flare-imminent regions are inherently only slightly 
brighter (higher specific intensity) than flare-quiet regions. This
result is a bit surprising, as one might expect that the magnetic
complexity strongly related to flare productivity would produce strong
corona-threading currents available to heat and preferentially brighten
flare-imminent regions significantly over similarly-sized but flare-quiet
sunspot groups \citep[see, {\it e.g.}][]{Asgari-Targhi_etal_2019}.
Such does not appear to be the case.

However, small structures that produce intense brightness variations
are more likely to impact distribution wings.  This can explain
the dominating performances of parameters based on the 
kurtosis of the running-difference distributions ($\kappa(\Delta {\rm I_{\ast}})$;
Tables~\ref{tbl:resultsCCL}-\ref{tbl:resultsMMS} and Figures
\ref{fig:radar_C1_24} --~\ref{fig:radar_M1_6}, see also 
Sections~\ref{sec:bestworst}, \ref{sec:interp_time}). 
The ability of the $\kappa(\Delta I_*)$ parameters to distinguish
flare-imminent from flare-quiet targets indicates that flare-imminent
regions display rapid variability in the E/UV images that is
small in spatial scale, as well.  These results are consistent with an
increased number of small-scale ongoing reconnection events related to
the increased magnetic complexity of these regions relative to AARPs that
are imminently flare-quiet.  

The skew of the running-difference has discriminating power for many
filters and more than one event definition.  Noting that, for example,
171\AA\ filter images are often used to detect coronal loops, we examined the parameter
distributions and in fact the data display both positive and negative
skew, but the positive-skew dominates and is slightly more pronounced
for the event populations.  In this context, the lack of performance
for $\mu(\Delta I_*)$ implies that overall, the brightness changes
on short timescales sum to zero.  Hence, $\varsigma(\Delta I_*)>0$ implies 
a small number of intense brightenings probably combined with a larger number
of less intense dimmings to produce an imbalance in the wings 
of the running-difference brightness distributions. 

Because we do not (yet) analyze the AARP data specifically in the context
of, for example, nearby open magnetic flux, we cannot comment on whether
we are detecting ``crinkles'' specifically \citep{SterlingMoore2001a}
or more generic enhanced small-scale activity.  However, we can conclude
that these results, based on moment analysis of time-series data, is likely only 
available because we are using full-resolution spatial sampling.

\subsection{Wavelength, Temperature, and Physical Regimes}
\label{sec:interp_wave}

The AIA filters do not uniquely sample single temperatures or physical
regimes \citep{aia_Lemen,Warren_etal_2012,Cheung_etal_2015}.  This fact
makes direct interpretation of the parameters in the context of plasma
temperature quite challenging if not potentially misleading, and
obviating the need for, {\it e.g.}, differential emission measure
analysis (forthcoming, see Section~\ref{sec:discussion}).
Still, analysis of the results as a function of filter
(Figures~\ref{fig:radar_C1_24}--\ref{fig:radar_M1_6}) shows patterns of
behavior that are notable in the context of the different regimes that
the filters do sample.

For some filters there is a significant difference in the \BSS\
across event definitions between the $\Sigma(I_*)$ parameters and
the higher-order \moments{$I_*}$ parameters.   This result implies
that the presence of emission is discriminating, but there is no
further information from the spatial distribution of the emission.
This trend is notably present in the  131, 171, 211, and 193 and
335\,\AA\ filter results.  In contrast, for the 94, 304, and 1600
\AA\ filters there is non-negligible performance for the higher-order
\moments{$I_*}$ parameters in addition to the $\Sigma(I_*)$, implying
that distinctive information about the spatial distribution (features)
can be present.  The common theme between the first set of filters is
that they are sensitive to hotter plasma than are the 304, and 1600 \AA\
filters \citep{aia_Lemen,ODwyer_etal_2010}.  These two filters are
not sensitive to flare-temperature plasma and while the 94\AA\ filter
intensity is in fact dominated by hot plasma, it does include a cooler
component \citep{Warren_etal_2012}.

The presence of hot plasma overall may be indicative of {\it past}
flare activity, and we must be reminded that data acquisition is
not separate from flare events (Figure~\ref{fig:events_schematic}).
The presence of a single flare does not usually directly impact (for
example) the inferred longer temporal behavior as parametrized by the
``$d/d{\rm t}$'' variables (see Figure~\ref{fig:aia_params_plots}),
although as mentioned earlier it can supply outlier events.  But the
$d\Sigma(I_*)/d{\rm t}$ parameters stand out as well as the $\Sigma(I_*)$
parameters.  To the extent that the
images in the filters that may be dominated by flare-temperature plasma,
this result signifies that its presence is an indicator of past and,
hence, {\it future} activity.  This result is reminiscent of the strong
performance (reflected in ``observer's wisdom'') of ``persistence'' as
a flare predictor \citep[see discussions in][]{flareprediction,ffc3_2}).

In line with this finding, we also see that parameters 
from the 94 and 131 \AA\ filters, sensitive to hot flare plasma, dominate the
top-performing parameters for the \MM\ event definitions compared to \CC. 
This result implies that there is increased activity/dynamics in hotter plasma prior to \MM\ flares, 
or that larger flares may be produced preferentially after smaller flares have 
energized the corona.  However, we note that as the exact order of the top parameters 
for \MML\ and especially \MMS\ is not very robust given the uncertainties, this 
result only provides a hint at the importance of the hottest channels for 
differentiating larger-flare-imminent regions.

Additionally, in some AIA filters and across event definitions (see
Figures~\ref{fig:radar_C1_24}\,--\,\ref{fig:radar_M1_6}), the mean
intensity $\mu(I_*)$ does not predict between the two populations well,
but the standard deviation $\sigma(I_*)$ does.  The spatial variation
of the brightness is broader (larger standard deviation)
for flare-imminent regions.  For a few filters, notably 94, 304, and 1600\AA,
this disparity extends to the higher-order moments of the intensity distribution,
with notably better performance by $\varsigma(I_*)$ and $\kappa(I_*)$ than
$\mu(I_*)$.  

Two of those latter filters are distinctly {\it not} sensitive
to flare-temperature coronal plasma.  \ion{He}{2} 304\,\AA\
is a relatively cool optically thick line sensitive to the chromosphere / upper
transition region, with a peak temperature response around 
0.05MK, albeit with challenging radiative transfer
characteristics \citep{Golding_etal_2017}.  It samples a different
physical regime than the other filters which image the upper corona
(see Figure~\ref{fig:aia_params_images}), especially in the context
of flares.  The \ion{C}{4} and ``continuum'' 1600\,\AA\ filter samples
the upper photosphere and transition region.  While flare ribbons are
often traced using 1600\,\AA\ emission, that emission is not particularly
hot \citep{Simoes_etal_2019} -- but the brightness in 1600\AA\ filter images
is also sensitive to the presence of magnetic structures and localized
areas of transient heating.  The 94\,\AA\ filter images are generally
dominated by hot active-region core plasma and flare plasma
\citep{aia_Lemen,Cheung_etal_2015}, but include a cooler-plasma
component \citep{Warren_etal_2012}, and additionally have
a notoriously low signal-to-noise ratio.

From all of this we can conclude that there is evidence of a
characteristic difference in the distribution of intensity between
flare-imminent and flare-quiet active regions.  In the high corona,
the features are more likely larger-scale, detectable by the standard
deviation of the distribution, whereas in the upper photosphere,
transition region, and chromosphere, the features are likely to include
smaller-scale features that impact the higher-order moments.

The temporal evolution of the moments of the brightness distributions,
also shows notable differences in patterns between filters that follow
the same trends as outlined above: $d\mu(I_*)/d{\rm t}$ shows 
no predictive capability across wavelength and event definition, 
$d\sigma(I_*)/d{\rm t}$ only for 94, 304 and to a small extent 335\AA,
then $d\varsigma(I_*)/d{\rm t}, d\kappa(I_*)/d{\rm t}$ show predictive
power for 94, 304, and 1600\AA\ but not for the other filters.  Again, this 
implies we detect evolution in the level of variability of small-scale intensity
changes, as could be related to general magnetic complexity and associated
on-going small reconnection events in the transition region and chromosphere.
This variability is not reflected in parameters derived from filters that sample
only hotter plasma, meaning we detect variations that are dominated by larger, less 
impulsively-varying structures.

The overall less-good performance of the 1600\AA\ parameters across event
definitions, specifically the ${\rm d}\varsigma(\Delta I_{\rm 1600})/{\rm
d}t$ and ${\rm d}\kappa(\Delta I_{\rm 1600})/{\rm d}t$ compared to the
strong results for the same parameters from filters that sample
coronal heights and temperatures, strengthens the case that parameters 
using EUV filters detect
small-scale reconnection events.  Such phenomena may be insufficiently
large or energetic enough to produce UV-radiation signatures in the
lower layers of the solar atmosphere.  At the chromospheric height and
temperatures detected in the 304\AA\ channel, however, and the higher
/ hotter channels, these small-scale high-frequency variations are
visible and bring power to differentiating between the populations,
across event definitions.

There are patterns in the results
(Figures~\ref{fig:radar_C1_24}--\ref{fig:radar_M1_6}) which imply that the
filters sensitive to more than one temperature detect different behaviors
from the different physical regimes they sample.  For example, filters
that are predominantly sensitive to active-region plasma temperatures (171,
193, 211\,\AA) show poor performance for $\sigma(I_*)$ whereas 1600\,\AA\
shows moderate performance in that parameter, as do 94\,\AA\ and 304\,\AA.
Similar behavior is seen for $d\kappa(I_{\rm 1600})/d{\rm t}$, whereas
$d\kappa(I_{\rm 171, 193, 211})/d{\rm t}$ show poor performance.
In contrast, the active-region plasma-dominated filters show moderate performance in
$\sigma(\Delta I_{\rm 171, 193, 211})$ whereas $\sigma(\Delta I_{\rm
1600})$ does not.  The 131\,\AA\ filter senses emission from both
flare-relevant \ion{Fe}{18} but also cooler transition-region \ion{Fe}{8};
the 304\,\AA\ line samples a mix of regimes; the 94\,\AA\ filter is
sensitive to the transition-region sensing \ion{Fe}{9}, \ion{Fe}{10}
emission as well as the flare-relevant \ion{Fe}{18}.  The performance
patterns for the 94\,\AA\ filter parameters, as compared to those from
the more selective hot- {\it vs.} cool-sensing filters, confirms that
both flare- and transition-region behaviors are being detected in the
94\,\AA\ filter, especially as we have not corrected for the ``warm''
component \citep[c.f.][]{Warren_etal_2012}.  The dominance of the 94\,\AA\
filter parameters in overall performance shows that multi-regime sampling
may enhance the breadth of information available on the flare-imminent
nature of solar active regions.

This analysis of NPDA results for the AIA filters and the implied
physical regimes they sample is not straightforward, that is very clear.
Rather than pushing the analysis further with regards to physical interpretation,
we acknowledge the need for, e.g., Differential Emission Measure analysis, 
which is beyond the scope of this article.

\section{Discussion} \label{sec:discussion}

We present here a large-sample statistical analysis of the behavior of the
solar chromosphere and corona as deduced from the parametrization of UV and
EUV images from AIA.  We specifically ask how these parametrizations behave
in flare-imminent active regions.  This study complements previous work
that focuses on the photospheric magnetic field \citep{dfa3,nci_daffs};
we find that there is some information available to
statistically, but not uniquely, differentiate between regions that will produce a flare event,
according to various event definitions, from those that will not.

Superficially, the work by
\citet{Nishizuka_etal_2017,Jonas_etal_2018,Alipour_etal_2019} appears
similar to the present study, given their use of AIA data in the context
of flare prediction.  However, there are very important differences.
First and foremost, this is {\it not} a study focused on empirical flare
prediction, but rather we ask whether there are physical characteristics
of flare-imminent active regions as viewed from chromospheric, transition
region, and coronal emission.  The data handling and preparation is
different, performed here with a strong emphasis on ensuring the ability
to perform quantitative physical analysis \citep{aarps}.  Lastly and most
importantly, by constructing the parameters specifically to investigate
physical behavior, including behavior on different temporal scales,
the results can lead to some physical interpretation.

The results show classification performance that varies from ``very
good'' through ``mediocre'' to ``poor'', depending on which combination
of event definition and metric is used.  The \BSS\ is similar to what
is achieved on similar-sized datasets when the question is posed for
parametrizations of the photosphere; this metric provides a summary
of how well the predicted probability for any given target reflects
the frequency of occurrence for other samples with the same measure.
High \BSS\ is extremely difficult to achieve as it is constructed against
the climatology, and class-imbalance -- while inconvenient, is a strong
influence for this metric.  It is a metric based on the probabilities and
thus the true distribution of the parameters, so that the ``mediocre''
and worse scores reflect the fact that substantial differences in
the distributions can be partially offset by the prior probabilities
(Figures~\ref{fig:npda_examples}, \ref{fig:npda_examples2}).

The \maxtss\ results are good, but caution must be used to understand
that this metric is optimized when the probability threshold used (or
incorporated into a cost function, for example) reflects the event
rate, again coming up against the class-imbalance reality of the Sun
\citep{Bloomfield_etal_2012,allclear,Kubo2019}.  Comparing the present
results to the very similar targets (although different latency periods),
sample sizes, and approach in \citet{nci_daffs}, the \maxtss\ results
are similar even though that study invoked multi-parameter NPDA.

The impressive scores here are the \ROCSS\ or $\mathcal{G}$, which summarize
the ROC plots and the correspondence between the value of a parameter,
its associated probability, and whether or not there was a corresponding event.
In this sense, we can definitively say that there is information in the coronal
images that is related to whether or not a region produces a flare event as we
define one, given the parameters we use.  

As the event rate decreases (Table~\ref{tbl:event_defs}), the best \BSS\ values get smaller
while the \maxtss, \ROCSS, and $\mathcal{G}$ values get larger. The
distributions of event-imminent versus event-quiet populations become
increasingly different with lower event rates, which is reflected in the
\maxtss, \ROCSS, and $\mathcal{G}$ values, but this is more than offset by the
increasing class imbalance that enters into the \BSS. Similar behavior is also
present in predictions made from parameters characterizing the photosphere 
\citep{allclear,nci_daffs,ffc3_1}.  Clearly, no single metric provides a
thorough evaluation of performance, and factors such as class imbalance
or event rate must also be considered when interpreting metrics, especially 
those for which thresholds or limits must be set.

We find that enhanced variability in EUV and UV intensity on short timescales
and small spatial scales is one of the strongest discriminators across event
definitions and AIA filters.  This enhancement is most likely of the form
of intense transient brightenings, whether small-scale and localized or 
rapid larger loop movement, rather than gradual loop movement or gradual heating/cooling,
as it preferentially enhances the wings (extremes) of the running-difference
image brightness distributions.  Of note here, spatial resolution matters 
in order for the parametrizations to detect these differences,
and these results validate our approach of retaining the full AIA
spatial sampling across the AARP fields of view \citep{aarps}.

On longer timescales, strong increases
(or decreases) in brightness moderately indicates impending flaring,
and while overall the presence of hot plasma is a good indicator, this
result is also consistent with the general correlation between active
region size and flare productivity.  The evolution of parameters describing
the corona can provide flare-imminent indicators, but with little preference
over ``static'' parameters. 

Of note, while coronal loop structures are readily detected through
an analysis of the spatial variations of emission in the 171, 211\,\AA\
filters.  the quantitative measure of these spatial variations ({\it e.g.}
$\sigma(I_{171,211})$) is not a good discriminator.  Also surprisingly
poor is the mean intensity and its longer-term trending, which implies
that there is minimal significant difference between magnetically complex
and magnetically simple active regions in terms of their average coronal
brightness and its temporal variation.

The differences in coronal, transition-region, and chromospheric E/UV
emission between flare-imminent and not-flare-imminent active regions has
broad implications for models of active-regions overall, and their upper
atmospheres in particular.  The approach outlined here and these results
provide constraints on the expected emission and kinematic behavior of
pre-event (and even post-event) active region upper atmospheres.

As pointed out in Section~\ref{sec:interp_wave}, simply analyzing the
behavior of the brightness and kinematics in AIA filters is tricky due
to their multi-thermal sensitivity.  We address this in an upcoming work
that uses differential emission measure analysis to disentangle densities
and temperatures across this AARP database \citep{nci_dem}.  Similarly,
a more complete picture will be built as we combine the AARP database
with the HARP magnetic field inputs; as of this work we simply begin the
process of statistically understanding the behavior of the chromospheric, 
transition region, and coronal regimes in the context of flare events using large-sample
data finally afforded by high-resolution continual imagery from \textit{SDO}/AIA.

\begin{acknowledgments}

The authors thank the referee for a thorough reading and insightful feedback
that helped improve the paper.
This work was made possible by funding primarily from NASA/GI Grant 80NSSC19K0285
with some initial exploration through AFRL SBIR Phase-I contract FA9453-14-M-0170,
and some final support from NASA/GI Grant 80NSSC21K0738 and NSF/AGS-ST Grant 2154653. 

\end{acknowledgments}

\facility{SDO (HMI and AIA), GOES (XRS)}
\software{SolarSoft \citep{solarsoft}, NCI \citep{nci_daffs}}

\bibliography{ms}{}

\begin{thebibliography}{}
\expandafter\ifx\csname natexlab\endcsname\relax\def\natexlab#1{#1}\fi
\providecommand{\url}[1]{\href{#1}{#1}}
\providecommand{\dodoi}[1]{doi:~\href{http://doi.org/#1}{\nolinkurl{#1}}}
\providecommand{\doeprint}[1]{\href{http://ascl.net/#1}{\nolinkurl{http://ascl.net/#1}}}
\providecommand{\doarXiv}[1]{\href{https://arxiv.org/abs/#1}{\nolinkurl{https://arxiv.org/abs/#1}}}

\bibitem[{{Alipour} {et~al.}(2019){Alipour}, {Mohammadi}, \&
  {Safari}}]{Alipour_etal_2019}
{Alipour}, N., {Mohammadi}, F., \& {Safari}, H. 2019, \apjs, 243, 20,
  \dodoi{10.3847/1538-4365/ab289b}

\bibitem[{{Argiroffi} {et~al.}(2008){Argiroffi}, {Peres}, {Orlando}, \&
  {Reale}}]{Argiroffi_etal_2008}
{Argiroffi}, C., {Peres}, G., {Orlando}, S., \& {Reale}, F. 2008, \aap, 488,
  1069, \dodoi{10.1051/0004-6361:200809355}

\bibitem[{{Asgari-Targhi} {et~al.}(2019){Asgari-Targhi}, {van Ballegooijen}, \&
  {Davey}}]{Asgari-Targhi_etal_2019}
{Asgari-Targhi}, M., {van Ballegooijen}, A.~A., \& {Davey}, A.~R. 2019, \apj,
  881, 107, \dodoi{10.3847/1538-4357/ab2e01}

\bibitem[{{Bamba} {et~al.}(2014){Bamba}, {Kusano}, {Imada}, \&
  {Iida}}]{Bamba_etal_2014}
{Bamba}, Y., {Kusano}, K., {Imada}, S., \& {Iida}, Y. 2014, \pasj, 66, S16,
  \dodoi{10.1093/pasj/psu091}

\bibitem[{{Barnes} {et~al.}(2014){Barnes}, {Birch}, {Leka}, \&
  {Braun}}]{trt_emerge3}
{Barnes}, G., {Birch}, A.~C., {Leka}, K.~D., \& {Braun}, D.~C. 2014, \apj, 786,
  19, \dodoi{10.1088/0004-637X/786/1/19}

\bibitem[{{Barnes} \& {Leka}(2006)}]{dfa2}
{Barnes}, G., \& {Leka}, K.~D. 2006, \apj, 646, 1303, \dodoi{10.1086/504960}

\bibitem[{{Barnes} {et~al.}(2007){Barnes}, {Leka}, {Schumer}, \&
  {Della-Rose}}]{SWJ}
{Barnes}, G., {Leka}, K.~D., {Schumer}, E.~A., \& {Della-Rose}, D.~J. 2007,
  Space Weather, 5, 9002, \dodoi{10.1029/2007SW000317}

\bibitem[{{Barnes} {et~al.}(2016){Barnes}, {Leka}, {Schrijver}, {Colak},
  {Qahwaji}, {Ashamari}, {Yuan}, {Zhang}, {McAteer}, {Bloomfield}, {Higgins},
  {Gallagher}, {Falconer}, {Georgoulis}, {Wheatland}, {Balch}, {Dunn}, \&
  {Wagner}}]{allclear}
{Barnes}, G., {Leka}, K.~D., {Schrijver}, C.~J., {et~al.} 2016, \apj, 829, 89,
  \dodoi{10.3847/0004-637X/829/2/89}

\bibitem[{{Bloomfield} {et~al.}(2012){Bloomfield}, {Higgins}, {McAteer}, \&
  {Gallagher}}]{Bloomfield_etal_2012}
{Bloomfield}, D.~S., {Higgins}, P.~A., {McAteer}, R.~T.~J., \& {Gallagher},
  P.~T. 2012, \apjl, 747, L41, \dodoi{10.1088/2041-8205}

\bibitem[{{Bobra} \& {Couvidat}(2015)}]{BobraCouvidat2015}
{Bobra}, M.~G., \& {Couvidat}, S. 2015, \apj, 798, 135,
  \dodoi{10.1088/0004-637X/798/2/135}

\bibitem[{{Bobra} {et~al.}(2014){Bobra}, {Sun}, {Hoeksema}, {Turmon}, {Liu},
  {Hayashi}, {Barnes}, \& {Leka}}]{hmi_sharps}
{Bobra}, M.~G., {Sun}, X., {Hoeksema}, J.~T., {et~al.} 2014, \solphys, 289,
  3549, \dodoi{10.1007/s11207-014-0529-3}

\bibitem[{{Breiman}(2001)}]{Breiman2001}
{Breiman}, L. 2001, Machine Learning, 45, 5, \dodoi{10.1023/A:1010933404324}

\bibitem[{{Cheung} {et~al.}(2015){Cheung}, {Boerner}, {Schrijver}, {Testa},
  {Chen}, {Peter}, \& {Malanushenko}}]{Cheung_etal_2015}
{Cheung}, M.~C.~M., {Boerner}, P., {Schrijver}, C.~J., {et~al.} 2015, \apj,
  807, 143, \dodoi{10.1088/0004-637X/807/2/143}

\bibitem[{{Cho} {et~al.}(2016){Cho}, {Lee}, {Chae}, {Wang}, {Ahn}, {Yang},
  {Lim}, \& {Maurya}}]{Cho_etal_2016}
{Cho}, K., {Lee}, J., {Chae}, J., {et~al.} 2016, \solphys, 291, 2391,
  \dodoi{10.1007/s11207-016-0963-5}

\bibitem[{{Cinto} {et~al.}(2020){Cinto}, {Gradvohl}, {Coelho}, \& {da
  Silva}}]{Cinto_etal_2020}
{Cinto}, T., {Gradvohl}, A.~L.~S., {Coelho}, G.~P., \& {da Silva}, A.~E.~A.
  2020, \solphys, 295, 93, \dodoi{10.1007/s11207-020-01661-9}

\bibitem[{{Dissauer} {et~al.}(2022{\natexlab{a}}){Dissauer}, {Leka}, {Barnes},
  \& {Wagner}}]{nci_dem}
{Dissauer}, K., {Leka}, K.~D., {Barnes}, G., \& {Wagner}, E.~L.
  2022{\natexlab{a}}, \apj, in preparation

\bibitem[{{Dissauer} {et~al.}(2022{\natexlab{b}}){Dissauer}, {Leka}, \&
  {Wagner}}]{aarps}
{Dissauer}, K., {Leka}, K.~D., \& {Wagner}, E.~L. 2022{\natexlab{b}}, \apj,
  accepted

\bibitem[{{Dissauer} {et~al.}(2022{\natexlab{c}}){Dissauer}, {Leka}, \&
  {Wagner}}]{aarp_data}
---. 2022{\natexlab{c}}, The NWRA AIA Active Region Patch Database, V1,
  \dodoi{TBD}

\bibitem[{{Efron} \& {Gong}(1983)}]{EfronGong1983}
{Efron}, B., \& {Gong}, G. 1983, Am. Stat., 37, 36

\bibitem[{{Freeland} \& {Handy}(1998)}]{solarsoft}
{Freeland}, S.~L., \& {Handy}, B.~N. 1998, \solphys, 182, 497,
  \dodoi{10.1023/A:1005038224881}

\bibitem[{{Garcia}(1994)}]{goes_xrs}
{Garcia}, H.~A. 1994, \solphys, 154, 275, \dodoi{10.1007/BF00681100}

\bibitem[{{Georgoulis} {et~al.}(2021){Georgoulis}, {Bloomfield}, {Piana},
  {Massone}, {Soldati}, {Gallagher}, {Pariat}, {Vilmer}, {Buchlin}, {Baudin},
  {Csillaghy}, {Sathiapal}, {Jackson}, {Alingery}, {Benvenuto}, {Campi},
  {Florios}, {Gontikakis}, {Guennou}, {Guerra}, {Kontogiannis}, {Latorre},
  {Murray}, {Park}, {von Stachelski}, {Torbica}, {Vischi}, \&
  {Worsfold}}]{flarecast}
{Georgoulis}, M.~K., {Bloomfield}, D.~S., {Piana}, M., {et~al.} 2021, Journal
  of Space Weather and Space Climate, 11, 39, \dodoi{10.1051/swsc/2021023}

\bibitem[{{Golding} {et~al.}(2017){Golding}, {Leenaarts}, \&
  {Carlsson}}]{Golding_etal_2017}
{Golding}, T.~P., {Leenaarts}, J., \& {Carlsson}, M. 2017, \aap, 597, A102,
  \dodoi{10.1051/0004-6361/201629462}

\bibitem[{{Harra} {et~al.}(2013){Harra}, {Matthews}, {Culhane}, {Cheung},
  {Kontar}, \& {Hara}}]{Harra_etal_2013}
{Harra}, L.~K., {Matthews}, S., {Culhane}, J.~L., {et~al.} 2013, \apj, 774,
  122, \dodoi{10.1088/0004-637X/774/2/122}

\bibitem[{{Hills}(1966)}]{hill66}
{Hills}, M. 1966, J.~R.~Statist.~Soc.~B, 28, 1

\bibitem[{{Hoeksema} {et~al.}(2014){Hoeksema}, {Liu}, {Hayashi}, {Sun},
  {Schou}, {Couvidat}, {Norton}, {Bobra}, {Centeno}, {Leka}, {Barnes}, \&
  {Turmon}}]{hmi_pipe}
{Hoeksema}, J.~T., {Liu}, Y., {Hayashi}, K., {et~al.} 2014, \solphys, 289,
  3483, \dodoi{10.1007/s11207-014-0516-8}

\bibitem[{{Imada} {et~al.}(2014){Imada}, {Bamba}, \&
  {Kusano}}]{ImadaBambaKusano2014}
{Imada}, S., {Bamba}, Y., \& {Kusano}, K. 2014, \pasj, 66, S17,
  \dodoi{10.1093/pasj/psu092}

\bibitem[{{Jolliffe} \& {Stephenson}(2012)}]{JolliffeStephenson2012}
{Jolliffe}, I.~T., \& {Stephenson}, D. 2012, Forecast Verification: A
  Practioner's Guide in Atmospheric Science, 2nd Edition (The Atrium, Southern
  Gate, Chichester, West Sussex PO19 8SQ, England: Wiley),
  \dodoi{10.1002/9781119960003}

\bibitem[{{Jonas} {et~al.}(2018){Jonas}, {Bobra}, {Shankar}, {Todd Hoeksema},
  \& {Recht}}]{Jonas_etal_2018}
{Jonas}, E., {Bobra}, M., {Shankar}, V., {Todd Hoeksema}, J., \& {Recht}, B.
  2018, \solphys, 293, \#48, \dodoi{10.1007/s11207-018-1258-9}

\bibitem[{{Joshi} {et~al.}(2011){Joshi}, {Veronig}, {Lee}, {Bong}, {Tiwari}, \&
  {Cho}}]{Joshi_etal_2011}
{Joshi}, B., {Veronig}, A.~M., {Lee}, J., {et~al.} 2011, \apj, 743, 195,
  \dodoi{10.1088/0004-637X/743/2/195}

\bibitem[{{Krista} \& {Chih}(2021)}]{deft}
{Krista}, L.~D., \& {Chih}, M. 2021, \apj, 922, 218,
  \dodoi{10.3847/1538-4357/ac2840}

\bibitem[{{Kubo}(2019)}]{Kubo2019}
{Kubo}, Y. 2019, Journal of Space Weather and Space Climate, 9, A17,
  \dodoi{10.1051/swsc/2019016}

\bibitem[{{Leka} \& {Barnes}(2003{\natexlab{a}})}]{params}
{Leka}, K.~D., \& {Barnes}, G. 2003{\natexlab{a}}, \apj, 595, 1277

\bibitem[{{Leka} \& {Barnes}(2003{\natexlab{b}})}]{dfa}
---. 2003{\natexlab{b}}, \apj, 595, 1296

\bibitem[{{Leka} \& {Barnes}(2007)}]{dfa3}
---. 2007, \apj, 656, 1173, \dodoi{10.1086/510282}

\bibitem[{{Leka} {et~al.}(2018){Leka}, {Barnes}, \& {Wagner}}]{nci_daffs}
{Leka}, K.~D., {Barnes}, G., \& {Wagner}, E.~L. 2018, Journal of Space Weather
  and Space Climate, 8, A25, \dodoi{10.1051/swsc/2018004}

\bibitem[{{Leka} {et~al.}(2022){Leka}, {Dissauer}, {Barnes}, \&
  {Wagner}}]{nci_aia_data}
{Leka}, K.~D., {Dissauer}, K., {Barnes}, G., \& {Wagner}, E.~L. 2022,
  {Replication Data for Properties of Flare-Imminent versus Flare-Quiet Active
  Regions from the Chromosphere through the Corona II: NonParametric
  Discriminant Analysis Results from NCI}, v1,  Harvard Dataverse,
  \dodoi{10.7910/DVN/WPN39J}

\bibitem[{{Leka} {et~al.}(2019{\natexlab{a}}){Leka}, {Park}, {Kusano},
  {Andries}, {Balch}, {Barnes}, {Bingham}, {Bloomfield}, {McCloskey},
  {Delouille}, {Falconer}, {Gallagher}, {Georgoulis}, {Hamad Nageem}, {Kubo},
  {Lee}, {Lee}, {Lobzin}, {Mun}, {Murray}, {Qahwaji}, {Sharpe}, {Steenburgh},
  {Steward}, \& {Terkildsen}}]{ffc3_1}
{Leka}, K.~D., {Park}, S.~H., {Kusano}, K., {et~al.} 2019{\natexlab{a}}, \apjs,
  243, 36, \dodoi{10.3847/1538-4365/ab2e12}

\bibitem[{{Leka} {et~al.}(2019{\natexlab{b}}){Leka}, {Park}, {Kusano},
  {Andries}, {Balch}, {Barnes}, {Bingham}, {Bloomfield}, {McCloskey},
  {Delouille}, {Falconer}, {Gallagher}, {Georgoulis}, {Hamad Nageem}, {Kubo},
  {Lee}, {Lee}, {Lobzin}, {Mun}, {Murray}, {Qahwaji}, {Sharpe}, {Steenburgh},
  {Steward}, \& {Terkildsen}}]{ffc3_2}
---. 2019{\natexlab{b}}, \apj, 881, 101, \dodoi{10.3847/1538-4357/ab2e11}

\bibitem[{{Lemen} {et~al.}(2012){Lemen}, {Title}, {Akin}, {Boerner}, {Chou},
  {Drake}, {Duncan}, {Edwards}, {Friedlaender}, {Heyman}, {Hurlburt}, {Katz},
  {Kushner}, {Levay}, {Lindgren}, {Mathur}, {McFeaters}, {Mitchell}, {Rehse},
  {Schrijver}, {Springer}, {Stern}, {Tarbell}, {Wuelser}, {Wolfson}, {Yanari},
  {Bookbinder}, {Cheimets}, {Caldwell}, {Deluca}, {Gates}, {Golub}, {Park},
  {Podgorski}, {Bush}, {Scherrer}, {Gummin}, {Smith}, {Auker}, {Jerram},
  {Pool}, {Soufli}, {Windt}, {Beardsley}, {Clapp}, {Lang}, \&
  {Waltham}}]{aia_Lemen}
{Lemen}, J.~R., {Title}, A.~M., {Akin}, D.~J., {et~al.} 2012, \solphys, 275,
  17, \dodoi{10.1007/s11207-011-9776-8}

\bibitem[{{Li} {et~al.}(2005){Li}, {Mickey}, \& {LaBonte}}]{Li_etal_2005}
{Li}, J., {Mickey}, D.~L., \& {LaBonte}, B.~J. 2005, \apj, 620, 1092,
  \dodoi{10.1086/427205}

\bibitem[{Machol(2022)}]{Machol_2022_PC}
Machol, J. 2022

\bibitem[{{Mason} \& {Hoeksema}(2010)}]{MasonHoeksema2010}
{Mason}, J.~P., \& {Hoeksema}, J.~T. 2010, \apj, 723, 634,
  \dodoi{10.1088/0004-637X/723/1/634}

\bibitem[{{McCloskey} {et~al.}(2018){McCloskey}, {Gallagher}, \&
  {Bloomfield}}]{McCloskey_etal_2018}
{McCloskey}, A.~E., {Gallagher}, P.~T., \& {Bloomfield}, D.~S. 2018, Journal of
  Space Weather and Space Climate, 8, A34, \dodoi{10.1051/swsc/2018022}

\bibitem[{{Nishizuka} {et~al.}(2017){Nishizuka}, {Sugiura}, {Kubo}, {Den},
  {Watari}, \& {Ishii}}]{Nishizuka_etal_2017}
{Nishizuka}, N., {Sugiura}, K., {Kubo}, Y., {et~al.} 2017, \apj, 835, 156,
  \dodoi{10.3847/1538-4357/835/2/156}

\bibitem[{{O'Dwyer} {et~al.}(2010){O'Dwyer}, {Del Zanna}, {Mason}, {Weber}, \&
  {Tripathi}}]{ODwyer_etal_2010}
{O'Dwyer}, B., {Del Zanna}, G., {Mason}, H.~E., {Weber}, M.~A., \& {Tripathi},
  D. 2010, \aap, 521, A21, \dodoi{10.1051/0004-6361/201014872}

\bibitem[{{Panos} \& {Kleint}(2020)}]{PanosKleint2020}
{Panos}, B., \& {Kleint}, L. 2020, \apj, 891, 17,
  \dodoi{10.3847/1538-4357/ab700b}

\bibitem[{{Pesnell} {et~al.}(2012){Pesnell}, {Thompson}, \& {Chamberlin}}]{sdo}
{Pesnell}, W.~D., {Thompson}, B.~J., \& {Chamberlin}, P.~C. 2012, \solphys,
  275, 3, \dodoi{10.1007/s11207-011-9841-3}

\bibitem[{{Plowman}(2016)}]{Plowman2016}
{Plowman}, J. 2016, Journal of Space Weather and Space Climate, 6, A8,
  \dodoi{10.1051/swsc/2016002}

\bibitem[{{Qiu} \& {Cheng}(2017)}]{QiuCheng2017}
{Qiu}, J., \& {Cheng}, J. 2017, \apjl, 838, L6,
  \dodoi{10.3847/2041-8213/aa6798}

\bibitem[{{Raboonik} {et~al.}(2017){Raboonik}, {Safari}, {Alipour}, \&
  {Wheatland}}]{Raboonik_etal_2017}
{Raboonik}, A., {Safari}, H., {Alipour}, N., \& {Wheatland}, M.~S. 2017, \apj,
  834, 11, \dodoi{10.3847/1538-4357/834/1/11}

\bibitem[{{Sawyer} {et~al.}(1986){Sawyer}, {Warwick}, \&
  {Dennett}}]{flareprediction}
{Sawyer}, C., {Warwick}, J.~W., \& {Dennett}, J.~T. 1986, {Solar Flare
  Prediction} (Boulder, CO: Colorado Assoc. Univ. Press)

\bibitem[{{Scherrer} {et~al.}(2012){Scherrer}, {Schou}, {Bush}, {Kosovichev},
  {Bogart}, {Hoeksema}, {Liu}, {Duvall}, {Zhao}, {Title}, {Schrijver},
  {Tarbell}, \& {Tomczyk}}]{hmi}
{Scherrer}, P.~H., {Schou}, J., {Bush}, R.~I., {et~al.} 2012, \solphys, 275,
  207, \dodoi{10.1007/s11207-011-9834-2}

\bibitem[{{Schonfeld} {et~al.}(2017){Schonfeld}, {White}, {Hock-Mysliwiec}, \&
  {McAteer}}]{Schonfeld_etal_2017}
{Schonfeld}, S.~J., {White}, S.~M., {Hock-Mysliwiec}, R.~A., \& {McAteer},
  R.~T.~J. 2017, \apj, 844, 163, \dodoi{10.3847/1538-4357/aa7b35}

\bibitem[{{Seki} {et~al.}(2017){Seki}, {Otsuji}, {Isobe}, {Ishii}, {Sakaue}, \&
  {Hirose}}]{Seki_etal_2017}
{Seki}, D., {Otsuji}, K., {Isobe}, H., {et~al.} 2017, \apjl, 843, L24,
  \dodoi{10.3847/2041-8213/aa7559}

\bibitem[{{Silverman}(1986)}]{Silverman86}
{Silverman}, B.~W. 1986, {Density Estimation for Statistics and Data Analysis}
  (London: Chapman and Hall)

\bibitem[{{Sim{\~o}es} {et~al.}(2019){Sim{\~o}es}, {Reid}, {Milligan}, \&
  {Fletcher}}]{Simoes_etal_2019}
{Sim{\~o}es}, P. J.~A., {Reid}, H. A.~S., {Milligan}, R.~O., \& {Fletcher}, L.
  2019, \apj, 870, 114, \dodoi{10.3847/1538-4357/aaf28d}

\bibitem[{{Sterling} \& {Moore}(2001{\natexlab{a}})}]{SterlingMoore2001b}
{Sterling}, A.~C., \& {Moore}, R.~L. 2001{\natexlab{a}}, \apj, 560, 1045,
  \dodoi{10.1086/322241}

\bibitem[{{Sterling} \& {Moore}(2001{\natexlab{b}})}]{SterlingMoore2001a}
---. 2001{\natexlab{b}}, \jgr, 106, 25227

\bibitem[{{Sterling} {et~al.}(2011){Sterling}, {Moore}, \&
  {Freeland}}]{SterlingMooreFreeland2011}
{Sterling}, A.~C., {Moore}, R.~L., \& {Freeland}, S.~L. 2011, \apjl, 731, L3,
  \dodoi{10.1088/2041-8205/731/1/L3}

\bibitem[{{Viall} \& {Klimchuk}(2012)}]{ViallKlimchuk2012}
{Viall}, N.~M., \& {Klimchuk}, J.~A. 2012, \apj, 753, 35,
  \dodoi{10.1088/0004-637X/753/1/35}

\bibitem[{{Viereck} \& {Machol}(2017)}]{2017AGUFMSH42A..06V}
{Viereck}, R.~A., \& {Machol}, J.~L. 2017, in AGU Fall Meeting Abstracts, Vol.
  2017, SH42A--06

\bibitem[{{Warren} {et~al.}(2012){Warren}, {Winebarger}, \&
  {Brooks}}]{Warren_etal_2012}
{Warren}, H.~P., {Winebarger}, A.~R., \& {Brooks}, D.~H. 2012, \apj, 759, 141,
  \dodoi{10.1088/0004-637X/759/2/141}

\bibitem[{{Welsch} {et~al.}(2009){Welsch}, {Li}, {Schuck}, \&
  {Fisher}}]{Welsch_etal_2009}
{Welsch}, B.~T., {Li}, Y., {Schuck}, P.~W., \& {Fisher}, G.~H. 2009, \apj, 705,
  821, \dodoi{10.1088/0004-637X/705/1/821}

\bibitem[{{Woods} {et~al.}(2017){Woods}, {Harra}, {Matthews}, {Mackay},
  {Dacie}, \& {Long}}]{Woods_etal_2017}
{Woods}, M.~M., {Harra}, L.~K., {Matthews}, S.~A., {et~al.} 2017, \solphys,
  292, 38, \dodoi{10.1007/s11207-017-1064-9}

\bibitem[{{Zhang} {et~al.}(2017){Zhang}, {Su}, \& {Ji}}]{Zhang_etal_2017}
{Zhang}, Q.~M., {Su}, Y.~N., \& {Ji}, H.~S. 2017, \aap, 598, A3,
  \dodoi{10.1051/0004-6361/201629477}

\end{thebibliography}
\bibliographystyle{aasjournal}


\end{document}